\documentclass{elsart}

\usepackage[square,comma]{natbib}
\usepackage{graphicx}
\usepackage{pxfonts}
\usepackage{amssymb}

\journal{}

\begin{document}

\thispagestyle{empty}
\begin{Large}
\textbf{DEUTSCHES ELEKTRONEN-SYNCHROTRON}

\textbf{\large{in der HELMHOLTZ-GEMEINSCHAFT}\\}
\end{Large}

DESY 07-031

March 2007

\begin{eqnarray}
\nonumber &&\cr \nonumber && \cr \nonumber &&\cr
\end{eqnarray}
\begin{eqnarray}
\nonumber
\end{eqnarray}
\begin{center}
\begin{Large}
\textbf{Undulator Radiation in a Waveguide}
\end{Large}
\begin{eqnarray}
\nonumber &&\cr \nonumber && \cr
\end{eqnarray}

\begin{large}
Gianluca Geloni, Evgeni Saldin, Evgeni Schneidmiller and Mikhail
Yurkov
\end{large}
\textsl{\\Deutsches Elektronen-Synchrotron DESY, Hamburg}
\begin{eqnarray}
\nonumber
\end{eqnarray}
\begin{eqnarray}
\nonumber
\end{eqnarray}
\begin{eqnarray}
\nonumber
\end{eqnarray}
ISSN 0418-9833
\begin{eqnarray}
\nonumber
\end{eqnarray}
\begin{large}
\textbf{NOTKESTRASSE 85 - 22607 HAMBURG}
\end{large}
\end{center}
%\end{widetext}
\clearpage
\newpage

\begin{frontmatter}



\title{Undulator Radiation in a Waveguide}


\author{Gianluca Geloni,}
\author{Evgeni Saldin,}
\author{Evgeni Schneidmiller}
\author{and Mikhail Yurkov}

\address{Deutsches Elektronen-Synchrotron (DESY), Hamburg,
Germany}

\begin{abstract}
We propose an analytical approach to characterize undulator
radiation near resonance, when the presence of the vacuum-pipe
considerably affects radiation properties. This is the case of the
far-infrared undulator beamline at the Free-electron LASer (FEL)
in Hamburg (FLASH), that will be capable of delivering pulses in
the TeraHertz (THz) range. This undulator will allow pump-probe
experiments where THz pulses are naturally synchronized to the VUV
pulse from the FEL, as well as the development of novel
electron-beam diagnostics techniques. Since the THz radiation
diffraction-size exceeds the vacuum-chamber dimensions,
characterization of infrared radiation must be performed
accounting for the presence of a waveguide. We developed a theory
of undulator radiation in a waveguide based on paraxial and
resonance approximation. We solved the field equation with a
tensor Green's function technique, and extracted figure of merits
describing in a simple way the influence of the vacuum-pipe on the
radiation pulse as a function of the problem parameters. Our
theory, that makes consistent use of dimensionless analysis,
allows treatment and physical understanding of many asymptotes of
the parameter space, together with their region of applicability.
\end{abstract}


\end{frontmatter}

% main text

\clearpage

\section{\label{sec:intro} Introduction}

At the start of this century we have seen a revolution in
synchrotron source intensities. This revolution stemmed from the
technique of Free-Electron Lasers (FELs), combined with recent
progress in accelerator technology, developed in connection with
high-energy linear colliders.

A new era of Synchrotron Radiation research has begun with first
user experiments on VUV-FEL, based on Self-Amplified Spontaneous
Emission (SASE) \cite{NEW1, NEW2}. These results have been
obtained at the TESLA (Tera Electronvolt Superconducting Linear
Accelerator) Test Facility (TTF) at Deutsches
Elektronen-SYnchrotron (DESY) at Hamburg, Germany. Radiation
pulses with a wavelength of $98$ nm, $40$ fs duration and $1.5$ GW
peak power where used at TTF, phase $1$ \cite{NEW3, NEW4}.

This facility is now called FLASH (Free-Electron LAser in
Hamburg), and operates as a user facility since August 2005
\cite{NEW5, NEW6}. Currently FLASH produces GW-level, laser-like
VUV (Vacuum Ultra-Violet) to EUV (Extreme Ultra-Violet) radiation
pulses in the wavelength range between $\lambda = 13$ nm and
$\lambda = 50$ nm, with a duration between $10$ fs and $50$ fs.
Effective operation of an FEL in the VUV range requires high
peak-current electron bunches (in the kA-range). Electrons are
initially produced in a laser-driven RF-gun and subsequently
compressed in magnetic chicanes to reach specifications. The
accelerator complex at FLASH produces ultra-short bunches
approaching sub-$100$ fs duration.

It is natural to take advantage of these ultra-short bunches in
order to provide coherent far-infrared (FIR) radiation. In fact,
intense, coherent FIR radiation pulses can be produced from
sub-$100$ fs electron bunches at a wavelength longer than, or
comparable with the bunch length. For coherently radiating
electrons, radiated energy is proportional to the square of the
electron number, in contrast to the incoherent case when energy in
the radiation pulse scales linearly with the number of electrons
involved in the process. The result is an enhancement in the
radiation intensity of many orders of magnitude. Reference
\cite{NEW7} describes a proposal for a FIR coherent source,
integrated in the FLASH user facility. In \cite{NEW7} installation
of an additional electromagnetic undulator after the VUV FEL is
proposed. The FIR source will use the spent electron beam coming
from the FEL process, and will allow to significantly extend the
scientific potential of FLASH without interfering with the main
(VUV) option in the FLASH operation.

Coherent FIR pulses will be intrinsically synchronized with the
VUV pulses. A first, natural application of this kind of photon
beams is for pump-probe experiments. Another application concerns
electron beam diagnostics. The femtosecond time-scale of electron
bunches at FLASH is beyond the range of standard
electronic-display instrumentation, and the development of
non-destructive methods for the measurement of longitudinal beam
current distribution is undoubtedly a challenging problem. In
\cite{NEW8} a diagnostics technique is proposed, based on the
measurement of FIR coherent radiation from electron bunches
passing through an undulator. This technique is non-destructive,
and characterization of bunches with strongly non-Gaussian shapes
(as in the FLASH case) is possible \cite{NEW9, GRI1}.

FLASH will soon be coupled with a FIR electromagnetic undulator
\cite{BORI}. At the time of writing, the project is entering into
the realization phase. Funding has been secured, and the undulator
has been built and delivered \cite{GRI2}. Undulator radiation
around the fundamental harmonic will always be used. The
fundamental harmonic will be tuned by adjusting the magnetic field
strength. The wavelength range ($\lambda = 60\div 200 ~\mu$m)
provided by this powerful radiation source (up to $10$ MW peak
power) will overlap with a large part of the THz-gap, extending
between $60~\mu$m and $600~\mu$m. This will allow both
applications for pump-probe experiments combining FIR and VUV
radiation \cite{NEW11}, and for non-destructive electron beam
diagnostics. In fact, measurements of the modulus of the electron
bunch form factor in the spectral range $\lambda = 10\div
200~\mu$m is sufficient to provide a precise reconstruction of the
electron bunch profile \cite{GRI1}.

In the case of the FIR undulator beamline at FLASH the electron
beam geometrical emittance is much smaller than the radiation
diffraction size. This means that, as pertains the
characterization of the THz pulses, the electron beam can safely
be modelled as a filament beam. Computer codes like SRW
\cite{CHU2} and SPECTRA \cite{TANA} can be used, in the
space-frequency domain, to study undulator radiation (UR) from
filament electron beams up to a wavelength when the influence of
the vacuum chamber is negligible. Alternatively, an analytical
formalism for describing near-zone Synchrotron Radiation (SR)
fields from undulators in terms of Fourier optics, was developed
in \cite{OURF}, also working in the free-space limit and in the
space-frequency domain. These methods help designing beamlines and
experiments. However, in the case of the FIR undulator beamline at
FLASH, wavelengths in the order of $200~\mu$m and a $4$ m-long
undulator yield a radiation diffraction size of order of a
centimeter. This rough estimate indicates that vacuum chamber
effects are expected to play an important role, since a circular
vacuum chamber with radius $R =1.8$ cm is foreseen. In this case,
conventional computer codes and analytical methods fail to predict
the correct radiation characteristics.

Summing up, in view of the practical application to the infrared
undulator line at FLASH, there is a need to develop a
comprehensive theory of undulator radiation in the presence of a
waveguide.

Optimization of the radiation transport system calls for a precise
characterization of THz pulses along the photon beamline. In the
present work we focus on the characterization of undulator
radiation in presence of a waveguide, extending the study
presented in \cite{OURF} to include the influence of a vacuum
chamber in our consideration. Our work is, therefore,
propaedeutical to the problem of optimizing the radiation
transport system after the undulator, but it does not directly
deal with it.

The task that one has to solve differs from the free-space case
only in the formulation of boundary conditions. As we will discuss
in Section \ref{sec:boun}, the paraxial approximation  applies as
in free-space but, on a perfectly conductive boundary, the
electric field must be orthogonal to the pipe surface. As in the
free-space case one can use a Green's function approach to solve
the field equations. The presence of different boundary conditions
complicates the solution of the paraxial equation for the field,
which can anyway be found explicitly, by accounting for the
tensorial nature of the Green's function. Our consideration is
quite general, and can be applied to UR sources as well as to
other long-wavelength radiation setups, like edge-radiation
setups. In both cases the paraxial approximation can be used. In
the UR case, the resonance approximation can be exploited too, in
addition to the paraxial approximation. Our theory is developed
under both paraxial and resonance approximation. We thus consider
a large number of undulator periods and a frequency range of
interest close to the fundamental harmonic, where the free-space
field exhibits horizontal polarization (for undulator field in the
vertical direction) and azimuthal symmetry. The simultaneous
application of paraxial and resonance approximation makes the UR
case richer and more difficult to be studied from a theoretical
viewpoint when compared with the edge-radiation case. In
particular, the region of applicability of both approximations
must be discussed and specified.

In this paper we will focus on UR only, leaving the study of the
edge-radiation case to a future, dedicated publication. In
free-space and under resonance approximation UR is horizontally
polarized. This is actually a replica of the undulator
polarization properties. Moreover, the field exhibits azimuthal
symmetry. These properties are lost when metallic boundaries are
introduced. It is worth mentioning that application of similarity
techniques helps to give a clear physical interpretation of
numerical results. In this paper we continue to use these
techniques, that we already applied to SR theory in free-space
(see e.g. \cite{OURF}). Despite the fact that equations for the
undulator source are significantly complicated by the presence of
a waveguide, we find that these complications result in the
appearance of a single dimensionless extra-parameter, namely the
waveguide diffraction parameter $\Omega = R^2/(\lambdabar L_w)$,
where $\lambdabar = \lambda/(2\pi)$ is the reduced radiation
wavelength, $L_w$ is the undulator length and, as already said,
$R$ is the waveguide radius. The physical interpretation of
$\Omega$ is the squared ratio between the waveguide radius and the
radiation diffraction size of UR in free-space.

Since we are practically interested in the FIR undulator beamline
at FLASH we put particular emphasis on planar undulators in the
presence of a pipe with circularly symmetric cross-section. An
explicit expression for the field is calculated as a superposition
of Transverse Electric (TE) and Transverse Magnetic (TM) modes.
Some figure of merit should be extracted from the full information
carried by the expression for the field about how the metallic
pipe influences radiation properties. We separately studied, for
horizontal and vertical polarization components, two-dimensional
intensity distributions on a transverse plane at arbitrary
distance from the undulator, for different choices of the problem
parameter. Also, we analyzed the total power as a function of the
waveguide diffraction parameter at perfect resonance. Conversely,
once the waveguide diffraction parameter is fixed, one can
investigate how the total power changes as a function of the
detuning from resonance. Finally, a comparison between the
magnitude of the horizontally and vertically polarized fields is
also proposed as a measure of the waveguide influence.

To the best of our knowledge, only a few articles
\cite{MOTZ,HAUS,AMIR} deal with the problem of radiation from a
wiggled electron in a waveguide. This problem is also discussed at
advanced textbook level in \cite{HART}. In reference \cite{AMIR}
one may find the following words: "Motz and Nakamura \cite{MOTZ}
(...) considered an infinitely long wiggler. As a consequence, the
outcome did not possess a realistic bandwidth. (...). In a (...)
article by Haus and Islam \cite{HAUS} (...) it is shown that in
the limit of a highly over-moded guide, a result similar to the
free-space expression is produced". Reference \cite{HAUS}
considers a rectangular waveguide, and focuses on "similarities
between the emission into the free and the bounded space" (cited
from \cite{AMIR}). However, we should say that reference
\cite{HAUS} solves the equations for the field in the rectangular
case in all generality, and with no restrictions on the undulator
parameter $K$. Reference \cite{AMIR} deals with differences with
respect to free-space emission, in the case of a planar waveguide
and planar undulator with small undulator parameter $K\ll 1$.
These restrictions limit the practical scope of that work. In
particular, extending the theory in \cite{AMIR} from the case of a
planar waveguide to other geometries is not straightforward, since
the explicit expression for the paraxial Green's function given in
\cite{AMIR} is only valid for the planar-waveguide case. The
planar-waveguide case has clear advantages from an educational
viewpoint, as it allows to reduce complexities to a minimum. In
particular, no restriction is made on the wiggling amplitude of
the electron motion. In fact, the electron oscillates in the
unbounded region between the parallel conducting plates. A
transparent physical picture arises, in terms of reflection on the
two metallic plates. However, such transparency comes at the cost
of a limited region of applicability of the theory in practical
cases of interest. In reference \cite{AMIR}, the resonance
approximation is exploited, and a spectral region for large
detuning from the first harmonic can be considered due to
simplifications intrinsic in the planar-waveguide geometry. In the
case of the FLASH infrared undulator a circular waveguide is going
to be used. The FLASH infrared undulator will operate for large
values of $K$, and radiation will be used near the resonance with
the first harmonic. We developed a theory to deal with this
situation by restricting our attention around resonance. Thus, we
eliminated restrictions for the waveguide geometry (and for the
undulator parameter $K$), but we introduced a condition about the
spectral range of interest, coinciding with the spectral range of
interest at the FLASH infrared undulator line. The wiggling
amplitude of the electron in the undulator is taken to be small
with respect to the dimension of the waveguide. This greatly
simplifies analytical calculations, and describes our practical
case of interest. As it will be seen in Section \ref{sec:wig},
within the region of parameter space where these conditions apply,
one may consider both cases when the waveguide influence is weak
(up to the free-space limit) or strong. Finally, the case of
radiation of an electron in a helical wiggler with a circular
waveguide is discussed in Section 9.3 of reference \cite{HART}. TM
modes are neglected (see Appendix A for details). Excluding TM
modes from consideration is held by us to be a misconception. In
contrast to this, we will demonstrate throughout the text that the
wiggler-induced motion couples with both TE and TM modes in a
circular (cylindrical) waveguide.

Our work is organized as follows. Besides this Introduction, in
the next Section \ref{sec:free} we review basic theory of
undulator radiation in free-space. Such a review is necessary,
because delicate physical assumptions used in the free-space case
continue to be exploited when a waveguide is present as well. In
the following Section \ref{sec:boun} we pose our problem,
discussing Maxwell's equations for a single electron moving in the
presence of metallic boundaries and within the paraxial
approximation. We derive a closed expression for the field with
the help of a tensor Green's function technique. At this stage the
geometry of the vacuum pipe is generic, as well as the trajectory
of the electron. In Section \ref{sec:circ} we fix the vacuum pipe
geometry, specializing our equations to the case of a vacuum pipe
with circular cross-section. We verify the correctness of our
results by deriving the free-space limit, studied in \cite{OURF}.
Then, in the following Sections \ref{sec:wig} we consider the
cases of a planar undulator, while we refer the interested reader
to Appendix A for the case of a helical undulator. Results
obtained in Section \ref{sec:wig} for the planar undulator case
are the main results in our paper, as they allow a complete
characterization of the field in the FIR undulator beamline at
FLASH. As said before, such characterization is indispensable for
any further analysis of the radiation transport system. We analyze
our results further in Section \ref{sec:res}, where we propose a
study of figure of merits, that will help designers and beamline
scientists to estimate the influence of the vacuum pipe on the
radiation characteristics. In Section \ref{sec:resi} we discuss
the influence of wall-resistance on our findings, demonstrating
their relevance. Finally, in Section \ref{sec:conc}, we come to
conclusions.

\section{\label{sec:free} Undulator radiation in free-space}

\subsection{\label{sub:gr} Green's function technique in free-space}

For any SR setup we can represent the electric field in time
domain $\vec{E}(\vec{r}, t)$ as a time-dependent function of an
observation point located at position
$\vec{r}=\vec{r}_{\bot}+z\vec{e}_z=x \vec{e}_x+y\vec{e}_y
+z\vec{e}_z$, where $\vec{e}_x$, $\vec{e}_y$ and $\vec{e}_z$ are
defined as (dimensionless) unit vectors along horizontal, vertical
and longitudinal direction in a given reference frame.  For
monochromatic waves of angular frequency $\omega$, the wave
amplitude has the form $\vec{E}(z,\vec{r}_\bot,t) =
\vec{\bar{E}}(z,\vec{r}_\bot) \exp[-i\omega t] + C.C.$, where
"C.C." indicates the complex conjugate of the preceding term and
$\vec{\bar{E}}$ describes the variation of the wave amplitude. The
vector $\vec{\bar{E}}$ actually represents the amplitude of the
electric field in the space-frequency domain. Accounting for
electromagnetic sources, i.e. in a region of space where current
and charge densities are present, the field in the space-frequency
obeys Helmholtz's equation, i.e. $c^2 \nabla^2 \vec{\bar{E}} +
\omega^2 \vec{\bar{E}} = 4 \pi c^2 \vec{\nabla} \bar{\rho} - 4 \pi
i \omega \vec{\bar{j}}$, where $\bar{\rho}(\vec{r},\omega)$ and
$\vec{\bar{j}}(\vec{r},\omega)$ are the Fourier transforms of the
charge density, $\rho(\vec{r},t)$, and of the current density,
$\vec{j}(\vec{r},t)$, while $c$ is the speed of light. We will
consider a single electron. Using the Dirac delta distribution, we
can write $\rho(\vec{r},t) = -e \delta(\vec{r}-\vec{r'}(t))$ and
$\vec{j}(\vec{r},t) = \vec{v}(t) \rho(\vec{r},t)$, where $(-e)$ is
the negative electron charge, $\vec{r'}(t)$ and $\vec{v}(t)$ are,
respectively, the position and the velocity of the particle at a
given time $t$, and $v_z$ its longitudinal velocity. After
calculation of the Fourier transform of these quantities and
substitution into Helmholtz's equation we obtain

\begin{eqnarray}
\left({\nabla}^2 + {2 i \omega \over{c}} {\partial\over{\partial
z}}\right) \vec{\widetilde{E}} &=& {4 \pi e\over{v_z(z)}}
\exp\left[{i \omega
\left({s(z)\over{v}}-{z\over{c}}\right)}\right]
\left[{i\omega\over{c^2}}\vec{v}(z) -\vec{\nabla}\right]
\delta\left(\vec{r}_\bot-\vec{r'}_\bot(z)\right),\label{incipit}
\end{eqnarray}
$s(z)$ being the curvilinear abscissa measured along the electron
trajectory, where we conventionally set $s(0)=0$. When the
longitudinal velocity of the electron, $v_z$, is close to the
speed of light $c$ (i.e. $\gamma_z^2 \gg 1$, where $\gamma_z(z)=
(1-v_z^2/c^2)^{-1/2}$ is the longitudinal Lorentz factor), the
Fourier components of the source are almost synchronized with the
electromagnetic wave travelling at the speed of light. In this
case the phase $\omega ({s(z)/{v}}-{z/{c}})$ is a slow function of
$z$ compared to the wavelength. For example, in the particular
case of motion on a straight section, one has $s(z) = z/v_z$, so
that $\omega ({s(z)/{v}}-{z/{c}}) = \omega z/(2\gamma_z^2 c)$, and
if $\gamma_z^2 \gg 1$ such phase grows slowly in $z$ with respect
to the wavelength. For a more generic motion, one similarly
obtains:

\begin{equation}
\omega \left({s(z_2)-s(z_1)\over{v}}-{z_2-z_1\over{c}}\right) =
\int_{z_1}^{z_2} d \bar{z} \frac{\omega}{2 \gamma_z^2(\bar{z})
c}~. \label{moregen}
\end{equation}
Mathematically, the phase in Eq. (\ref{moregen}) enters in the
Green's function solution of Eq. (\ref{incipit}) as a factor in
the integrand. As we integrate along $z'$, the factor
$\omega(s(z')/v - z'/c)$ leads to an oscillatory behavior of the
integrand over a certain integration range in $z'$. Such range can
be identified with the value of $z_2-z_1$ for which the right hand
side of Eq. (\ref{moregen}) is of order unity, and it is naturally
defined as the radiation formation length $L_f$ of the system at
frequency $\omega$. Of course there exists some freedom in the
choice of such definition: "order of unity" is not a precise
number, and reflects the fact that there is no abrupt threshold
between "oscillatory" and "non-oscillatory" behavior of the
integrand in the solution of Eq. (\ref{incipit}). In the following
we define the formation length $L_f$ as the interval $z_2-z_1$
such that the right hand side of Eq. (\ref{moregen}) is strictly
equal to unity. It is easy to see by inspection of Eq.
(\ref{moregen}) that if $v_z$ is sensibly smaller than $c$ (but
still of order $c$), i.e. $v_z\sim c$ but $1/\gamma_z^2 \sim 1$,
then $L_f \sim \lambdabar$, where we introduced the reduced
wavelength $\lambdabar = {\lambda}/({2\pi})$, and $\lambda = 2\pi
c/\omega$. On the contrary, when $v_z$ is very close to $c$, i.e.
$1/\gamma_z^2 \ll 1$, the right hand side of Eq. (\ref{moregen})
is of order unity for $L_f = z_2-z_1 \gg \lambdabar$. When the
radiation formation length is much longer than $\lambdabar$, the
electric field envelope $\vec{\widetilde{E}}_\bot =
\vec{\bar{E}}_\bot \exp{[-i\omega z/c]}$ does not vary much along
$z$ on the scale of $\lambdabar$, that is $\mid
\partial_z \widetilde{E}_{x,y}\mid \ll \omega/c \mid
\widetilde{E}_{x,y}\mid$. Therefore, the second order derivative
with respect to $z$ in the $\nabla^2$ operator on the left hand
side of Eq. (\ref{incipit}) is negligible with respect to the
first order derivative with respect to $z$. As a result, Eq.
(\ref{incipit}) can be simplified as

\begin{eqnarray}
\mathcal{D} \left[\vec{\widetilde{E}}_\bot(z,\vec{r}_\bot)\right]
= \vec{f}(z, \vec{r}_\bot) ~.\label{field1}
\end{eqnarray}
The differential operator $\mathcal{D}$ in Eq. (\ref{field1}) is
defined by

\begin{eqnarray}
\mathcal{D} \equiv \left({\nabla_\bot}^2 + {2 i \omega \over{c}}
{\partial\over{\partial z}}\right) ~,\label{Oop}
\end{eqnarray}
where ${\nabla_\bot}^2$ is the Laplacian operator over transverse
cartesian coordinates. The vector $\vec{f}(z, \vec{r}_\bot)$ is
specified by the trajectory of the source electron,
$\vec{r'}_\bot(z)$, and is written as

\begin{eqnarray}
\vec{f} = && \frac{4 \pi e}{c}   \exp\left[{i \int_{0}^{z} d
\bar{z} \frac{\omega}{2 \gamma_z^2(\bar{z}) c}}\right]
\left[\frac{i\omega}{c^2}\vec{v}_\bot(z) -\vec{\nabla}_\bot
\right]\delta\left(\vec{r}_\bot-\vec{r'}_\bot(z)\right)~.\cr &&
\label{fv}
\end{eqnarray}
Here we considered transverse components of $\vec{\widetilde{E}}$
only and we substituted $v_z(z)$ with $c$, based on the fact that
$1/\gamma_z^2 \ll 1$. Eq. (\ref{field1}) is Maxwell's equation in
paraxial approximation.

In the following, with some abuse of language, we will refer to
the slowly varying envelope of $\vec{\bar{E}}_\bot$, i.e. to
$\vec{\widetilde{E}}_\bot$, simply as the field.

The Green's function for Eq. (\ref{field1}), namely the solution
corresponding to a unit point source, depends on boundary
conditions. In free-space it must obey

\begin{eqnarray}
G(z-z';\vec{r}_{\bot}-\vec{r'}_\bot) = -{1\over{4\pi (z-z')}}
\exp\left[ i\omega{\mid \vec{r}_{\bot}
-\vec{r'}_\bot\mid^2\over{2c (z-z')}}\right] \label{green}~,
\end{eqnarray}
assuming $z-z' > 0$. When $z-z' < 0$ the paraxial approximation
does not hold, and the paraxial wave equation, Eq. (\ref{field1}),
should be substituted, in the space-frequency domain, by the more
general Helmholtz's equation. However, the radiation formation
length for $z - z'<0$ is very short with respect to the case $z -
z' >0$, i.e. there is effectively no radiation for observer
positions $z-z' <0$. As a result, in this paper we will consider
only $z - z'> 0$. It follows that the observer is located
downstream of the sources. This leads to the  final result:

\begin{eqnarray}
\vec{\widetilde{E}}_\bot(z, \vec{r}_{\bot},\omega) &=& -{i \omega
e\over{c^2}} \int_{-\infty}^{z} dz' \frac{1}{z-z'}
\left[{\vec{v}(z')\over{c}}
-{\vec{r}_{\bot}-\vec{r'}_\bot(z')\over{z-z'}}\right]\cr &&\times
\exp\left\{i\omega\left[{\mid \vec{r}_{\bot}-\vec{r'}_\bot
\mid^2\over{2c (z-z')}}+ \int_{0}^{z'} d \bar{z} \frac{1}{2
\gamma_z^2(\bar{z}) c}\right] \right\} ~. \label{generalfin2}
\end{eqnarray}
Eq. (\ref{generalfin2}) is valid at any observation position $z$
such that the paraxial approximation is valid, i.e. up to
distances between the observer and the electromagnetic sources
comparable with the radiation wavelength. One may recognize two
terms in Eq. (\ref{generalfin2}). The first in $\vec{v}(z')$ can
be traced back to the current term in Eq. (\ref{incipit}), while
the second, in $\vec{r}_{\bot}-\vec{r'}_\bot(z')$, corresponds to
the gradient term in Eq. (\ref{incipit}).

The far zone is defined as the region of observation where the
direction from any trajectory point to the observer can be
considered constant. In the far zone, Eq. (\ref{generalfin2})
reduces to

\begin{eqnarray}
\vec{\widetilde{{E}}}_\bot(z, \vec{\theta},\omega)&& = -{i \omega
e\over{c^2}z} \int_{-\infty}^{z} dz' \left({\vec{v}(z')\over{c}}
-{\vec{\theta}}\right) \cr && \times{\exp{\left[i \int_{0}^{z'} d
\bar{z} \frac{\omega}{2 \gamma_z^2(\bar{z}) c}+ \frac{i \omega}{{2
c}}\left( z ~\theta^2-2 \vec{\theta}\cdot\vec{
{r}'}_{\bot}(z')+{z' \theta^2}\right)\right]}} ~,
\label{generalfin}
\end{eqnarray}
where $\vec{\theta}= \vec{r}_\bot/z$ defines the observation
direction, and $\theta \equiv |\vec{\theta}|$.

An ultra-relativistic electron radiates non-negligibly, in the
far-zone, up to observation angles of order $\theta^2 \lesssim
\lambdabar/L_f$ (see \cite{OURF}). Moreover,  for
ultra-relativistic electrons, $L_f \gg \lambdabar$. The angular
region $\theta^2 \lesssim \lambdabar/L_f \ll 1$ formally coincides
with the region of applicability of the paraxial approximation.
However, since radiation is negligible elsewhere, it follows that
paraxial approximation can be applied to describe radiation from
an ultra-relativistic electron at any observation angle of
interest.

If we now decompose the electric field distribution in the far
zone as a superposition of plane waves (angular spectrum), the
angle of propagation of each plane wave is represented by the
ratio between the transverse wave vector $\vec{k}_\bot$ and the
longitudinal wave number $k_z$, that is $\vec{k}_\bot/k_z$. Note
that in free-space, $\vec{k}_\bot$ and $k_z$ are allowed to vary
continuously across the reciprocal space.  Non-negligible plane
wave components of the angular spectrum are those seen at angles
$\theta^2 \lesssim \lambdabar/L_f \ll 1$, i.e.  propagating at
angles ${k}_\bot^2/k_z^2 \ll 1$. Now, since $k_z \simeq k =
1/\lambdabar$ this may also be stated by requiring ${k}_\bot^2
c^2/\omega^2 \ll 1$. This result will be useful in Section
\ref{sec:boun}.

\subsection{\label{sub:urp} Undulator radiation in paraxial approximation. Far zone}

Let us apply the method outlined in Section \ref{sub:gr} to the
case of planar undulator. In general, the term \textit{undulator
radiation field}  means only a part of the total field seen by an
observer from a realistic setup, because  one should account for
contributions from the entire trajectory of the particle. We may
follow any textbook like \cite{WIED} in deriving well-known
relations in the far-zone. For the electron transverse velocity we
assume

\begin{eqnarray}
v_x(z') = -{c \theta_s} \sin(k_w z') = -\frac{c \theta_s}{2
i}\left[\exp(ik_w z')-\exp(-i k_w z') \right]~. \label{vxpl}
\end{eqnarray}
Here  $k_w=2\pi/\lambda_w$, and $\lambda_w$ is the undulator
period. Moreover, $\theta_s=K/\gamma$, where $K$ is the deflection
parameter defined as

\begin{eqnarray}
K = \frac{e\lambda_w H_w}{2 \pi m_e c^2}~,\label{Kpar}
\end{eqnarray}
$m_e$ being the electron mass at rest and $H_w$ being the maximal
magnetic field of the undulator on axis.

The longitudinal Lorentz factor $\gamma_z(z)$ is a function of the
position down the undulator, so that the phase in Eq.
(\ref{moregen}) calculated at $z_1=0$ and $z_2=z$ gives:

\begin{eqnarray}
\int_0^z \frac{\omega}{2c\gamma_z^2(\bar{z}) } d\bar{z}=
\frac{\omega}{2 c \bar{\gamma}_z^2}z - \frac{\omega \theta_s^2}{8
k_w c} \sin(2 k_w z)~,\label{phasep}
\end{eqnarray}
where the average longitudinal Lorentz factor $\bar{\gamma}_z$ is
defined as

\begin{equation}
\bar{\gamma}_z = \frac{\gamma}{\sqrt{1+K^2/2}}~. \label{bargz}
\end{equation}
We write the undulator length as $L_w = N_w \lambda_w$, where
$N_w$ is the number of undulator periods. With the help of Eq.
(\ref{generalfin}) we obtain an expression, valid in the far zone:

\begin{eqnarray}
{\vec{\widetilde{E}}}_\bot&=& {i \omega e\over{c^2 z}}
\int_{-L_w/2}^{L_w/2} dz' {\exp\left[i
\Phi_T\right]\exp\left[i\frac{\omega \theta^2 z}{2c}\right]}
\left[{K\over{\gamma}} \sin\left(k_w z'\right)\vec{e}_x
+\vec{\theta}\right]~. \label{undurad}
\end{eqnarray}
Here

\begin{eqnarray}
\Phi_T &=& \left({\omega \over{2 c \bar{\gamma}_z^2}}+ {\omega
\theta^2 \over{2  c }}\right) z' -
{K\theta_x\over{\gamma}}{\omega\over{k_w c}}\cos(k_w z') -
{K^2\over{8\gamma^2}} {\omega\over{k_w c}} \sin(2 k_w z')
~.\label{phitundu}
\end{eqnarray}

\begin{figure}
\begin{center}
\includegraphics*[width=140mm]{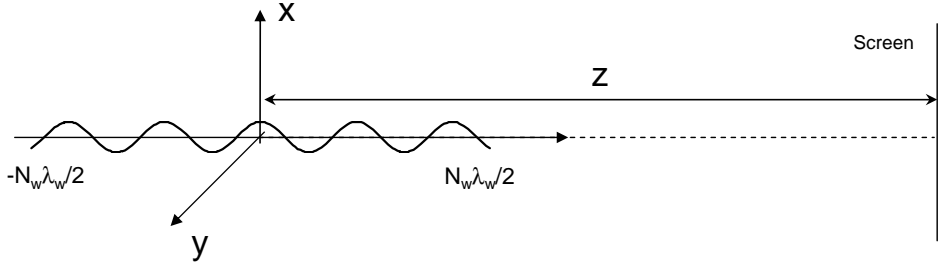}% Here is how to import EPS art
\caption{\label{undugeo} Geometry for undulator radiation. }
\end{center}
\end{figure}

The choice of the integration limits in Eq. (\ref{undurad})
implies that the reference system has its origin in the center of
the undulator as in Fig. \ref{undugeo}. As for Eq.
(\ref{generalfin2}) and Eq. (\ref{generalfin}), one recognizes two
terms in Eq. (\ref{undurad}). The first in $\sin(k_w z')$ can be
traced back to the first (current) term in Eq.
(\ref{generalfin2}), while the second, in ${\vec{\theta}}$,
corresponds to the second (gradient) term in Eq.
(\ref{generalfin2}).

Eq. (\ref{undurad})  has been derived under the paraxial
approximation, that can always be applied for ultrarelativistic
systems characterized by a large parameter $\gamma^2 \gg 1$.

\subsection{\label{sub:res} Resonant approximation in the far zone}

Usually, it does not make sense to calculate the intensity
distribution from Eq. (\ref{undurad}) alone, without extra-terms
(both interfering and not) from the other parts of the electron
trajectory. This means that one should have complete information
about the electron trajectory and calculate extra-terms to be
added to Eq. (\ref{undurad}) in order to have the total field from
a given setup. Yet, we can find \textit{particular situations} for
which the contribution from Eq. (\ref{undurad}) is dominant with
respect to others. In this case Eq. (\ref{undurad}), alone, has
independent physical meaning.

One of these situations is  when the resonance approximation is
valid. This approximation does not replace the paraxial one, based
on $\gamma^2 \gg 1$, but it is used together with it. It takes
advantage of another parameter that is usually large, i.e. the
number of undulator periods $N_w \gg 1$. In this case, the
integral in $dz'$ in Eq. (\ref{undurad}) exhibits simplifications,
independently of the frequency of interest due to the long
integration range with respect to the scale of the undulator
period.

In the particular case when the frequency of interest is near the
fundamental resonance frequency

\begin{eqnarray}
\omega_r = 2 k_w c \bar{\gamma}_z^2~, \label{res}
\end{eqnarray}
or other harmonics \textit{odd} multiples of $\omega_r$,
extra-simplifications can be exploited, allowing one to neglect
the gradient term in ${\vec{\theta}}$ in Eq. (\ref{undurad}), as
well as the constrained particle motion in the Green's function,
i.e. the second term in $\cos(k_w z')$ in the phase Eq.
(\ref{phitundu}), that corresponds to the term in
$\vec{\theta}\cdot\vec{ {r}'}_{\bot}(z')$ in Eq.
(\ref{generalfin}). This leads to horizontally polarized radiation
and to azimuthal symmetry of the field. It should be stressed that
odd harmonics constitute a particular case. Neglecting the
gradient term and the constrained particle's motion in the Green's
function does not coincide with the application of the resonance
approximation, understood as exploitation of the large parameter
$N_w\gg 1$. For example, resonance approximation can be used to
study \textit{even} harmonics but, as shown in  \cite{HAR2}, in
that case the gradient term and the constrained particle's motion
in the Green's function must be retained.

In this paper we will be interested in frequencies near the first
(fundamental) harmonic $\omega_r$, so that the above mentioned
extra-simplifications can be exploited. Let us show how this can
be done, and let us discuss how radiation characteristics
(polarization and symmetry) are related to the possibility of
neglecting gradient term and constrained particle's motion in the
Green's function. First, we can specify "how near" $\omega$ is to
$\omega_r$ by introducing a detuning parameter $C$, defined as

\begin{equation}
C = {\omega \over{2\bar{\gamma}_z^2c}}-k_w =
\frac{\Delta\omega}{\omega_r} k_w~.\label{detC}
\end{equation}
Here $\omega = \omega_r + \Delta \omega$. Eq. (\ref{phitundu}) can
thus be written as

\begin{eqnarray}
\Phi_T &=& \left(k_w + C + {\omega \theta^2 \over{2  c }}\right)
z' - {K\theta_x\over{\gamma}}{\omega\over{k_w c}}\cos(k_w z') -
{K^2\over{8\gamma^2}} {\omega\over{k_w c}} \sin(2 k_w z')
~,\label{phitundu2}
\end{eqnarray}
so that, in all generality, the field in Eq. (\ref{undurad}) can
be written as

\begin{eqnarray}
&&{\vec{\widetilde{E}}}_\bot= \exp\left[i\frac{\omega \theta^2
z}{2c}\right] \frac{i \omega e}{c^2 z} \int_{-L_w/2}^{L_w/2}
dz'\left\{\frac{K}{2 i \gamma}\left[\exp\left(2 i k_w
z'\right)-1\right]\vec{e}_x +\vec{\theta}\exp\left(i k_w
z'\right)\right\} \cr &&\times \exp\left[i \left(C + {\omega
\theta^2 \over{2 c }}\right) z' -
{K\theta_x\over{\gamma}}{\omega\over{k_w c}}\cos(k_w z')  -
{K^2\over{8\gamma^2}} {\omega\over{k_w c}} \sin(2 k_w z') \right]
~. \cr &&\label{undurad2}
\end{eqnarray}
As first proposed in \cite{ALFE} one may use the Anger-Jacobi
expansion:

\begin{equation}
\exp\left[i a \sin(\psi)\right] = \sum_{p=-\infty}^{\infty} J_p(a)
\exp\left[ip\psi\right]~, \label{alfeq}
\end{equation}
where $J_p(\cdot)$ indicates the Bessel function of the first kind
of order $p$, to write the integral in Eq. (\ref{undurad2}) in a
different way:

\begin{eqnarray}
&&{\vec{\widetilde{E}}}_\bot= \exp\left[i\frac{\omega \theta^2
z}{2c}\right] \frac{i \omega e}{c^2 z} \sum_{m,n=-\infty}^\infty
J_m(u) J_n(v) \exp\left[\frac{i \pi n}{2}\right] \cr && \times
\int_{-L_w/2}^{L_w/2} dz'\exp\left[i \left(C + {\omega \theta^2
\over{2 c }}\right) z'\right] \left\{\frac{K}{2 i \gamma}
\left[\exp\left(2 i k_w z'\right)-1\right]\vec{e}_x
+\vec{\theta}\exp\left(i k_w z'\right)\right\} \cr &&\times
\exp\left[i (n+2m) k_w z'\right] ~,\label{undurad3}
\end{eqnarray}
where

\begin{equation}
u = - \frac{K^2 \omega}{8 \gamma^2 k_w c}~~~~\mathrm{and}~~~v = -
\frac{K \theta_x \omega}{\gamma k_w c}~. \label{uv}
\end{equation}
Up to now we just re-wrote Eq. (\ref{undurad}) in a different way.
Eq. (\ref{undurad}) and Eq. (\ref{undurad3}) are equivalent. Of
course, definition of $C$ in Eq. (\ref{detC}) is suited to
investigate frequencies around the fundamental harmonic but no
approximation is taken besides the paraxial approximation.

Whenever

\begin{equation}
C  + \frac{\omega \theta^2}{{2 c}} \ll k_w \label{eqq} ~,
\end{equation}
the first phase term in $z'$ under the integral sign in Eq.
(\ref{undurad3}) is varying slowly on the scale of the undulator
period $\lambda_w$. As a result, simplifications arise when $N_w
\gg 1$, because fast oscillating terms in powers of  $\exp[i k_w
z']$ effectively average to zero. When these simplifications are
taken,  resonance approximation is applied, in the sense that one
exploits the large parameter $N_w \gg 1$. This is possible under
condition (\ref{eqq}). Note that (\ref{eqq}) restricts the range
of frequencies for positive values of $C$ independently of the
observation angle ${\theta}$, but for any value $C<0$ (i.e. for
wavelengths longer than $\lambdabar_r = c/\omega_r$) there is
always some range of $\theta$ such that Eq. (\ref{eqq}) can be
applied. Altogether, application of the resonance approximation is
possible for frequencies around $\omega_r$ and lower than
$\omega_r$. Once any frequency is fixed, (\ref{eqq}) poses
constraints on the observation region where the resonance
approximation applies. Similar reasonings can be done for
frequencies around higher harmonics with a more convenient
definition of the detuning parameter $C$.

Within the resonance approximation we further select frequencies
such that

\begin{eqnarray}
\frac{|\Delta \omega|}{\omega_r} \ll 1~,~~~~ \mathrm{i.e.}~~|C|
\ll k_w ~.\label{resext}
\end{eqnarray}
Note that this condition on frequencies automatically selects
observation angles of interest $\theta^2 \ll 1/\gamma_z^2$. In
fact, if one considers observation angles outside the range
$\theta^2 \ll 1/\gamma_z^2$, condition (\ref{eqq}) is not
fulfilled, and the integrand in Eq. (\ref{undurad3}) exhibits fast
oscillations on the integration scale $L_w$. As a result, one
obtains zero transverse field, $\vec{\widetilde{E}}_\bot = 0$,
with accuracy $1/N_w$. Under the constraint imposed by
(\ref{resext}), independently of the value of $K$ and for
observation angles of interest $\theta^2 \ll 1/\gamma_z^2$, we
have

\begin{equation}
|v|={K|\theta_x|\over{\gamma}}{\omega\over{k_w c}} =
\left(1+\frac{\Delta \omega}{\omega_r}\right) \frac{2 \sqrt{2}
K}{\sqrt{2+K^2}} \bar{\gamma}_z |\theta_x| \lesssim
\bar{\gamma}_z |\theta_x| \ll 1~. \label{drop}
\end{equation}
This means that, independently of $K$, $|v| \ll 1$ and we may
expand $J_n(v)$ in Eq. (\ref{undurad3}) according to $J_n(v)
\simeq [2^{-n}/\Gamma(1+n)]~v^n$, $\Gamma(\cdot)$ being the Euler
gamma function

\begin{eqnarray}
\Gamma(z) = \int_0^\infty dt~t^{z-1} \exp[-t] ~.\label{geule}
\end{eqnarray}
Similar reasonings can be done for frequencies around higher
harmonics with a different definition of the detuning parameter
$C$. However, around odd harmonics, the before-mentioned
expansion, together with the application of the resonance
approximation for $N_w \gg 1$ (fast oscillating terms in powers of
$\exp[i k_w z']$ effectively average to zero), yields
extra-simplifications.

Here we are dealing specifically with the first harmonic.
Therefore, these extra-simplifications apply. They enforce a
stronger version of the resonance approximation allowing one to
neglect both the constrained motion in the Green's function and
the gradient term in the expression for the field (respectively,
the term in $\cos(k_w z')$ in the phase of Eq. (\ref{undurad2})
and the term in $\vec{\theta}$ in Eq. (\ref{undurad2})). First,
non-negligible terms in the expansion of $J_n(v)$ are those for
small values of $n$, since $J_n(v) \sim v^n$, with $|v|\ll 1$. The
value $n=0$ gives a non-negligible contribution $J_0(v) \sim 1$.
Then, since the integration in $d z'$ is performed over a large
number of undulator periods $N_w\gg 1$, all terms of the expansion
in Eq. (\ref{undurad3}) but those for $m=-1$ and $m=0$ average to
zero due to resonance approximation. Note that surviving
contributions are proportional to $K/\gamma$, and can be traced
back to the current term in $\vec{e}_x$ only, while the gradient
term in $\vec{\theta}$ in Eq. (\ref{undurad3}) averages to zero
for $n=0$. Values $n=\pm 1$ already give negligible contributions.
In fact, $J_{\pm 1}(v) \sim v$. Then, the term in $\vec{e}_x$ in
Eq. (\ref{undurad3}) is $v$ times the term with $n=0$ and is
immediately negligible, regardless of the values of $m$. The
gradient term in $\vec{\theta}$ would survive averaging when $n=1,
~m=-1$ and when $n=-1, ~m=0$. However, it scales as $\vec{\theta}
v$. Now, using condition (\ref{resext}) we see that, for
observation angles of interest $\theta^2 \ll 1/\gamma_z^2$,
$|\vec{\theta}|~ |v| \sim (\sqrt{2}~ K~/\sqrt{2+K^2}~)
~\bar{\gamma}_z \theta^2 \ll K/\gamma$. Therefore, the gradient
term is negligible with respect to the current term in $\vec{e}_x$
for $n=0$, that scales as $K/\gamma$. All terms corresponding to
larger values of $|n|$ are negligible.

Summing up, all terms of the expansion in Eq. (\ref{alfeq}) but
those for $n=0$ and $m=-1$ or $m=0$ give negligible contribution.
After definition of

\begin{eqnarray}
A_{JJ} = J_0\left(\frac{\omega K^2}{8 k_w c \gamma^2}\right) -
J_1\left(\frac{\omega K^2}{8 k_w c \gamma^2}\right) ~,\label{AJJ}
\end{eqnarray}
that can be calculated at $\omega = \omega_r$ since $|C| \ll k_w$,
we have

\begin{eqnarray}
&&{\vec{\widetilde{E}}}_\bot= - \frac{K \omega e A_{JJ}}{2 \gamma
c^2 z} \exp\left[i\frac{\omega \theta^2 z}{2c}\right]
\int_{-L_w/2}^{L_w/2} dz' \exp\left[i \left(C + {\omega \theta^2
\over{2  c }}\right) z' \right] \vec{e}_x~,\label{undurad5}
\end{eqnarray}
yielding the well-known free-space field distribution:

\begin{eqnarray}
{\vec{\widetilde{E}}}_\bot&&(z, \vec{\theta}) = -\frac{K \omega e
L_w A_{JJ} }{2 \gamma c^2 z} \exp\left[i\frac{\omega \theta^2
z}{2c}\right] \mathrm{sinc}\left[\frac{L_w}{2}\left(C+\frac{\omega
\theta^2}{2c} \right)\right] \vec{e}_x~,\cr && \label{generalfin4}
\end{eqnarray}
where $\mathrm{sinc}(\cdot) \equiv \sin(\cdot)/(\cdot)$.
Therefore, the field is horizontally polarized and azimuthal
symmetric. It should be clear that result in Eq.
(\ref{generalfin4}) is valid for arbitrary values of $K$, that may
assume values much larger than unity, as in the long-wavelength
range of the FIR undulator line at FLASH.

Condition (\ref{eqq}) defines the applicability region of the
resonance approximation in free-space in terms of the detuning
parameter $C$ and the observation angle $\vec{\theta}$. Condition
(\ref{resext}) defines the range of the detuning parameter $C$
where a stronger version of the resonance approximation applies
for odd harmonics. It imposes a stronger constraint on the
detuning parameter $C$ but it eliminates constraints on the
observation angles because, as said before, observation angles
outside the region of interest $\theta^2 \ll 1/\gamma_z^2$ simply
return zero field. Thus, Eq. (\ref{generalfin4}) can be used
without restrictions on $\theta$. Sometimes we will say that
(\ref{resext}) enforces the "strong" resonance approximation, when
the extra-simplifications we just discussed apply.

The constrained motion term in the Green's function and the
gradient term in the equation for the field are strictly related
to polarization and azimuthal symmetry of the field. When the
constrained motion in the Green's function can be neglected, also
the gradient term can be neglected and the field is horizontally
polarized and azimuthal symmetric. There are situations when this
cannot be done. For example, when studying the long-wavelength
region when (\ref{resext}) is not valid, the field looses these
properties. However, if (\ref{eqq}) holds, resonance approximation
can still be applied, in the sense that the large parameter $N_w
\gg 1$ can be exploited to simplify the integration in $d z'$. In
the case of even harmonics, that has been studied in reference
\cite{HAR2}, one can use restrictions (\ref{resext}) and expand
the Bessel function similarly to what has been done here. However,
in this case, both constrained motion term in the Green's function
and gradient term in the equation for the field are not
negligible. Therefore, the field is not horizontally polarized,
nor azimuthal symmetric.

\subsection{\label{sub:nearz}  Undulator radiation in the near zone}

As has been described in \cite{OURF}, SR fields can be treated in
terms of Fourier optics. In particular, radiation from an
ultra-relativistic particle in an undulator can be interpreted as
radiation from a virtual source placed in the undulator center,
i.e. at $z=0$, which produces a laser-like beam. The concept of
virtual source has already been proposed (see \cite{GOV1},
\cite{GOV2} and \cite{KRIN}). In \cite{OURF} we developed this
concept up to an analytical description of the virtual source that
is similar, in many aspects, to the waist of a laser beam. In the
case of undulator radiation in free-space it exhibits a plane
wavefront. It is completely specified, for any given polarization
component, by a real-valued field amplitude distribution and can
be determined from the knowledge of the far-zone distribution in
Eq. (\ref{generalfin4}) through the relation (see \cite{OURF}):

\begin{eqnarray}
{ \widetilde{E}}_\bot(0,\vec{r}_{\bot} )&=& \frac{i \omega {z}}{2
\pi c} \int d\vec{\theta}\exp{\left[-\frac{i \omega |\vec{
\theta}|^2}{2 c} z\right]}{ \widetilde{E}}_\bot(\vec{\theta})
\exp\left[\frac {i \omega}{c} \vec{r}_{\bot}\cdot
\vec{\theta}\right] ~, \label{virfiemody}
\end{eqnarray}
where the integration in $d\vec{\theta}$ is understood to be
performed over the entire plane spanned by $\vec{\theta}$. The
integral in Eq. (\ref{virfiemody}) can be calculated analytically
at perfect resonance ($C=0$) yielding the field distribution at
the source:

\begin{eqnarray}
{ \widetilde{E}}_\bot(0,\vec{r}_{\bot} ) &=& -\frac{i A_{JJ} e
\theta_s \omega}{2 c^2} \left[\pi-2
\mathrm{Si}\left(\frac{r_\bot^2 \omega }{L_w c
}\right)\right]~,\cr &&\label{horpolfreep2}
\end{eqnarray}
where $\mathrm{Si}(\cdot)$ indicates the sin integral function

\begin{eqnarray}
\mathrm{Si}(z) = \int_0^z dt ~\frac{\sin(t)}{t}\label{sifde}
~.\end{eqnarray}
The quantity in Eq. (\ref{horpolfreep2}) is a scalar because, as
we have seen, the field is horizontally polarized. Numerical
investigations are possible off-resonance, for $C\ne 0$. Once the
field at the virtual source is known, the field at other
longitudinal positions, both in the far and in the near zone, up
to distances to the sources comparable with the radiation
wavelength, can be obtained with the help of the Fresnel
propagation formula:

\begin{equation}
{ \widetilde{E}_\bot}( z,\vec {r}_{\bot}) = \frac{i \omega}{2 \pi
c z} \int d \vec{ {r}'}_{\bot}~
\widetilde{E}_{\bot}(0,\vec{r'}_{\bot}) \exp{\left[\frac{i \omega
\left|{\vec{ {r}}}_{\bot}-\vec{ {r}'_{\bot}}\right|^2}{2 c
z}\right]}~, \label{fieldpropback}
\end{equation}
yielding, at any position $z$

\begin{equation}
\widetilde{E}_{\bot}\left({z},r_\bot\right) = \frac{K \omega e
A_{JJ}}{2 c^2 \gamma} \left[ \mathrm{Ei} \left(\frac{i \omega
r_\bot^2
  }{2{z} c - L_w c}\right)- \mathrm{Ei} \left(\frac{i \omega
r_\bot^2   }{2{z} c + L_w c}\right) \right]~. \label{Esum}
\end{equation}
Here $\mathrm{Ei}(\cdot)$ indicates the exponential integral
function

\begin{eqnarray}
\mathrm{Ei}(z) = -\int_{-z}^\infty dt
~\frac{\exp[-t]}{t}\label{Eifde} ~,\end{eqnarray}
where the principal value of the integral is taken. The far-zone
limit of this expression is reached for $z\gg L_w$ and reads

\begin{eqnarray}
\widetilde{{E}}_{\bot}(z, \theta)&=& -\frac{K \omega e L_w A_{JJ}
}{2 \gamma c^2 z} \exp\left[i\frac{\omega \theta^2 z}{2c}\right]
\mathrm{sinc}\left[\frac{L_w \omega \theta^2}{4 c}\right]
~,\label{undurad4bis}
\end{eqnarray}
that coincides with Eq. (\ref{generalfin4}) at $C=0$.

\section{\label{sec:boun} Paraxial Green's function with boundary conditions}

In the previous Section \ref{sec:free} we considered the problem
of characterizing radiation from an ultrarelativistic electron
moving on an arbitrary trajectory in free-space using Paraxial
Maxwell's equations.

\begin{figure}
\begin{center}
\includegraphics*[width=100mm]{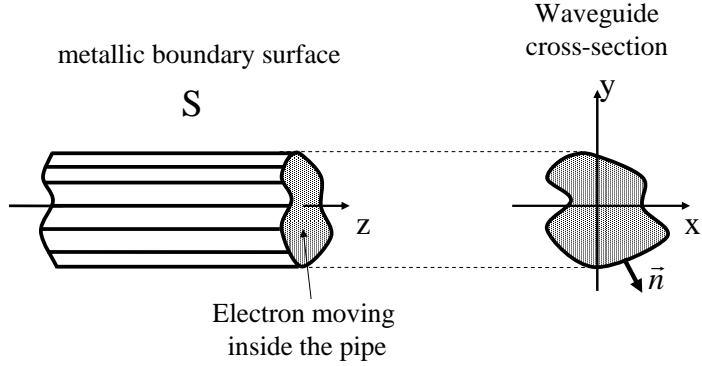}% Here is how to import EPS art
\caption{Geometry of the problem. The electron is moving inside
the pipe. \label{geobou}}
\end{center}
\end{figure}

We now wish to generalize that treatment, to describe radiation
from an ultrarelativistic electron moving on a given trajectory
inside a metallic pipe of arbitrary cross-section. An example is
depicted in Fig. \ref{geobou}, where $\vec{n}$ is a vector field
defined on the boundary surface $S$, such that $|\vec{n}|=1$. At
any point, $\vec{n}$ is orthogonal to $S$ and points outwards.

As we have discussed at the end of Section \ref{sub:gr} by
expanding the far-zone field distribution in free-space in terms
of plane waves, the paraxial approximation applies when the
transverse wave number $k_\bot$ obeys ${k}_\bot^2 c^2/\omega^2 \ll
1$. This is no restriction for ultra-relativistic systems with
$1/\gamma^2 \ll 1$. In fact, non-negligible components of the
angular spectrum are those for ${k}_\bot^2 c^2/\omega^2 \ll 1$. We
can take advantage of this relation also in case a waveguide of
typical geometrical dimension $R$ is present (for a circular
waveguide $R$ indicates the radius). The difference with respect
to the free-space case is in a discrete number of modes,
corresponding to a discrete range of values that $k_\bot$ may
assume. The concept of radiation formation length must be ascribed
to each propagating mode, but for ultra-relativistic electrons
$L_f \gg \lambdabar$ holds for each mode. One concludes that
paraxial approximation holds when ${k}_\bot^2 c^2/\omega^2 \ll 1$
for each mode. The same conclusion can be drawn by remembering
that the paraxial approximation is based on the fact that the
electron velocity must be close to the phase velocity of light. In
the waveguide, the phase velocity of light is larger than $c$,
i.e. $v_\mathrm{ph} = c(1-c^2 k_\bot^2/\omega^2)^{-1/2}$. As a
result, in order for the electron velocity to be near to
$v_\mathrm{ph}$ one must require that the electron velocity is
close to $c$ \textit{and} additionally that ${k}_\bot^2
c^2/\omega^2 \ll 1$. This means that the paraxial approximation is
applicable for frequencies well above the cutoff frequency of a
given mode, $\omega \gg \omega_\mathrm{cutoff} \simeq c k_\bot$.
When $\omega \gg \omega_\mathrm{cutoff}$ the waveguide is said to
be overmoded. Thus,  the paraxial approximation is applicable when
the waveguide is overmoded. Since the smallest cutoff frequency
(that for the fundamental mode) is about $\omega_\mathrm{cutoff}
\sim c/R$, it also follows that the waveguide is overmoded for
$\lambdabar \ll R$. We assume that the last condition is verified
through this paper.

Since $\lambdabar \ll R$ the paraxial approximation can be applied
and is convenient to introduce, as for the free-space case in
Section \ref{sec:free}, the slowly varying envelope of the Fourier
transform of the transverse electric field, "the field"
$\vec{\widetilde{E}}_\bot(z,\vec{r}_\bot)$. Here
$(z,\vec{r}_\bot)$ indicates a point inside the waveguide, as
shown in Fig. \ref{geobou}. The paraxial equation for the field is
identical to that for the free-space problem, Eq. (\ref{field1}),
where the differential operator $\mathcal{D}$ is defined in Eq.
(\ref{Oop}), and $\vec{f}(z, \vec{r}_\bot)$ in Eq. (\ref{fv}).

We only need to account for the presence of the vacuum pipe by
specifying proper boundary conditions. In the case of perfect
metal (we will treat wall resistance effects in Section
\ref{sec:res}), the electric field must be orthogonal to the
boundary surface $S$, i.e.

\begin{equation}
\left(n_x \widetilde{E}_y - n_y \widetilde{E}_x\right)_{\Big|_S}=
\left(\vec{n} \times \vec{\widetilde{E}}_\bot\right)_{\Big|_S} = 0
\label{bou1}
\end{equation}
and

\begin{equation}
\left(\widetilde{E}_{z}\right)_{\Big|_S}= 0 ~.\label{bou2}
\end{equation}
Maxwell's equation for the curl of the magnetic field implies that
Eq. (\ref{bou2}) is equivalent to

\begin{equation}
\left[\left(\vec{\nabla} \times \vec{\widetilde{H}}
\right)_{z}\right]_{\Big|_{S}}= 0~. \label{bou2b}
\end{equation}
Here $\vec{\widetilde{H}}$, analogously to $\vec{\widetilde{E}}$,
is the slowly varying envelope of the magnetic field in the
space-frequency domain, and will be referred to as the "magnetic
field" . Since paraxial approximation applies, the transverse
components of the magnetic field $\widetilde{H}_x$ and
$\widetilde{H}_y$ are related to the electric field by
$\widetilde{E}_x \simeq \widetilde{H}_y$ and $\widetilde{E}_y
\simeq - \widetilde{H}_x$ so that Eq. (\ref{bou2b}) finally yields

\begin{equation}
\left(\frac{\partial \widetilde{E}_x}{\partial x} + \frac{\partial
\widetilde{E}_y}{\partial y}\right)_{\Big|_{S}} =
\left(\vec{\nabla}_\bot \cdot
\vec{\widetilde{E}}_\bot\right)_{\Big|_S} = 0 \label{bou2c}
\end{equation}
Eq. (\ref{bou1}) and Eq. (\ref{bou2c}) are boundary conditions in
the paraxial approximation. To the best of our knowledge, they are
only discussed in \cite{FELB}. At first glance the second
condition, Eq. (\ref{bou2c}), is not obvious. In paraxial
approximation Eq. (\ref{bou2}) looks automatically satisfied, so
that one might be led to use Eq. (\ref{bou1}) only. Actually, Eq.
(\ref{bou1}) poses conditions on the field amplitude while Eq.
(\ref{bou2c}) poses conditions on the field derivatives, so they
are both necessary.

Summing up, we should find solution to the problem

\begin{equation}
\left\{
\begin{array}{l}
\mathcal{D} \left[\vec{\widetilde{E}}_\bot(z,\vec{r}_\bot)\right]
= \vec{f}(z, \vec{r}_\bot)
\\
\left(\vec{n} \times \vec{\widetilde{E}}_\bot\right)_{\Big|_S} = 0
\\
\left(\vec{\nabla}_\bot \cdot
\vec{\widetilde{E}}_\bot\right)_{\Big|_S} = 0~,
\end{array}\right.\label{pb1}
\end{equation}
where the operator $\mathcal{D}$ and the vector function $\vec{f}$
are defined in Eq. (\ref{Oop}) and Eq. (\ref{fv}) respectively. We
proceed by using a Laplace transform technique, that allows one to
dispose of the partial derivative with respect to $z$ in
$\mathcal{D}$ in the first equation in (\ref{pb1}).

First we define Laplace $\mathcal{L}$ and inverse Laplace
$\mathcal{L}^{-1}$ transform of a function $g(z,\vec{r}_\bot)$ as

\begin{equation}
\widehat{g}(p,\vec{r}_\bot) \equiv
\mathcal{L}\left[g(z,\vec{r}_\bot)\right](p) = \int_0^{\infty}
g(z,\vec{r}_\bot) \exp\left[-pz\right] dz \label{ltr}
\end{equation}
with $\mathrm{Re}[p]>0$, and

\begin{equation}
g(z,\vec{r}_\bot) \equiv
\mathcal{L}^{-1}\left[\widehat{g}(p,\vec{r}_\bot)\right](z) =
\int_{\delta-i \infty}^{\delta+i \infty}
\widehat{g}(p,\vec{r}_\bot) \exp\left[pz\right] dp~, \label{latr}
\end{equation}
where $\delta$ is a real number larger than all the real parts of
the singularities of $\widehat{g}(p)$.

Consistently applying a Laplace transformation to the equation set
(\ref{pb1}) we obtain a redefinition of the problem in terms of
the Laplace transform of the field
$\vec{\widehat{E}}_\bot(p,\vec{r}_\bot) \equiv
\mathcal{L}\left[\vec{\widetilde{E}}_\bot(z,\vec{r}_\bot)\right](p)$:

\begin{equation}
\left\{
\begin{array}{l}
\widehat{\mathcal{D}}
\left[\vec{\widehat{E}}_\bot(p,\vec{r}_\bot)\right] = ({2 i \omega
}/{c})~ \vec{\widetilde{E}}_\bot(0,\vec{r}_\bot)
+\vec{\widehat{f}}(p, \vec{r}_\bot)
\\
\left(\vec{n} \times \vec{\widehat{E}}_\bot\right)_{\Big|_S} = 0
\\
\left(\vec{\nabla}_\bot \cdot
\vec{\widehat{E}}_\bot\right)_{\Big|_S} = 0~,
\end{array}\right.\label{pbl1}
\end{equation}
where $\vec{\widehat{f}}$ is the Laplace transform of $\vec{f}$
and

\begin{eqnarray}
\widehat{\mathcal{D}} \equiv \left({\nabla_\bot}^2 + {2 i \omega p
\over{c}}\right) ~.\label{LOop}
\end{eqnarray}
Note that the presence of the initial condition
$\vec{\widetilde{E}}_\bot(0,\vec{r}_\bot)$ in Eq. (\ref{pbl1})
refers to the possibility of introducing an external field into
the system. In what follows we set
$\vec{\widetilde{E}}_\bot(0,\vec{r}_\bot) = 0$.

Now, suppose that we find a tensor of components
$\widehat{G}^{\alpha}_\beta$ such that

\begin{eqnarray}
\widehat{E}^\alpha = \int \widehat{G}^\alpha_\beta(\vec{r}_\bot,
\vec{r'}_\bot,p) \widehat{f}^\beta(\vec{r'}_\bot,p)
~d\vec{r'}_\bot~, \label{Gl}
\end{eqnarray}
where we indicated tensor and vector components with Greek
indexes\footnote{Note that high and low indexes are
interchangeable, as the metric tensor is represented by the
identity matrix.}. In this case, the inverse Laplace transform of
$\widehat{G}^{\alpha}_\beta$, that will be written as
${G}^{\alpha}_\beta$, is the tensor Green's function for the
problem (\ref{pb1}), inclusive of the proper boundary conditions.
As a result\footnote{Formally we are translating the initial point
$z=0$, where $\vec{\widetilde{E}}_\bot(0,\vec{r}_\bot) = 0$, at
$z\longrightarrow -\infty$.}

\begin{eqnarray}
\widetilde{E}^{\alpha}(\vec{r}_\bot,z) = \int_{-\infty}^{z} dz'
\int d\vec{r'}_\bot~ G^\alpha_\beta\left(\vec{r}_\bot,
\vec{r'}_\bot,z-z'\right) f^\beta\left(\vec{r'}_\bot,z'\right) ~,
\label{EG}
\end{eqnarray}
where, as in Section \ref{sub:gr}, we integrate up to $z'=z$
because the radiation formation length for $z - z'<0$ is very
short with respect to the case $z - z' >0$. Summing up,  we first
have to find $\widehat{G}$, then to apply a Laplace inverse
transform in order to get $G$ and, finally, to solve for
$\widetilde{E}^\alpha$.

We start by specifying the eigenvalue problem associated with the
problem set (\ref{pbl1}), that is

\begin{equation}
\left\{
\begin{array}{l}
\widehat{\mathcal{D}} \left[\vec{F}_j(\vec{r}_\bot)\right] =
\Lambda_j \vec{F}_j(\vec{r}_\bot)
\\
\left(\vec{n} \times \vec{F}_j\right)_{\Big|_S} = 0
\\
\left(\vec{\nabla}_\bot \cdot \vec{F}_j\right)_{\Big|_S} = 0~.
\end{array}\right.\label{pbeig}
\end{equation}
Here the index $j$ is understood to identify any possible vector
solving the equations set in (\ref{pbeig}). Posing $\lambda_j
\equiv 2i\omega p/c-\Lambda_j$ yields an equivalent eigenvalue
problem involving Helmholtz's equation:

\begin{equation}
\left\{
\begin{array}{l}
\nabla_\bot^2 \vec{F}_j(\vec{r}_\bot) + \lambda_j
\vec{F}_j(\vec{r}_\bot) = 0
\\
\left(\vec{n} \times \vec{F}_j\right)_{\Big|_S} = 0
\\
\left(\vec{\nabla}_\bot \cdot \vec{F}_j\right)_{\Big|_S} = 0~.
\end{array}\right.\label{pbeigH}
\end{equation}
It can be verified that boundary conditions are homogeneous, so
that the domain of the Laplacian operator is, in our case, the
vector space of twice differentiable (square integrable) functions
obeying boundary conditions in (\ref{pbeigH}). The Laplacian
operator defined in this way can be shown to be self-adjoint with
respect to the inner product defined as

\begin{equation}
\left\langle \vec{f}~, ~\vec{g}~\right\rangle \equiv \int_S
\vec{f}^{~*} \cdot \vec{g} ~~d\vec{r}_\bot ~.\label{inn}
\end{equation}
This means that for any chosen $\vec{f}$ and $\vec{g}$ in the
domain of the Laplacian operator we have

\begin{equation}
\left\langle \vec{f}~,~ \nabla_\bot^2 \vec{g}~\right\rangle^* =
\left\langle \vec{g}~, ~\nabla_\bot^2 \vec{f}~~\right\rangle~,
\label{selfadj}
\end{equation}
where asterisks indicate complex conjugation. As a result,
eigenvalues $\lambda_j$ are real. Eigenvectors $\vec{F}_j$
referring to different eigenvalues  are orthogonal and span over
the entire domain. The spectrum is discrete, and eigenvectors can
be chosen real. We normalize the eigenvectors $\vec{F}_j$ so that

\begin{eqnarray}
\left\langle \vec{F}_j, \vec{F}_j\right\rangle = 1~. \label{normF}
\end{eqnarray}
As a result we can write the decomposition

\begin{equation}
\widehat{G}^\alpha_{~\beta} = \sum_j \frac{F_j^{~\alpha}
F^{~*}_{j~\beta}}{2 i \omega p/c-\lambda_j} ~,\label{LGdec}
\end{equation}
that is analogous to the approach for solution of a
Sturm-Liouville problem. Since eigenvectors can be chosen real we
will do so, and from now on we will avoid complex conjugation
everywhere.

Helmholtz's theorem (also named the fundamental theorem of vector
calculus) states that any vector field (in our case, the
two-dimensional vector field $\vec{F}_j$) that is twice
continuously differentiable and vanishing rapidly enough at
infinity, can be split in the sum of two vector fields; the first,
$\vec{V}_i$, irrotational, i.e. $\vec{\nabla} \times \vec{V}_i=0$
and the second, $\vec{V}_d$, divergenceless, i.e. $\vec{\nabla}
\cdot \vec{V}_d = 0$. Considering two independent scalar (and
real) functions $\psi_j^{\mathrm{TE}}$ and $\psi_j^{\mathrm{TM}}$,
we take $\vec{V}_i = \partial_x \psi_j^{\mathrm{TM}} \vec{e}_x +
\partial_y \psi_j^{\mathrm{TM}} \vec{e}_y$ and $\vec{V}_d =
\partial_y \psi_j^{\mathrm{TE}} \vec{e}_x - \partial_x
\psi_j^{\mathrm{TE}} \vec{e}_y$. We thus obtain the following
representation of $\vec{F}_j$ in terms of scalar functions
$\psi_j^{\mathrm{TE}}$ and $\psi_j^{\mathrm{TM}}$:

\begin{eqnarray}
\vec{F}_j  = \left[\vec{e}_x \left(\partial_y
\psi_j^{\mathrm{TE}}+\partial_x \psi_j^{\mathrm{TM}} \right)+
\vec{e}_y \left(-\partial_x \psi_j^{\mathrm{TE}} +
\partial_y \psi_j^{\mathrm{TM}}\right)\right]~.
\label{defpsi}
\end{eqnarray}
In the following it will be clear that $\psi_j^{\mathrm{TE}}$ and
$\psi_j^{\mathrm{TM}}$ are physically related, respectively, to
transverse electric and transverse magnetic field modes. At the
present time, however, these functions are only introduced for
mathematical convenience. In fact, by substitution in
(\ref{pbeigH}), the vectorial eigenvalue problem for $\vec{F}_j$
can be reformulated in terms of the following scalar problem

\begin{equation}
\left\{
\begin{array}{l}
\nabla_\bot^2 \psi_j^\mathrm{TE,TM}(\vec{r}_\bot) +
\lambda_j^\mathrm{TE,TM} ~\psi_j^\mathrm{TE,TM}(\vec{r}_\bot) = 0
\\
\vec{n} \cdot \left(\vec{\nabla}_\bot
\psi_j^\mathrm{TE}\right)_{\Big|_S} = 0
\\
\left(\psi_j^\mathrm{TM}\right)_{\Big|_S} = 0~.
\end{array}\right.\label{pbeigpsiH}
\end{equation}
Boundary conditions in Eq. (\ref{pbeigpsiH}) do not allow for
hybrid TE-TM modes, as in the case of walls with finite resistance
(see Section \ref{sec:resi}). In other words, solutions
$\psi_j^\mathrm{TE}$ and $\psi_j^\mathrm{TM}$ are not degenerate.
This can be seen by inspection of Eq.
(\ref{pbeigpsiH})\footnote{In fact, suppose there exist two
non-zero solutions $\psi_j^\mathrm{TE}$ and $\psi_j^\mathrm{TM}$
sharing a common eigenvalue $\lambda$. One would have
$\nabla_\bot^2 (\psi_j^\mathrm{TE}-\psi_j^\mathrm{TM})=0$.
However, on $S$, $\psi_j^\mathrm{TM} =0$ and thus, on $S$,
$\nabla_\bot^2 \psi_j^\mathrm{TE}=0$. For this to be in agreement
with Helmholtz's equation in (\ref{pbeigpsiH}), one must have
$\lambda=0$, hence $\nabla_\bot^2 \psi_j^\mathrm{TE}=0$ and
$\nabla_\bot^2 \psi_j^\mathrm{TM}=0$ everywhere inside the
waveguide. Then, since $\psi_j^\mathrm{TM}=0$ on $S$ it also must
be zero everywhere, and degeneracy is impossible.}.

It follows that the particular choice of boundary conditions
allows a choice of eigenvectors $\vec{F}_j$ different from that in
Eq. (\ref{defpsi}), that gives the same results but allows
separation of TE and TM modes:

\begin{eqnarray}
\left\{
\begin{array}{l}
\vec{F}_j^{\mathrm{TE}}  = \vec{e}_x \partial_y
\psi_j^{\mathrm{TE}}- \vec{e}_y \partial_x
\psi_j^{\mathrm{TE}} \\
\vec{F}_j^{\mathrm{TM}}  = \vec{e}_x \partial_x
\psi_j^{\mathrm{TM}}+ \vec{e}_y
\partial_y \psi_j^{\mathrm{TM}}
\end{array}\right.
~. \label{defF2}
\end{eqnarray}
We formally adopt this new eigenvector set, because this is
compatible with the following normalization condition for
$\psi_j^{\mathrm{TE,TM}}$:

\begin{eqnarray}
\int_S d\vec{r}_\bot \left|\vec{\nabla}
\psi_j^{\mathrm{TE,TM}}\right|^2 = 1 ~.\label{normpsi}
\end{eqnarray}
The reason why the eigenvector set (\ref{defF2}) and the
normalization condition Eq. (\ref{normpsi}) are desirable is only
formal: this choice yields traditional results known in literature
for the eigenvalue problem (\ref{pbeigpsiH}) when a particular
geometry of the waveguide is fixed, as in the next Section
\ref{sec:circ}.

Solution of the problem (\ref{pbeigpsiH}) depends on the setup
geometry and cannot be further specified unless the surface $S$ is
given. Later on we will consider the case of a circular waveguide.
At this time we assume that the problem is solved in some
particular geometry, and that $\psi_j^\mathrm{TE,TM}$ are fixed.
As a result $\vec{F}_j$ are known through Eq. (\ref{defpsi}) and
the tensor $\widehat{G}^{\alpha \beta}$ is completely specified.
In order to find the tensor Green's function, we should find the
inverse Laplace transform of $\widehat{G}^{\alpha \beta}$, that is

\begin{equation}
G^{\alpha\beta}(z,\vec{r}_\bot) \equiv
\mathcal{L}^{-1}\left[\widehat{G}^{\alpha\beta}(p,\vec{r}_\bot)\right](z)
= \int_{\delta-i \infty}^{\delta+i \infty}
\widehat{G}^{\alpha\beta}(p,\vec{r}_\bot) \exp\left[pz\right] dp~.
\label{latrG}
\end{equation}
Here $\delta$ is a real number positioned, on the complex plane,
on the right-hand side of all poles $p_j~$. These are located at

\begin{equation}
p_j = - ~i \frac{c\lambda_j}{2 \omega}~,\label{polec}
\end{equation}
as it can be seen by inspecting Eq. (\ref{LGdec}).

From Eq. (\ref{LGdec}), and from the fact that
$\vec{F}_j=\vec{F}_j(\vec{r}_\bot)$, we see that the dependence of
$\widehat{G}$ on $p$ is only in the denominator of the right hand
side of Eq. (\ref{LGdec}). As a result, $\widehat{G}
\longrightarrow 0$ uniformly as $p \longrightarrow \infty$. It
follows that Jordan's lemma applies and we obtain

\begin{eqnarray}
G^{\alpha\beta} = \sum_j \left\{\mathrm{Res}\left[\widehat{G}
\exp(p_j z) \right]\right\}^{\alpha\beta}
\label{Jor1}\end{eqnarray}
where

\begin{eqnarray}
\left(\mathrm{Res}\left[\widehat{G} \exp[p_j z]
\right]\right)^{\alpha\beta} &=& \lim_{p~\rightarrow -{i c
\lambda_j}/({2\omega})} \left\{\left(p+ \frac{i c
\lambda_j}{2\omega} \right)\widehat{G}^{\alpha\beta}(p) \exp[p_j
z]\right\} \cr && = \frac{c}{2i\omega}
F^\alpha\left(\vec{r}_\bot\right)
F^\beta\left(\vec{r'}_\bot\right)\exp\left[-\frac{i c
\lambda_j}{2\omega}z\right]\label{Jor2}
\end{eqnarray}
Substituting Eq. (\ref{defpsi}) in Eq. (\ref{Jor2}), and Eq.
(\ref{Jor2}) in Eq. (\ref{Jor1}) we obtain the following explicit
expression for the tensor Green's function

\begin{eqnarray}
G\left(\vec{r}_\bot,\vec{r'}_\bot,z\right) &&= \frac{c}{2 i
\omega} \sum_j \Bigg\{\exp\left[-\frac{i c \lambda^\mathrm{TE}_j
}{2\omega}z\right]\left[\vec{e}_x~
\partial_{y~}
\psi^\mathrm{TE}_j\left(\vec{r}_\bot\right)-\vec{e}_y~
\partial_x \psi^\mathrm{TE}_j\left(\vec{r}_\bot\right)
\right] \cr&& \otimes\left[\vec{e}_x~
\partial_{y'}
\psi^\mathrm{TE}_j\left(\vec{r'}_\bot\right)-\vec{e}_y~
\partial_{x'} \psi^\mathrm{TE}_j\left(\vec{r'}_\bot\right)\right]\Bigg\}\cr && +
\frac{c}{2 i \omega} \sum_j \Bigg\{\exp\left[-\frac{i c
\lambda^\mathrm{TM}_j }{2\omega}z\right]\left[\vec{e}_x~
\partial_{x}
\psi^\mathrm{TM}_j\left(\vec{r}_\bot\right)+\vec{e}_y~
\partial_{y~} \psi^\mathrm{TM}_j\left(\vec{r}_\bot\right)
\right]\cr&&\otimes\left[\vec{e}_x~
\partial_{x'}
\psi^\mathrm{TM}_j\left(\vec{r'}_\bot\right)+\vec{e}_y~
\partial_{y'} \psi^\mathrm{TM}_j\left(\vec{r'}_\bot\right)\right]\Bigg\}~,\label{Gfex}
\end{eqnarray}
where notation $\otimes$ indicates the tensor product. It should
be noted that the result in Eq. (\ref{Gfex}) coincides with that
obtained in \cite{FELB}\footnote{Numerical factor and sign
differences are due to different definitions for the Green's
function used in \cite{FELB} compared to the present paper. In
that reference a Green's function is first calculated starting
with the rigorous solution for the eigenfunctions of the
waveguide, without applying paraxial approximation. Only in a
second step such approximation is applied. In this paper instead,
we directly derive the paraxial Green's function for an overmoded
waveguide using approximate boundary conditions on the waveguide
walls (obtained by application of the paraxial approximation) and
a Laplace-transform technique. We thus directly solve the initial
value problem in paraxial approximation. The present derivation is
shorter and more straightforward than that in \cite{FELB}.}. Once
the tensor Green's function $G$ is known, the slowly varying
envelope of the electric field is given by

\begin{eqnarray}
\widetilde{E}^\alpha &=& \frac{4\pi e}{c} \int_{-\infty}^{z} dz'
\left\{ \frac{i\omega}{c^2} v_\bot^\beta(z')
G^\alpha_\beta\left(\vec{r}_\bot,\vec{r'}_\bot(z'), z-z'
\right)+\partial'_\beta
G^\alpha_\beta\left(\vec{r}_\bot,\vec{r'}_\bot(z'), z-z'\right)
\right\}\cr &&\times \exp\left[\frac{i\omega}{2c} \int_0^{z'}
\frac{d\bar{z}}{\gamma^2_z(\bar{z})} \right]~,\label{efielG}
\end{eqnarray}
where we made explicit use of Eq. (\ref{moregen}), and derivatives
$\partial'_\beta$ are taken with respect to $\vec{r'}_\bot$ .

\section{\label{sec:circ} Green's function for a circular waveguide}

We now restrict our analysis to a particular geometry of the
waveguide. Namely, we select a circular cross-section of radius
$R$. In order to explicitly compute the tensor Green's function in
Eq. (\ref{Gfex}) we need to solve the problem in (\ref{pbeigpsiH})
with this particular choice of geometry.

For both TE and TM modes, we need to solve Helmholtz's equation.
Because of the particular symmetry in the case of a circular
waveguide, it is convenient to use transverse polar coordinates so
that $\vec{r}_\bot$ is identified by radial distance
$r=\sqrt{x^2+y^2}$ and angle $\phi = \arctan(y/x)$. The equation
set in (\ref{pbeigpsiH}) yields

\begin{eqnarray}
\left[\frac{\partial^2}{\partial r^2}+\frac{1}{r}
\frac{\partial}{\partial
r}+\frac{1}{r^2}\frac{\partial^2}{\partial
\phi^2}\right]\psi_j^\mathrm{TE,TM} + \lambda_j^\mathrm{TE,TM~}
\psi_j^\mathrm{TE,TM} = 0~, \label{Hpolar}
\end{eqnarray}
with boundary conditions

\begin{eqnarray}
\left(\frac{\partial \psi_j^\mathrm{TE}}{\partial r}
\right)_{\Big|_{r=R}}=
0~~~~\mathrm{and}~~~~~\left(\psi_j^\mathrm{TM}\right)_{\Big|_{r=R}}=0~.
\label{Hpolarboun}
\end{eqnarray}
Solutions of Eq. (\ref{Hpolar}) are

\begin{eqnarray}
\left(
\begin{array}{l}
\psi_{mk1}^\mathrm{TE,TM}
\\
\psi_{mk2}^\mathrm{TE,TM}
\end{array}
\right) = A_{mk}^\mathrm{TE,TM}
J_m\left(\sqrt{\lambda_{mk}^\mathrm{TE,TM}}~~ r~\right) \left(
\begin{array}{l}
\sin(m\phi)
\\
\cos(m\phi)
\end{array}
\right) ~,\label{solHpolar}
\end{eqnarray}
where the eigenvalues $\lambda_{mk}^\mathrm{TE,TM}$ and the
normalization coefficients $A_{mk}^\mathrm{TE,TM}$ are specified
by the TE and TM boundary conditions in (\ref{pbeigpsiH}).

Switching to notation $\mu_{mk} \equiv \sqrt{\lambda_{mk}^{TE}}
~R$ and $\nu_{mk} \equiv \sqrt{\lambda_{mk}^{TM}} ~R$ one defines
$\mu_{mk}$ and $\nu_{mk}$ respectively as the roots of equations

\begin{eqnarray}
J'_m\left(\mu_{mk}\right) = 0 \label{TEeig}
\end{eqnarray}
and equations

\begin{eqnarray}
J_m\left(\nu_{mk}\right) = 0 ~.\label{TMeig}
\end{eqnarray}
The correspondent normalization coefficients for the TE and for
the TM modes are given by

\begin{eqnarray}
A_{mk}^\mathrm{TE} =
\sqrt{\frac{a_m}{\pi}}\frac{1}{\sqrt{\mu^2_{mk} - m^2 }
J_m(\mu_{mk})} \label{TEcouplA}
\end{eqnarray}
and

\begin{eqnarray}
A_{mk}^\mathrm{TM} = \sqrt{\frac{a_m}{\pi}}\frac{1}{\nu_{mk}~
J_{m-1}(\nu_{mk})} \label{TMcouplA}
\end{eqnarray}
with $a_0 = 1$ and $a_m = 2$ for $m \geq 1$.

This means that fixing an eigenfunction $F_j$ corresponds to the
specification of two positive integers $m\geq 0$ and $k>0$, and
either a $\sin(\cdot)$ or a $\cos(\cdot)$ solution.

We should now substitute Eq. (\ref{solHpolar}) in the expression
for the Green's function, Eq. (\ref{Gfex}). Before doing so,
however, we need to express partial derivatives of functions
$\psi_j^\mathrm{TE,TM}$ with respect to $x$ and $y$ in terms of
partial derivatives with respect to $r=\sqrt{x^2+y^2}$ and $\phi =
\arctan(y/x)$. This can be done with the help of relations

\begin{equation}
\frac{\partial}{\partial x} = \cos(\phi) \frac{\partial}{\partial
r} - \sin(\phi)\frac{1}{r}\frac{\partial}{\partial \phi}
\label{dpsidxyrf1}
\end{equation}
and

\begin{equation}
\frac{\partial}{\partial y} = \sin(\phi) \frac{\partial}{\partial
r} + \cos(\phi) \frac{1}{r} \frac{\partial}{\partial \phi}
~.\label{dpsidxyrf2}
\end{equation}
Using Eq. (\ref{dpsidxyrf1}) and Eq. (\ref{dpsidxyrf2})  and
taking advantage of:

\begin{eqnarray}
J_m\left(\xi\right) = \frac{\xi}{2 m}\left[J_{m-1}\left(\xi\right)
+J_{m+1}\left(\xi\right) \right]\cr && \label{Jrel1}
\end{eqnarray}
and

\begin{eqnarray}
\frac{d J_m\left(\xi\right)}{d \xi} = \frac{1}{2
}\left[J_{m-1}\left(\xi\right) -J_{m+1}\left(\xi\right)
\right]\label{Jrel2}
\end{eqnarray}
we obtain, from Eq. (\ref{Gfex}), the following expression for the
tensor Green's function in the case of a circular waveguide:

\begin{eqnarray}
G&&\left(\vec{r}_\bot,\vec{r'}_\bot,z-z'\right) = \frac{c}{2 i
\omega} \sum_{m=0}^{\infty}\sum_{k=1}^{\infty}
\left(A^{TE}_{mk}\right)^2 \left(\frac{\mu_{mk}}{2 R }\right)^2
\exp\left[-\frac{i c (z-z')}{2\omega R^2 } \mu_{mk}^2\right] \cr
&& \times \left\{\left[\begin{array}{c} J_{m-1}(\mu_{mk}r/R)
\cos[(m-1)\phi]+J_{m+1}(\mu_{mk}r/R) \cos[(m+1)\phi]
\\
-J_{m-1}(\mu_{mk}r/R) \sin[(m-1)\phi]+J_{m+1}(\mu_{mk}r/R)
\sin[(m+1)\phi]
\end{array}\right] \right. \cr && \left. \otimes \left[\begin{array}{c} J_{m-1}(\mu_{mk}r'/R)
\cos[(m-1)\phi']+J_{m+1}(\mu_{mk}r'/R) \cos[(m+1)\phi']
\\
-J_{m-1}(\mu_{mk}r'/R) \sin[(m-1)\phi']+J_{m+1}(\mu_{mk}r'/R)
\sin[(m+1)\phi']
\end{array}\right] \right. \cr && \left. +
\left[\begin{array}{c} -J_{m-1}(\mu_{mk}r/R)
\sin[(m-1)\phi]-J_{m+1}(\mu_{mk}r/R) \sin[(m+1)\phi]
\\
-J_{m-1}(\mu_{mk}r/R) \cos[(m-1)\phi]+J_{m+1}(\mu_{mk}r/R)
\cos[(m+1)\phi]
\end{array}\right] \right. \cr && \left. \otimes
\left[\begin{array}{c} -J_{m-1}(\mu_{mk}r'/R)
\sin[(m-1)\phi']-J_{m+1}(\mu_{mk}r'/R) \sin[(m+1)\phi']
\\
-J_{m-1}(\mu_{mk}r'/R) \cos[(m-1)\phi']+J_{m+1}(\mu_{mk}r'/R)
\cos[(m+1)\phi']
\end{array}\right]\right\}
\cr && + \frac{c}{2 i \omega}
\sum_{m=0}^{\infty}\sum_{k=1}^{\infty} \left(A^{TM}_{mk}\right)^2
\left(\frac{\nu_{mk}}{2 R }\right)^2 \exp\left[-\frac{i c
(z-z')}{2\omega R^2 } \nu_{mk}^2\right] \cr && \times
\left\{\left[\begin{array}{c} J_{m-1}(\nu_{mk}r/R)
\sin[(m-1)\phi]-J_{m+1}(\nu_{mk}r/R) \sin[(m+1)\phi]
\\
J_{m-1}(\nu_{mk}r/R) \cos[(m-1)\phi]+J_{m+1}(\nu_{mk}r/R)
\cos[(m+1)\phi]
\end{array}\right] \right. \cr && \left. \otimes \left[\begin{array}{c} J_{m-1}(\nu_{mk}r'/R)
\sin[(m-1)\phi']-J_{m+1}(\nu_{mk}r'/R) \sin[(m+1)\phi']
\\
J_{m-1}(\nu_{mk}r'/R) \cos[(m-1)\phi']+J_{m+1}(\nu_{mk}r'/R)
\cos[(m+1)\phi']
\end{array}\right] \right. \cr && \left. +
\left[\begin{array}{c} J_{m-1}(\nu_{mk}r/R)
\cos[(m-1)\phi]-J_{m+1}(\nu_{mk}r/R) \cos[(m+1)\phi]
\\
-J_{m-1}(\nu_{mk}r/R) \sin[(m-1)\phi]-J_{m+1}(\nu_{mk}r/R)
\sin[(m+1)\phi]
\end{array}\right] \right. \cr && \left. \otimes
\left[\begin{array}{c} J_{m-1}(\nu_{mk}r'/R)
\cos[(m-1)\phi']-J_{m+1}(\nu_{mk}r'/R) \cos[(m+1)\phi']
\\
-J_{m-1}(\nu_{mk}r'/R) \sin[(m-1)\phi']-J_{m+1}(\nu_{mk}r'/R)
\sin[(m+1)\phi']
\end{array}\right]\right\}~.
\label{Gfexcir}
\end{eqnarray}
In order to verify the correctness of Eq. (\ref{Gfexcir}) we study
the free-space limit, that corresponds to the limit for large
values of the waveguide radius $R \longrightarrow \infty$.

Since we are interested in characterizing the fields over a finite
transverse direction, and since sources have a finite transverse
size, the limit $R \longrightarrow \infty$ allows one to
substitute Bessel functions in Eq. (\ref{TEcouplA}) and Eq.
(\ref{TMcouplA}) with asymptotic expressions for $k \gg 1$. First
remember that

\begin{eqnarray}
J_m(\xi) \approx \sqrt{\frac{2}{\pi \xi}} \cos\left(\xi-\frac{\pi
m}{2}-\frac{\pi}{4}\right)~,~~~~\xi\gg 1~. \label{asympJl}
\end{eqnarray}
One sees that Eq. (\ref{TEeig}) and Eq. (\ref{TMeig}) are
satisfied by

\begin{eqnarray}
\mu_{mk} = \pi\left(\frac{m}{2} + k
+\frac{1}{4}\right)~,~~\nu_{mk} = \pi\left(\frac{m}{2} + k
+\frac{3}{4}\right)~, \label{munumk}
\end{eqnarray}
yielding

\begin{eqnarray}
J^2_m(\mu_{mk})\simeq \frac{2}{\pi
\mu_{mk}}~,~~~J^2_{m-1}(\nu_{mk})\simeq \frac{2}{\pi \nu_{mk}}~.
\label{JJsq}
\end{eqnarray}
As a result we obtain

\begin{eqnarray}
\left(A_{mk}^\mathrm{TE}\right)^2 = \frac{a_m}{2
\mu_{mk}}~,~~\left(A_{mk}^\mathrm{TM}\right)^2 =  \frac{a_m}{2
\nu_{mk}} ~.\label{limitATETM}
\end{eqnarray}
Substituting Eq. (\ref{limitATETM}) in Eq. (\ref{Gfexcir}), posing
$\xi = \pi k/R$ and replacing the sum over $k$ in Eq.
(\ref{Gfexcir}) with an integral over $d\xi$ we obtain, after
cumbersome but straightforward calculations:

\begin{eqnarray}
G&&\left(r,r',\phi-\phi',z-z'\right) = \frac{c}{4 i \omega \pi}
\sum_{m=-\infty}^{\infty} \int_0^\infty \d\xi \xi
\exp\left[-\frac{i c (z-z')}{2\omega  }\xi^2\right] \cr && \times
J_m(\xi r) J_{m}(\xi r')\left(\begin{array}{cc}
~\cos[m(\phi-\phi')]~ & ~\sin[m(\phi-\phi')]
\\
-\sin[m(\phi-\phi')]~ & ~\cos[m(\phi-\phi')]
\end{array}
\right)~. \label{Gabase}
\end{eqnarray}
Eq. (\ref{Gabase}) can also be written as

\begin{eqnarray}
G&&\left(r,r',\phi-\phi',z-z'\right) = \frac{c}{4 i \omega \pi}
\int_0^\infty \d\xi \xi \exp\left[-\frac{i c (z-z')}{2\omega
}\xi^2\right] \cr && \times J_0(\xi r) J_{0}(\xi
r')\left(\begin{array}{cc} 1~ & ~0 \\ 0~ & ~1\end{array} \right) +
\frac{c}{2 i \omega \pi} \sum_{m=1}^{\infty} \int_0^\infty \d\xi
\xi \exp\left[-\frac{i c (z-z')}{2\omega }\xi^2\right] \cr &&
\times J_m(\xi r) J_{m}(\xi r')\left(\begin{array}{cc}
~\cos[m(\phi-\phi')]~ & ~0
\\
0~ & ~\cos[m(\phi-\phi')]
\end{array}
\right)~. \label{Gabase2}
\end{eqnarray}
As before, $z>z'$.  The integrals in $d \xi$ can be performed
yielding

\begin{eqnarray}
G&&\left(r,r',\phi-\phi',z-z'\right) = -\frac{1}{4 \pi (z-z')}
J_0\left(\frac{\omega r r'}{c (z-z')}\right)\cr &&\times
\exp\left[\frac{i(r^2+r'^2)\omega}{2c(z-z')}\right]
\left(\begin{array}{cc} 1~ & ~0 \\
0~ & ~1\end{array} \right) - \frac{1}{2   \pi} \sum_{m=1}^{\infty}
i^{-m} J_m\left(\frac{\omega r r'}{c (z-z')}\right) \cr &&\times
\exp\left[\frac{i(r^2+r'^2)\omega}{2c(z-z')}\right]
\left(\begin{array}{cc} ~\cos[m(\phi-\phi')]~ & ~0
\\ 0~ & ~\cos[m(\phi-\phi')] \end{array}\right)~. \label{Gabase3}
\end{eqnarray}
Using the Anger-Jacobi expansion

\begin{eqnarray}
J_0(\zeta) + 2 \sum_{m=1}^{\infty} i^{-m} J_m(\zeta) \cos[m
(\phi-\phi')]=\exp[-i \zeta \cos(\phi-\phi')] \label{expesp}
\end{eqnarray}
we find that in free-space the tensor Green's function Eq.
(\ref{Gfexcir}) reduces, for both vertical and horizontal
polarization components, to the scalar Green's function in Eq.
(\ref{green}):

\begin{eqnarray}
G = -{1\over{4\pi (z-z')}} \exp\left[ i\omega{\mid \vec{r}_{\bot}
-\vec{r'}_\bot\mid^2\over{2c (z-z')}}\right] \label{Gscal}~,
\end{eqnarray}
as it must be.

\section{\label{sec:wig} Radiation of wiggled electron in circular waveguide}

We now want to apply Eq. (\ref{Gfexcir}) and Eq. (\ref{efielG}) to
study the case of an undulator in presence of a circular waveguide
in the case of a planar undulator. As we have seen before, the
transverse electron velocity is taken along the horizontal axis
and is specified in Eq. (\ref{vxpl}). The longitudinal Lorentz
factor $\gamma_z(z)$ is not constant, due to oscillatory motion of
electrons along the longitudinal axis. As a result, Eq.
(\ref{phasep}) holds. One must substitute Eq. (\ref{vxpl}) and Eq.
(\ref{phasep}) in Eq. (\ref{efielG}) to calculate the field
components with the help of Eq. (\ref{Gfexcir}).

Eq. (\ref{efielG}) has been derived under the paraxial
approximation. In Section \ref{sec:free} we discussed the region
of applicability of the paraxial approximation in free-space,
concluding that it can always be applied for ultra-relativistic
systems when $1/\gamma^2 \ll 1$. In Section \ref{sec:boun} we
extended our consideration to the case when a waveguide is
present. We concluded that, for an ultra-relativistic system, the
paraxial approximation holds when the waveguide is overmoded, i.e.
$R\gg \lambdabar$. In the case of radiation of wiggled electrons
this condition is necessary but not sufficient. This can be seen
straightforwardly for our practical case of interest $K \gtrsim
1$, with the help of a geometrical argument. From Eq. (\ref{vxpl})
it may be seen that electrons wiggle with a wiggling amplitude

\begin{eqnarray}
r_w = \frac{K}{\gamma} \lambdabar_w = \frac{2 K \lambdabar
\bar{\gamma}_z }{\sqrt{1+K^2/2}} ~.\label{rw}
\end{eqnarray}
We require $R > r_w \gtrsim \bar{\gamma}_z \lambdabar \gg
\lambdabar$. In fact, when $R \gg \lambdabar$ we can still discuss
in general about an overmoded waveguide and, thus, about paraxial
approximation, but in order to discuss about radiation from a
wiggled electron we must impose that the waveguide radius is
larger than the amplitude of the wiggling motion. It is
interesting to note that a different reasoning holds for $K\ll 1$,
leading to a similar conclusion. Namely, for $K\ll 1$ we can
define an inertial frame moving with the same average longitudinal
velocity of the electron, characterized by the average
longitudinal Lorentz factor $\bar{\gamma}_z$. In this frame, the
wiggling motion is non-relativistical and the undulator field is
similar to that of an incoming plane wave with wavelength
$\lambda_w/\gamma$. We are actually dealing with a
Thompson-scattering problem, where the electron radiates as a
dipole at frequency $\omega = c \gamma/\lambdabar_w$. Such
frequency must be higher than the cutoff frequency
$\omega_\mathrm{cutoff} \sim c/R$ so that $R \gtrsim
\lambdabar_w/\gamma$. For frequencies near resonance this
requirement translates to $R \gtrsim \gamma \lambdabar$. As a
result we may state, independently of $K$, that it makes sense to
deal with the case $R\gtrsim \bar{\gamma}_z \lambdabar \gg
\lambdabar$.

As has been seen in Section \ref{sec:free} in the case of
undulator radiation in free-space, the resonance approximation
constitutes a further approximation that can be applied together
with the paraxial one. We concluded that resonance approximation
allows simplifications by exploitation of the large parameter $N_w
\gg 1$. We also discussed how, for odd harmonics and  when
condition (\ref{resext}) holds, further simplifications take
place, allowing one to neglect gradient term in the field equation
and constrained motion in the Green's function, giving a stronger
type of resonance approximation. This strong resonance
approximation can be applied not only in free-space but also in
the presence of a waveguide, exactly as the paraxial approximation
can. In the case of a circular waveguide with radius $R$ Eq.
(\ref{Gfexcir}) must be taken as Green's function instead of the
free-space Green's function, Eq. (\ref{Gscal}). In the limit for
$N_w \gg 1$ and when certain constraints on the parameter space
are verified, similarly as for the free-space case, one may
neglect the gradient term in Eq. (\ref{efielG}), as well as the
constrained particle motion in the Green's function Eq.
(\ref{Gfexcir}).

Let us discuss these simplifications in detail. We may use
Anger-Jacobi expansion to obtain from Eq. (\ref{efielG}) a
representation of the field that is the analogous of Eq.
(\ref{undurad3}) in free-space. Namely, for any value $k$, the
current terms in the integrand in $d z'$ in Eq. (\ref{efielG}) are
proportional to

\begin{eqnarray}
F_c(z')&=& \frac{K}{\gamma} \exp[i (2 p+1\pm 1) k_w z']
\exp\left[i\left(C+\frac{c \zeta^2}{2\omega R^2} \right)z'\right]
J_p\left(u\right) \cr && \times J_{m \pm 1}\left(\frac{\zeta
r'(z')}{R}\right) \mathcal{S}_{m\pm1}(z') \cr &&\label{icz1}
\end{eqnarray}
while the gradient terms are proportional to

\begin{eqnarray}
F_g(z')&=& \frac{c \zeta}{\omega R}\exp[i (2p+1) k_w z']
\exp\left[i\left(C+\frac{c \zeta^2}{2\omega R^2 }\right) z'\right]
J_p\left(u\right)\cr &&\times J_{m}\left(\frac{\zeta
r'(z')}{R}\right)\mathcal{S}_{m}(z')~,\cr &&\label{igz1}
\end{eqnarray}
where $\zeta = \mu_{mk}$ or $\zeta=\nu_{mk}$, $\mathcal{S}_{l}$
can be either  $\cos[l \phi'(z')]$ or $\sin[l \phi'(z')]$ and
parameter $u$ has been defined in Eq. (\ref{uv}). As one may see
by inspection, the role of the index $p$ in Eq. (\ref{icz1}) and
Eq. (\ref{igz1}) is the same as the role of $m$ in Eq.
(\ref{undurad3}), while the role of the index $m$ in Eq.
(\ref{icz1}) and Eq. (\ref{igz1}) is similar to the role of $n$ in
Eq. (\ref{undurad3}).

Comparison between Eq. (\ref{icz1}) or Eq. (\ref{igz1}) and the
integrand in Eq. (\ref{undurad3}) shows a striking similarity
between the role of the observation angle $\theta$ in free-space
and, for each mode, of the quantity $c \zeta/(\omega R)$. This
relation should not be surprising. As we have seen at the
beginning of Section \ref{sub:gr}, after expanding the far-zone
field distribution in free-space in terms of plane waves one sees
that the transverse wave numbers of interest $k_\bot$ obey
${k}_\bot^2 c^2/\omega^2 \ll 1$, i.e.  the paraxial approximation
can always be applied. In fact, the plane wave components of the
angular spectrum for which paraxial approximation is applicable,
are also the only relevant components of the spectrum,  and are
those that can be seen at observation angles  $\theta^2 \ll 1$.
The quantity $c \zeta/(\omega R)$ is the analogous of the
free-space propagation angle of a spacial Fourier component,
$\vec{k}_\bot c/\omega$, that is allowed to vary continuously
across the reciprocal space. The difference with respect to the
free-space case is in the presence of a discrete number of modes,
corresponding to a discrete range of values that $c \zeta/(\omega
R)$ may assume. All we have to do to show the applicability of the
resonance approximation is to rely on the analogy between $\theta$
and $c \zeta/(\omega R)$, and to follow the same derivation for
the free-space case.

First, as already said, Eq. (\ref{icz1}) and Eq. (\ref{igz1}) are
analogous to the integrand of  Eq. (\ref{undurad3}).

Second, condition (\ref{eqq}) can be stated, for each mode, as

\begin{equation}
C  +\frac{c \zeta^2}{2\omega R^2 }  \ll k_w \label{eqqR} ~,
\end{equation}
that allows to exploit the large parameter $N_w\gg1$ by averaging
powers of $\exp[i k_w z']$ over a large number of undulator
periods.

Third, condition (\ref{resext}) allowed to neglect constrained
motion in the Green's function and gradient term in the field
equation in free-space. This enforced the strong resonance
approximation around the fundamental (and other odd) harmonic. In
the presence of a waveguide condition (\ref{resext}) remains
unvaried:

\begin{eqnarray}
\frac{|\Delta \omega|}{\omega_r} \ll 1~,~~~~ \mathrm{i.e.}~~|C|
\ll k_w ~.\label{zetar}
\end{eqnarray}
Note that this condition on frequencies automatically selects
propagation angles of interest $c^2 \zeta^2/(2 \omega R^2) \ll
1/\bar{\gamma}_z^2$. In fact, one can consider propagation angles
outside the range $c \zeta^2/(2 \omega R^2) \ll
1/\bar{\gamma}_z^2$, but in this case Eq. (\ref{icz1}) and Eq.
(\ref{igz1}) exhibit fast oscillations on the integration scale
$L_w$ because condition (\ref{eqqR}) is no more fulfilled. As a
result, with accuracy $1/N_w$, one obtains zero transverse field,
$\vec{\widetilde{E}}_\bot = 0$. These remarks are analogous to the
free-space case, discussed in Section \ref{sub:res}. Also
similarly to the free-space case, the argument in the Bessel
functions $J_{m\pm 1}(\cdot)$ and $J_m(\cdot)$ in Eq. (\ref{icz1})
and Eq. (\ref{igz1}) is small for propagation angles of interest
$c \zeta^2/(2 \omega R^2) \ll 1/\bar{\gamma}_z^2$ :

\begin{equation}
\frac{\zeta r'(z')}{R}  \lesssim \frac{\zeta r_w}{R} = \frac{2
\sqrt{2} K}{\sqrt{2+K^2}}\frac{c \zeta }{R \omega } \bar{\gamma}_z
\ll 1 \label{zetarb} ~.
\end{equation}
This means that,  independently of $K$, we may expand the Bessel
functions $J_{m\pm 1}(\cdot)$ and $J_m(\cdot)$ in Eq. (\ref{icz1})
and (\ref{igz1}), exactly as we did in Section \ref{sub:res} with
Eq. (\ref{undurad3}). As for the free-space case, this expansion
can be performed also around even harmonics. However, for the
first (and odd) harmonic extra-simplifications hold. Using the
same kind of reasonings as in Section \ref{sub:res} and making
explicit use of condition (\ref{zetar}) we see that, after
integration in $d z'$ along a large number of undulator periods
$N_w \gg 1$, the only surviving contributions are of current type
for $m=1$ and either $p=-1$ or $p=0$.  It reads:

\begin{eqnarray}
F_c(z')&=&  \frac{K}{\gamma}J_p(u)\exp\left[i\left(C+\frac{ c
}{2\omega R^2 }\zeta^2\right)z'\right] ~.\label{icz1b}
\end{eqnarray}
When condition (\ref{zetar}) is valid, i.e. $|C|\ll k_w$, we can
neglect, for the first harmonic, gradient term and constrained
motion in the Green's function, exactly as in the free-space case.
Note that Eq. (\ref{icz1b}) is valid also for propagation angles
outside the region of interest $c \zeta^2/(2 \omega R^2) \ll
1/\bar{\gamma}_z^2$. In this case in fact, once the integration in
$d z'$ is performed, one obtains a negligible contribution to the
field, with accuracy $1/N_w$.

Let us consider the propagation direction of interest $c
\zeta^2/(2 \omega R^2) \ll 1/\bar{\gamma}_z^2$. It can be seen
that, in order to obtain non-zero contributions, we may take
$\zeta \sim 1$\footnote{\label{expl}From the phase factor in Eq.
(\ref{efielG}) we see that there is a maximal integration length
of interest, characteristic of any system, after which the
integrand of Eq. (\ref{efielG}) exhibits oscillatory behavior.
This length is the formation length for the system in free space,
$L_f$. Once $L_f$ is known, from the phase factor in $(z-z')$ in
Eq. (\ref{Gfexcir}) follows that there is an upper limit to the
values of $\zeta$ giving non-negligible contributions (with a
certain fixed accuracy) to the field. In fact, above a certain
value of $\zeta$ the exponential functions in $(z-z')$ in Eq.
(\ref{Gfexcir}) show strongly oscillating behavior in the
integration variable $z'$. This upper value is estimated as the
ratio between the pipe radius $R$ and the radiation diffraction
size $\sqrt{\lambdabar L_f}$, that is $R/\sqrt{\lambdabar L_f}$.
We conclude that values of interest for $\zeta$ are up to order
$R/\sqrt{\lambdabar L_f}$. This also mean that, in calculations,
the maximal number $k_\mathrm{max}$ of modes of interest should be
taken as as $k_\mathrm{max} \gg R/\sqrt{\lambdabar L_f}$.}. Thus,
our condition on the propagation directions of interest becomes
$R^2 \gg (\bar{\gamma}_z \lambdabar)^2$. This condition should be
stated together with the previously found constraint $R^2 \gtrsim
(\bar{\gamma}_z \lambdabar)^2 \gg \lambdabar^2$, meaning that
paraxial approximation can be applied and that we can discuss
about wiggled electron motion. Note that we always deal with the
small parameter $(\lambdabar \bar{\gamma}_z)^2 /(\lambdabar L_w)
\sim 1/N_w \ll 1$, where $\lambdabar L_w$ is the squared radiation
diffraction size. It follows that we can discuss about both strong
($R^2 \lesssim \lambdabar L_w$) and weak ($R^2 \gg \lambdabar
L_w$) waveguide influence within the required condition $R^2 \gg
(\lambdabar \bar{\gamma}_z)^2 \gg \lambdabar^2$. It is also
interesting to remind that in cases of practical interest (this is
the case for the undulator beamline at FLASH) we have $K \gtrsim
1$, so that $r_w \simeq 2 \lambdabar \bar{\gamma}_z$.

Summing up, Eq. (\ref{efielG}) gives

\begin{eqnarray}
\widetilde{E}^\alpha = \frac{2 \pi  e \omega \theta_s A_{JJ}}{c^2}
\int_{-L_w/2}^{L_w/2} dz' \exp[i C z']
G^\alpha_1{\Big|_{r'(z')=0}} ~,\label{fieldplcart}
\end{eqnarray}
where $A_{JJ}$ is defined in Eq. (\ref{AJJ}), and $G$ is given in
Eq. (\ref{Gfexcir}). We obtain

\begin{eqnarray}
\widetilde{E}_{x}(r,\phi,z) = -\frac{i \omega e \theta_s
A_{JJ}}{c^2} \sum_{k=1}^{\infty} && \left\{ \mathcal{A}^{\mu}_k(z)
\left[J_o\left(\mu_{1k} \frac{r}{R} \right) + J_2\left(\mu_{1k}
\frac{r}{R} \right) \cos( 2 \phi)\right] + \right.\cr &&\left.
\mathcal{A}^{\nu}_k(z) \left[J_o\left(\nu_{1k} \frac{r}{R}
\right)- J_2\left(\nu_{1k} \frac{r}{R} \right) \cos( 2
\phi)\right]\right\} \label{fieldexp}
\end{eqnarray}
and

\begin{eqnarray}
\widetilde{E}_{y}(r,\phi,z) =  - \frac{i \omega e \theta_s
A_{JJ}}{c^2} \sin(2 \phi) \sum_{k=1}^{\infty} && \left\{
\mathcal{A}^{\mu}_k(z) J_2\left(\mu_{1k} \frac{r}{R} \right)  -
\mathcal{A}^{\nu}_k(z) J_2\left(\nu_{1k} \frac{r}{R} \right)
\right\} ~.\label{fieldeyp}
\end{eqnarray}
where $\mathcal{A}^{\mu}_k(z)$ and $\mathcal{A}^{\nu}_k(z)$ are
given by

\begin{eqnarray}
\mathcal{A}^{\mu}_k(z) = \frac{ C^\mu_k \exp[-i C^\mu_k z]
}{(\mu^2_{1k}-1) J_1^2(\mu_{1k})} L_w
\mathrm{sinc}\left[\frac{L_w}{2}\left(C^{\mu}_k
+C\right)\right]\label{Amu}
\end{eqnarray}
and

\begin{eqnarray}
\mathcal{A}^{\nu}_k(z) = \frac{  C^\nu_k \exp[-i C^\nu_k z]
}{\nu^2_{1k} J_0^2(\nu_{1k})} L_w
\mathrm{sinc}\left[\frac{L_w}{2}\left(C^{\nu}_k +C\right)\right]~,
\label{Anu}
\end{eqnarray}
where

\begin{eqnarray}
C^\mu_k = \frac{\mu_{1k}^2 c}{2 \omega R^2}~,~~~ C^\nu_k =
\frac{\nu_{1k}^2 c}{2 \omega R^2} ~.\label{cmunu}
\end{eqnarray}
Note that each mode propagates according to an exponential law in
$z$, as specified in Eq. (\ref{Amu}) and Eq. (\ref{Anu}). The
field can be propagated forwards or backwards, changing the sign
of $z$. The field at position $z=0$ can be interpreted as
radiation from a virtual source placed in the undulator center and
producing a laser-like beam, similarly as in the free-space case.

As before, it is instructive to study the free-space limit.
Proceeding as in the previous Section \ref{sec:circ}, i.e. using
asymptotic expressions for $J_1^2(\mu_{1k})$ and
$J_0^2(\nu_{1k})$, and substituting the summation over $k$ with an
integral in $d \xi$, with $\xi = \pi k/R$, we obtain the following
free-space limit for the horizontal field component:

\begin{eqnarray}
\widetilde{E}_x(z,r,C) = -\frac{i A_{JJ} e \theta_s L_w}{2 c}
\int_0^\infty d\xi ~\xi
\mathrm{sinc}\left[\frac{L_w}{2}\left(C+\frac{\xi^2 c
}{2\omega}\right)\right] J_o(r\xi) \exp\left[-\frac{i \xi^2 c z
}{2\omega}\right]~.\cr &&\label{horpolfreep}
\end{eqnarray}
The integral in Eq. (\ref{horpolfreep}) can be solved analytically
at perfect resonance ($C=0$) and for $z=0$:

\begin{eqnarray}
\widetilde{E}_{x}(0,r,0) &=& -\frac{i A_{JJ} e \theta_s \omega}{2
c^2} \left[\pi-2 \mathrm{Si}\left(\frac{r^2 \omega }{L_w c
}\right)\right]~.\label{horpolfreep2b}
\end{eqnarray}
Position $z=0$ is in the center of the undulator. The meaning of
Eq. (\ref{horpolfreep2b}) is, in fact, that of the field
distribution of a virtual source located in the middle of the
setup discussed in Eq. (\ref{horpolfreep2}) of Section
\ref{sub:nearz}. Similar reasoning applied to the vertical
component gives back zero, that is in agreement with the fact that
the free-space field from a planar undulator (with vertical
magnetic field) is horizontally polarized.

In free-space, under the applicability of the strong resonance
approximation (condition (\ref{resext})), one obtains a
horizontally polarized field that is also azimuthal symmetric.
This is not the case when one is working under the applicability
of analogous approximation in the presence of a waveguide
(condition (\ref{zetar})). The presence of a waveguide alters
boundary conditions of the problem, that are now formulated mixing
horizontal and vertical polarization components (see Eq.
(\ref{bou1})), thus destroying horizontal polarization and
azimuthal symmetry.

To conclude this Section we mention that in Appendix A we present
a derivation of the field in the case of a helical undulator.
Comparison with the planar undulator case, that we just treated,
yields immediately

\begin{eqnarray}
\widetilde{E}^\alpha =
\frac{A_{JJ}}{2}\left(\widetilde{E}^\alpha_{+}+\widetilde{E}^\alpha_{-}\right)~,
\label{epmpl}
\end{eqnarray}
where $\widetilde{E}^\alpha_{\pm}$ are the fields from electrons
rotating in opposite directions in a helical undulator, as defined
in Appendix A. Eq. (\ref{epmpl}) means that the field from a
planar undulator can be seen as a superposition of fields from two
helical undulators where electrons follow clockwise and
counterclockwise trajectories.

\section{\label{sec:res} Analysis of results (Exemplifications)}

We now analyze our main results, Eq. (\ref{fieldexp}) and Eq.
(\ref{fieldeyp}). To this purpose, it is convenient to introduce
the following normalized units:

\begin{eqnarray}
&&\vec{\hat{E}}_\bot = \left(-\frac{c^2}{A_{JJ} \omega e
\theta_s}\right) \vec{\widetilde{E}}_\bot \cr && \hat{C} = C L_w =
2\pi N_w \frac{\Delta\omega}{\omega_r} \cr && \hat{z} =
\frac{z}{L_w} \cr && \hat{r} = \frac{r}{\sqrt{L_w\lambdabar}} \cr
&& \Omega = \frac{R^2}{L_w\lambdabar} \cr && \hat{C}^{\mu}_k =
{C}^{\mu}_k L_w =  \frac{\mu_{1k}^2}{2 \Omega} \cr &&
\hat{C}^{\nu}_k = {C}^{\nu}_k L_w =  \frac{\nu_{1k}^2}{2 \Omega}
~. \label{norunit}
\end{eqnarray}
We may write Eq. (\ref{fieldexp}) and Eq. (\ref{fieldeyp}) in
normalized units as

\begin{eqnarray}
\hat{E}_{x}(\hat{r},\phi,\hat{z}) = i \sum_{k=1}^{\infty}
&&\left\{ \mathcal{A}^{\mu}_k(\hat{z}) \left[J_o\left(
\frac{\mu_{1k}\hat{r}}{\sqrt{\Omega}} \right) + J_2\left(
\frac{\mu_{1k}\hat{r}}{\sqrt{\Omega}} \right) \cos( 2 \phi)\right]
\right.\cr &&\left. + \mathcal{A}^{\nu}_k(\hat{z})
\left[J_o\left(\frac{\nu_{1k}\hat{r}}{\sqrt{\Omega}} \right)-
J_2\left(\frac{\nu_{1k}\hat{r}}{\sqrt{\Omega}}\right) \cos( 2
\phi)\right]\right\} \label{fieldexpn}
\end{eqnarray}
and

\begin{eqnarray}
\hat{E}_{y}(\hat{r},\phi,\hat{z}) =i \sum_{k=1}^{\infty} &&
\left\{ \mathcal{A}^{\mu}_k(\hat{z})
J_2\left(\frac{\mu_{1k}\hat{r}}{\sqrt{\Omega}} \right) -
\mathcal{A}^{\nu}_k(\hat{z})
J_2\left(\frac{\nu_{1k}\hat{r}}{\sqrt{\Omega}} \right) \right\}
\sin(2 \phi)~.\label{fieldeypn}
\end{eqnarray}
$\mathcal{A}^{\mu}_k(z)$ and $\mathcal{A}^{\nu}_k(z)$ can
similarly be expressed as

\begin{eqnarray}
\mathcal{A}^{\mu}_k(\hat{z}) = \frac{ \hat{C}^\mu_k \exp[-i
\hat{C}^\mu_k \hat{z}] }{(\mu^2_{1k}-1) J_1^2(\mu_{1k})}
\mathrm{sinc}\left[\frac{1}{2}\left(\hat{C}^{\mu}_k
+\hat{C}\right)\right] \label{Amum0}
\end{eqnarray}

and

\begin{eqnarray}
\mathcal{A}^{\nu}_k(\hat{z}) = \frac{ \hat{C}^\nu_k \exp[-i
\hat{C}^\nu_k \hat{z}] }{\nu^2_{1k} J_0^2(\nu_{1k})}
\mathrm{sinc}\left[\frac{1}{2}\left(\hat{C}^{\nu}_k
+\hat{C}\right)\right] ~. \label{Anun0}
\end{eqnarray}
Note that the $\mathrm{sinc}(\cdot)$ functions in the expressions
for $\mathcal{A}^{\mu,\nu}_k$ is a direct consequence of our model
of the undulator, where the magnetic field is instantaneously
switched on and off, at positions $\hat{z}=-1/2$ and $\hat{z}=1/2$
respectively. In fact, the $\mathrm{sinc}({\cdot})$ functions are
Fourier transforms of a rectangular function with respect to
$\hat{C}^\mu_{k} +\hat{C}$ or $\hat{C}^\nu_{k} +\hat{C}$,
modelling the instantaneous switch on and switch off of the
undulator field. The presence of high frequency components in the
rectangular function implies the presence of contributions with
high values of $k$. In its turn, the sum over $k$ in Eq.
(\ref{Gfexcir}) can be interpreted as a superposition of plane
waves propagating at angles $\mu_{1k}/\sqrt{\Omega}$ (or
$\nu_{1k}/\sqrt{\Omega}$) in units of the diffraction angle
$\sqrt{\lambdabar/L_w}$. Thus, higher values of $k$ correspond to
the introduction of high spatial frequency components in the
expressions for the field, Eq. (\ref{fieldexpn}) and Eq.
(\ref{fieldeypn}).

However, we should account for the fact that our theory applies
with a finite accuracy related to the use of the resonance
approximation. When using this approximation we neglect
contributions to the field with an accuracy of order $1/N_w$. The
instantaneous switch on and switch off is also valid with this
accuracy. This means that, in the analysis of our results, it does
not make sense to consider high spatial frequency contributions
due to abrupt switching of the undulator fields on a scale shorter
than the undulator period $\lambda_w$, because these are outside
of the accuracy of the paraxial approximation. We may introduce a
spatial frequency filter in our expression for the field by
replacing Eq. (\ref{fieldplcart}) with

\begin{eqnarray}
\widetilde{E}^\alpha = \frac{2 \pi  e \omega \theta_s A_{JJ}}{c^2}
\int_{-\infty}^{\infty} dz' \exp[i C z'] S(z')
G^\alpha_1{\Big|_{r'(z')=0}} ~,\label{fieldplcart2}
\end{eqnarray}
where the function $S(z')$ introduces some smoothing of the
rectangular undulator profile on a scale of $\lambda_w$. In
normalized units Eq. (\ref{fieldplcart2}) reads

\begin{eqnarray}
\hat{E}^\alpha = - 2\pi \int_{-\infty}^{\infty} dz' \exp[i \hat{C}
\hat{z}'] S(\hat{z}') G^\alpha_1{\Big|_{\hat{r}'(\hat{z}')=0}}
~.\label{fieldplcartn}
\end{eqnarray}
We model $S(\hat{z}')$ as a constant function along the undulator
length with exponentially decaying edges on a typical distance
$\Delta$:

\begin{eqnarray}
S(\hat{z}') =  \left\{\begin{array}{cl}
\exp\left[-(\hat{z}'+1/2)^2/(2\Delta^2)\right] &~~~~\mathrm{for}~
\hat{z}'<-1/2
\\ 1&~~~~\mathrm{for}~ -1/2<\hat{z}'<1/2
\\ \exp\left[-(\hat{z}'-1/2)^2/(2
\Delta^2)\right] &~~~~\mathrm{for}~ \hat{z}'>1/2
\end{array}\right.~.\label{S1n}
\end{eqnarray}
It follows that Eq. (\ref{Amum0}) and Eq. (\ref{Anun0}) should be
replaced by

\begin{eqnarray}
\mathcal{A}^{\mu}_k(\hat{z}) = \frac{ \hat{C}^\mu_k \exp[-i
\hat{C}^\mu_k \hat{z}] }{(\mu^2_{1k}-1) J_1^2(\mu_{1k})}
\mathcal{F}\left\{S(\hat{z}'),\left(\hat{C}^\mu_k
+\hat{C}\right)\right\}\label{Amum}
\end{eqnarray}

and

\begin{eqnarray}
\mathcal{A}^{\nu}_k(\hat{z}) = \frac{ \hat{C}^\nu_k \exp[-i
\hat{C}^\nu_k \hat{z}] }{\nu^2_{1k} J_0^2(\nu_{1k})}
\mathcal{F}\left\{S(\hat{z}'),\left(\hat{C}^\nu_k
+\hat{C}\right)\right\}~, \label{Anun}
\end{eqnarray}
where $\mathcal{F}\{S(\hat{z}'),(\hat{C}^{\mu,\nu}_k +\hat{C})\}$
is the Fourier transform of the function $S$ with respect to
$(\hat{C}^{\mu,\nu}_k +\hat{C})$. The introduction of the function
$S(\hat{z}')$ introduces a suppression of higher frequency
components of $\mathcal{F}\{S\}$, and corresponds to a suppression
of higher spacial frequencies in the field distribution, i.e. to a
smoother field distribution. For a fixed value of $\Omega$ and
$\hat{z}$ we qualitatively expect a smoother field distribution
for finite values of ${\Delta}$ compared to the hard-edge limit
for ${\Delta}\ll 1$. Note that the level of high spacial
frequencies depends in a complicated way on $\Omega$, because the
perfectly metallic waveguide effectively acts as a mirror. The
value of $\Delta$ should be actually chosen to cut off high
spatial frequencies that are outside the region of applicability
of the resonance approximation, i.e. $\Delta \sim 1/N_w$. Thus,
the correct value of $\Delta$ depends, case by case, on the type
of setup considered. The Fourier transform of the function
$S(\hat{z}')$ in Eq. (\ref{S1n}) to be inserted into Eq.
(\ref{Amum}) and into Eq. (\ref{Anun}) reads

\begin{eqnarray}
\mathcal{F}&&\left\{S(\hat{z}'),\left(\hat{C}^{\nu,\mu}_k
+\hat{C}\right)\right\} =
\mathrm{sinc}\left[\frac{1}{2}\left(\hat{C}^{\nu,\mu}_k
+\hat{C}\right)\right]+ \sqrt{2\pi}~ {\Delta}
\exp\left[-\frac{{\Delta}^2}{2}\left(\hat{C}^{\nu,\mu}_k
+\hat{C}\right)^2\right] \cr && \times
\left\{\cos\left[\frac{1}{2}\left(\hat{C}^{\nu,\mu}_k
+\hat{C}\right)\right]-\sin\left[\frac{1}{2}\left(\hat{C}^{\nu,\mu}_k
+\hat{C}\right)\right]
\mathrm{erfi}\left[\frac{{\Delta}}{\sqrt{2}}\left(\hat{C}^{\nu,\mu}_k
+\hat{C}\right)\right]\right\} ~.\label{FTSn}
\end{eqnarray}
where the imaginary error function $\mathrm{erfi}(\cdot)$ is
defined as

\begin{eqnarray}
\mathrm{erfi}\left(z\right) = \frac{1}{i} \mathrm{erf}(i z) =
\frac{2}{i\sqrt{\pi}} \int_{0}^{i z} \exp{[-t^2]} dt
~.\label{erfidef}
\end{eqnarray}

\begin{figure}
\begin{center}
\includegraphics*[width=120mm]{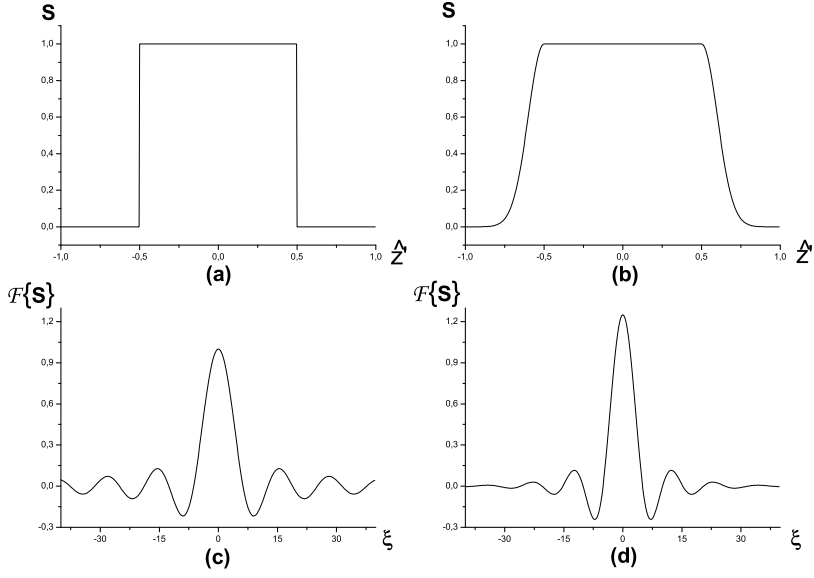}% Here is how to import EPS art
\caption{Comparison between hard-edge undulator case and
high-frequency filtering. Function $S(\hat{z}')$ in the hard-edge
case with ${\Delta}=0$ (a) and in the case when filtering is
applied with ${\Delta}=0.1$ (b). Their Fourier transforms
$\mathcal{F}\{S\}$ with respect to $\xi = \hat{C}^{\mu,\nu}_k
+\hat{C}$ are plot in (c) for the hard edge case and in (d) when
filtering is present. \label{SFTS}}
\end{center}
\end{figure}

Fig. \ref{SFTS} presents a comparison between functions $S$ and
$\mathcal{F}\{S\}$ for ${\Delta}=0.1$ and ${\Delta}=0$, the latter
case describing an undulator with hard edges. A filtering effect
can be clearly seen, suppressing higher frequency components of
$\mathcal{F}\{S\}$.

\begin{figure}
\begin{center}
\includegraphics*[width=120mm]{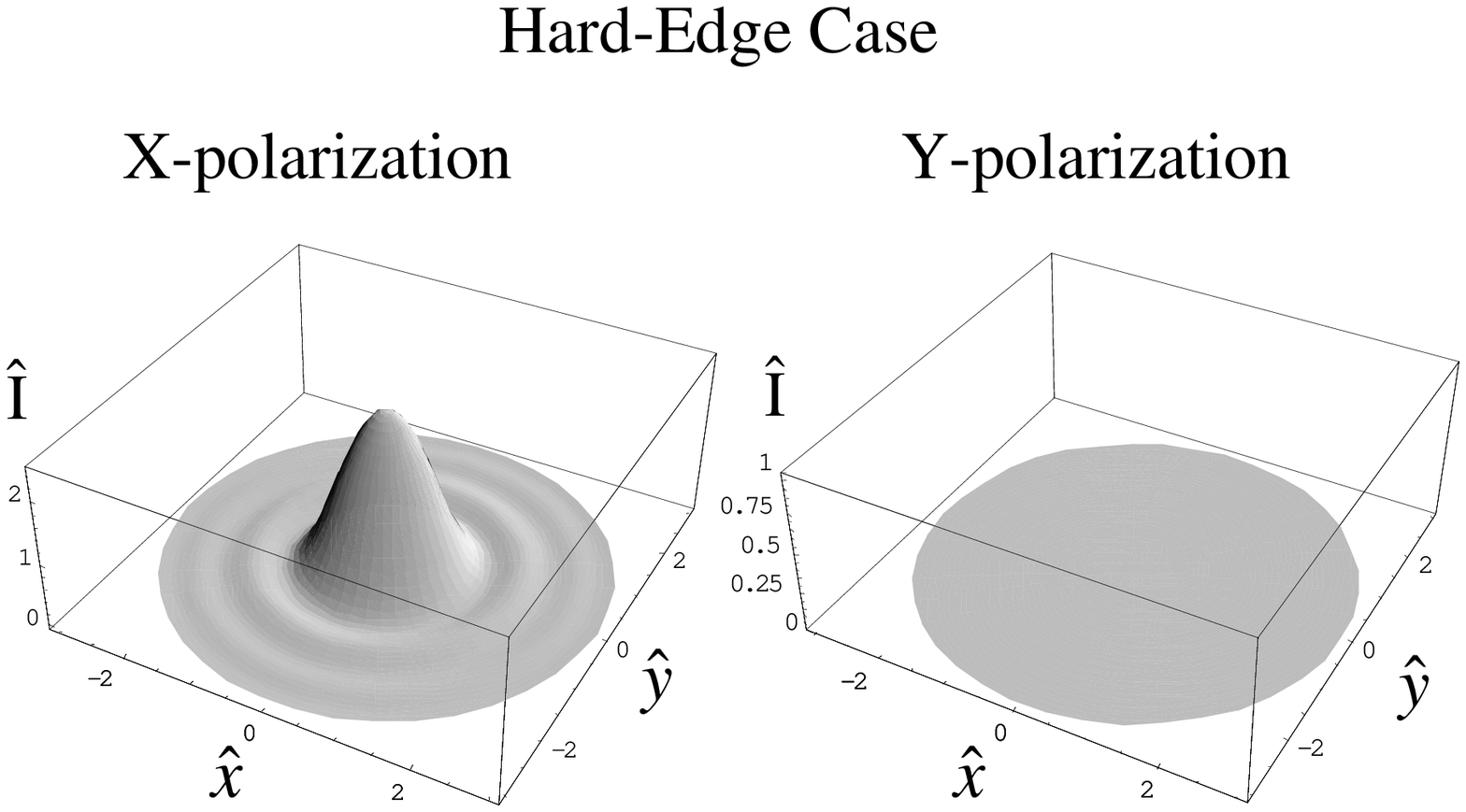}% Here is how to import EPS art
\caption{Three-dimensional view of the virtual source in the
free-space limit for ${\Delta}=0$ (hard-edge case) and $\hat{C}=0$
(perfect resonance) . \label{VS3D}}
\end{center}
\end{figure}

Eq. (\ref{fieldexpn}) and Eq. (\ref{fieldeypn}) completely
describe radiation from a planar undulator in the presence of a
circular waveguide. The limit for $\Omega \gg 1$ can be taken as
free-space limit, and allows one to compute the free-space case as
a finite sum of modes, that is in an alternative way with respect
to the calculation of integrals in \cite{OURF}. Position
$\hat{z}=0$ corresponds to the virtual source discussed at the end
of Section \ref{sub:nearz}. The concept of laser-like radiation
beam can in fact be naturally extended to the case a waveguide is
present. Only, different boundary conditions have to be accounted
for. Mathematically this means that propagation of the laser-like
source must be performed with the proper Green's function, Eq.
(\ref{Gfexcir}) multiplied by the numerical factor $-2 i
\omega/c$, rather than with the free-space Green's function, Eq.
(\ref{Gscal}) (also multiplied by $-2 i \omega/c$). As said before
one must carefully select the number of modes used for
computation, according to $k \gg \sqrt{\Omega}$. In Fig.
\ref{VS3D} we show a three-dimensional view of the virtual source
in the free-space limit for ${\Delta}=0$ and $\hat{C}=0$. We plot
$\hat{I} =|\hat{E}|^2$ for both horizontal (x) and vertical (y)
polarization components as a function of $\hat{x}$ and $\hat{y}$,
defined by $\vec{\hat{r}} = \hat{x}~ \vec{e}_x+ \hat{y}~
\vec{e}_y$.

\begin{figure}
\begin{center}
\includegraphics*[width=120mm]{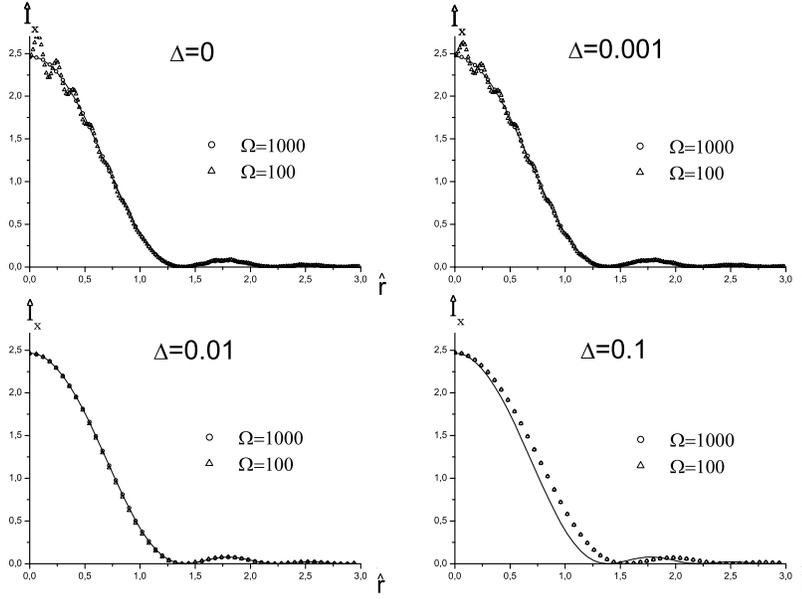}% Here is how to import EPS art
\caption{Intensity profiles of the virtual source ($\hat{z}=0$) at
large values of $\Omega = R^2/(\lambdabar L_w)$ for different
values of $\Delta$ and $\hat{C}=0$ (perfect resonance). This 2D
plot is obtained cutting the 3D intensity profile at $\hat{y}=0$
(i.e. at $\phi=0$). The solid line is obtained with the help of
Eq. (\ref{horpolfreep2}). \label{Ohigh}}
\end{center}
\end{figure}
\begin{figure}
\begin{center}
\includegraphics*[width=130mm]{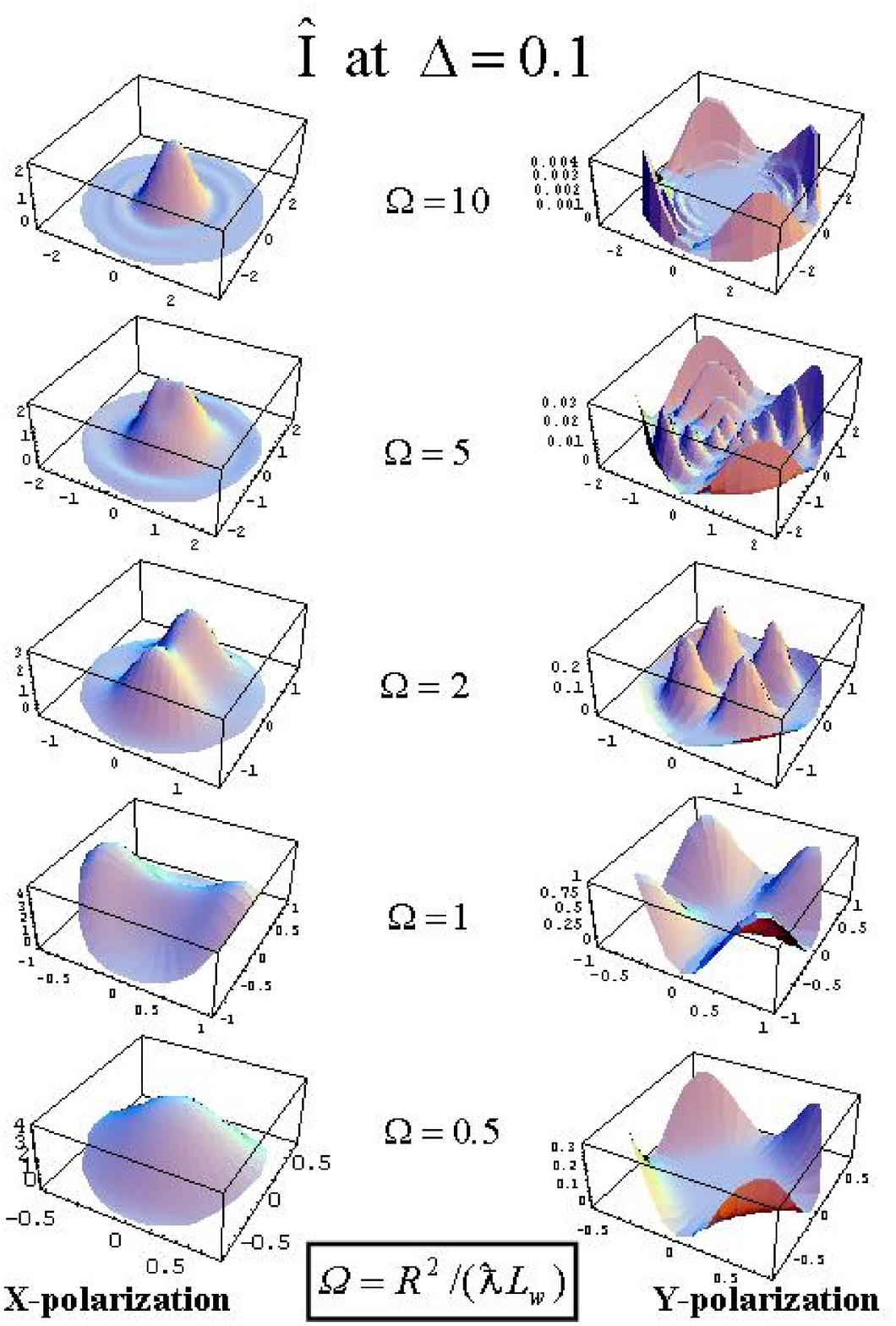}% Here is how to import EPS art
\caption{Intensity profiles of the virtual source ($\hat{z}=0$) at
different values of $\Omega=R^2/(\lambdabar L_w)$ for $\Delta =
0.1$ and $\hat{C}=0$ (perfect resonance). \label{delta0p1}}
\end{center}
\end{figure}
\begin{figure}
\begin{center}
\includegraphics*[width=130mm]{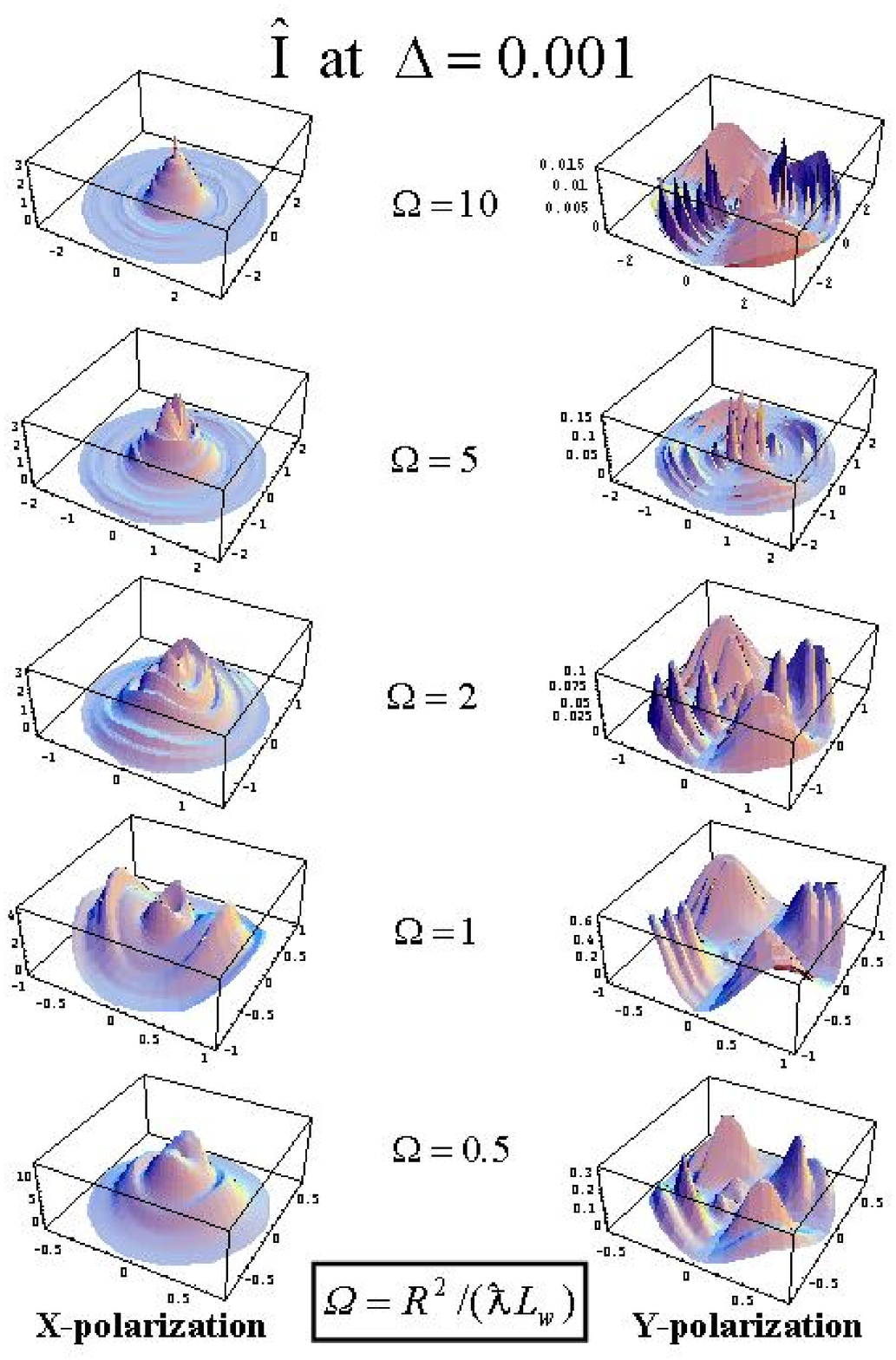}% Here is how to import EPS art
\caption{Intensity profiles of the virtual source ($\hat{z}=0$) at
different values of $\Omega = R^2/(\lambdabar L_w)$ for $\Delta =
0.001$ and $\hat{C}=0$ (perfect resonance). \label{delta0p001}}
\end{center}
\end{figure}

We set $\hat{C}=0$ (perfect resonance). In Fig. \ref{Ohigh} we
show a comparison between the analytic expression for the
intensity distribution $\hat{I}$ at the virtual source position,
that is obtained from a the normalized version of Eq.
(\ref{horpolfreep2}), and numerical expressions obtained through
Eq. (\ref{fieldexpn}) for different values of ${\Delta}$ at
$\Omega=1000$ and $\Omega=100$. A 2D plot is obtained by cutting
the 3D intensity profile at $\hat{y}=0$ (i.e. at $\phi=0$). Only
the horizontal polarization component is important in this case.
The importance of high-frequency filtering becomes clear in the
case for $\Omega=100$ and ${\Delta}=0$ or ${\Delta}=0.001$. In
general, the smaller the value of $\Omega$, the more filtering
becomes relevant. Differences between results for ${\Delta}=0$ and
${\Delta}=0.001$ should be taken as exemplification of the
filtering process. However, they are irrelevant from a practical
viewpoint. For example, if the number of undulator periods is of
order $100$, high frequency components for $\Delta = 0.001$ or
$\Delta =0$ will not be distinguishable and, in fact, fall outside
of the region of applicability of the resonance approximation.

The behavior of the intensity profile is shown in Fig.
\ref{delta0p1} for $\Delta = 0.1$ and in Fig. \ref{delta0p001} for
$\Delta = 0.001$, where $\hat{I} =|\hat{E}|^2$ is plot for both
the horizontal (Eq. (\ref{fieldexpn})) and vertical (Eq.
(\ref{fieldeypn})) polarization components at $\hat{z}=0$. The
number of modes used in all plots are in all cases much larger
than $\sqrt{\Omega}$. Furthermore, we verified that results do not
change by changing the number of modes used in the computation,
provided that condition $k_\mathrm{max} \gg \sqrt{\Omega}$ is
fulfilled. Comparison of Fig. \ref{delta0p1} and Fig.
\ref{delta0p001} clearly shows the effect of different edge
dimensions. Note that in the free-space case, the horizontally
polarized field is azimuthal symmetric.  The situation becomes
more complicated when the influence of the waveguide begins to be
important, i.e. for values of $\Omega$ comparable with unity. In
this case the field presents both horizontal and vertical
polarization components, and azimuthal symmetry is lost, as is
easy to see from figures and by inspection of Eq.
(\ref{fieldexpn}) and Eq. (\ref{fieldeypn}).

The knowledge of Eq. (\ref{fieldexpn}) and Eq. (\ref{fieldeypn})
solves the problem of characterizing planar undulator radiation in
the presence of a circular vacuum pipe. Depending on the
application one may use these equations in different ways. For
example, one can propagate radiation in the presence of a vacuum
pipe with a certain radius up to a given distance down the
beamline and subsequently propagate the electric field
distribution in free-space with the help of Eq. (\ref{Gscal}) (or
with the help of Fourier codes like ZEMAX \cite{ZEMA}). One may
also account for changes of the waveguide radius by further
solving the initial value problem for the field with the help of
the proper Green's function, that will be proportional to Eq.
(\ref{Gfexcir}). In principle one may even account for different
geometries of the pipe, passing from a section of the beamline to
another one, by investigating different eigenvalue problems with
respect to that for a circular pipe considered here. Many are the
possibility of application of our theory. However, due to loss of
azimuthal symmetry it is not easy to directly investigate the
field distribution, because plots of the field forcefully become
three-dimensional, as we have just seen.

It is thus worth to focus on finding figure of merits describing
the influence of the pipe, that can give designers and scientists
some measure of the influence of the waveguide in the stage of
beamline design or planning of experiments.

One figure of merit of interest is the ratio between the power
density for a specific value of $\Omega$ integrated over the
waveguide cross-section and the angle-integrated power density in
free space:

\begin{eqnarray}
\hat{W} = {\int d\vec{\hat{r}}_\bot
\left|\vec{\hat{E}}_\bot(\Omega)\right|^2}\Bigg/\left({\int
d\vec{\hat{r}}_\bot \left|\lim_{\Omega \rightarrow \infty}
\vec{\hat{E}}_\bot\right|^2}\right)~. \label{Wn}
\end{eqnarray}
When calculating the denominator in Eq. (\ref{Wn}) we should
introduce the same high-frequency cutoff, $\Delta$, introduced for
the numerator. This means that we substitute Eq. (\ref{Amum}) and
Eq. (\ref{Anun}) (with $S(\hat{z}')$ defined in Eq. (\ref{S1n}))
in Eq. (\ref{fieldexpn}) and Eq. (\ref{fieldeypn}), we take the
limit for $\Omega \longrightarrow \infty$ and we integrate in $d
\vec{\hat{r}}_\bot$ on a transverse plane at any fixed position
$\hat{z}$, the result being independent of $\hat{z}$. In
particular we choose to integrate the far-zone distribution and we
make use of the definition $\vec{\hat{\theta}} \equiv
\vec{\hat{r}}_\bot/\hat{z}$. Calling $D$ the denominator in Eq.
(\ref{Wn}) we obtain:

\begin{eqnarray}
D &=& {\int d\vec{\hat{r}}_\bot \left|\lim_{\Omega
\rightarrow\infty} \vec{\hat{E}}_\bot\right|^2} = \frac{\pi}{2}
\int_0^\infty d\hat{\theta} \hat{\theta}
\left\{\mathrm{sinc}\left(\frac{\hat{\theta}^2}{4}\right)
\right.\cr &&\left. +\sqrt{2\pi} \Delta \exp\left[-\frac{\Delta^2
\hat{\theta}^4}{8}\right]
\left[\cos\left(\frac{\hat{\theta}^2}{4}\right)-\mathrm{erfi}\left(\frac{
\hat{\theta}^2 \Delta}{2\sqrt{2}}
\right)\sin\left(\frac{\hat{\theta}^2}{4}\right)\right] \right\}^2
~. \label{DENWn0}
\end{eqnarray}
The case for $\Delta=0$ can be calculated analytically as:

\begin{eqnarray}
D = \frac{\pi}{2} \int_0^\infty d\hat{\theta} \hat{\theta}
\mathrm{sinc}^2\left(\frac{\hat{\theta}^2}{4}\right) =
\frac{\pi^2}{2}~, \label{DENWn}
\end{eqnarray}
that also follows from a normalized version of Eq.
(\ref{undurad4bis}). Numerical integration gives $D \simeq 5.02$
for $\Delta = 0.01$ and $D \simeq 5.81 $ for $\Delta = 0.1$, that
are cases considered in this paper. It remains to calculate

\begin{eqnarray}
\int d\vec{\hat{r}}_\bot \left|\vec{\hat{E}}_\bot(\Omega)\right|^2
=  \int_0^{2\pi} d\phi \int_0^{\sqrt{\Omega}} d\hat{r} \hat{r}
\left(\left|\vec{\hat{E}}_x(\Omega)\right|^2+\left|\vec{\hat{E}}_y(\Omega)\right|^2\right)
\label{NUMWn}
\end{eqnarray}
In order to calculate the square modulus in Eq. (\ref{NUMWn}) one
has to compute the square modulus of an infinite sum in $k$. It
can be shown that cross terms of this sum involving both TE and TM
modes or different values of $k$ vanish. One thus finds the result

\begin{eqnarray}
\int d\vec{\hat{r}}_\bot \left|\vec{\hat{E}}_\bot(\Omega)\right|^2
&=& 2 \pi \sum_{k=1}^{k=\infty}
\left\{\left|\mathcal{A}^\mu_k\right|^2\int_0^{\sqrt{\Omega}}
d\hat{r} \hat{r} \left[J_o^2\left(\frac{\mu_{1k}
\hat{r}}{\sqrt{\Omega}}\right)+J_2^2\left(\frac{\mu_{1k}
\hat{r}}{\sqrt{\Omega}}\right)\right] \right\} \cr && + 2 \pi
\sum_{k=1}^{k=\infty}\left\{
\left|\mathcal{A}^\nu_k\right|^2\int_0^{\sqrt{\Omega}} d\hat{r}
\hat{r} \left[J_o^2\left(\frac{\nu_{1k}
\hat{r}}{\sqrt{\Omega}}\right)+J_2^2\left(\frac{\nu_{1k}
\hat{r}}{\sqrt{\Omega}}\right)\right] \right\} \label{NUMWn2}
\end{eqnarray}
Integrals in Eq. (\ref{NUMWn2}) can be calculated analytically.
Further using the fact that $\nu_{1k}$ is the $k$-th root of the
Bessel function $J_1(\cdot)$ we obtain

\begin{eqnarray}
\hat{W} &=& \frac{\pi\Omega}{D} \sum_{k=1}^{\infty}
\left\{\left|\mathcal{A}^\mu_k\right|^2
\left[J_o^2(\mu_{1k})+J_1(\mu_{1k})\left[J_1(\mu_{1k})-J_3(\mu_{1k})\right]
+J_2^2(\mu_{1k}) \right]\right\} \cr && + \frac{\pi \Omega}{D}
\sum_{k=1}^{k=\infty}\left\{ \left|\mathcal{A}^\nu_k\right|^2
\left[J_o^2(\nu_{1k}) +J_2^2(\nu_{1k}) \right]
\right\}~,\label{Wnfinal}
\end{eqnarray}
where $\mathcal{A}^{\mu}_k(z)$ and $\mathcal{A}^{\nu}_k(z)$ are
given in Eq. (\ref{Amum}) and Eq. (\ref{Anun}), while $D$ is
defined in Eq. (\ref{DENWn0}). One may now study, for a given
value $\Delta\sim 1/N_w$, i.e. for a certain number of undulator
periods, the dependence of $\hat{W}$ on the waveguide parameter
$\Omega$ or on the detuning parameter $\hat{C}$.

\begin{figure}
\begin{center}
\includegraphics*[width=130mm]{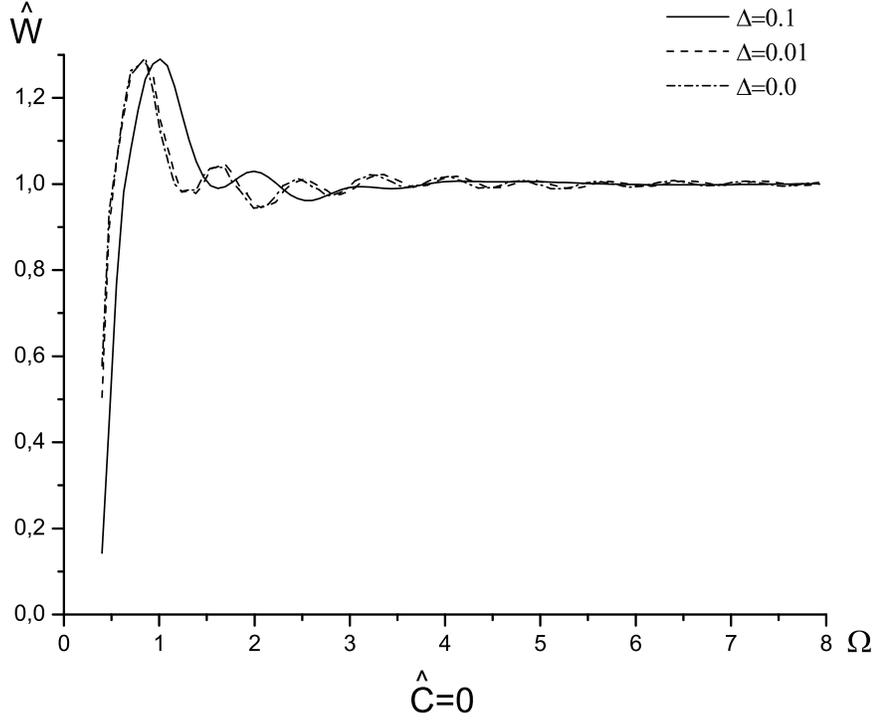}% Here is how to import EPS art
\caption{Plot of $\hat{W}$ as in Eq. (\ref{Wnfinal}) as a function
of $\Omega = R^2/(\lambdabar L_w)$ for $\hat{C}=0$ (perfect
resonance) at different values of $\Delta$. \label{WC0}}
\end{center}
\end{figure}

First we set $\hat{C}=0$ and we plot, in Fig. \ref{WC0}, the
function $\hat{W}(\Omega)$ for different values of $\Delta$
ranging from $0$ to $0.1$. It is seen that edge effects, that are
very important in the intensity distribution, can be seen as a
translation in $\Omega$ when it comes to the figure of merit
$\hat{W}(\Omega)$. Note that the dependence on $\Omega$ in the
coefficients $\left|\mathcal{A}^{\mu,\nu}_k\right|^2$ is of the
form $\sin^2\left[\mu_{1k}^2/(4\Omega)\right]$ or
$\sin^2\left[\nu_{1k}^2/(4\Omega)\right]$ for $\hat{C}=0$, $\Delta
= 0$ and $\hat{z}=0$. This expression is then multiplied by
$\Omega$ in Eq. (\ref{Wnfinal}). Thus $\hat{W}$ must go to zero as
$\Omega \longrightarrow 0$. However, for small values of $\Omega$,
only a few waveguide modes are excited. One thus expects
oscillations in $\hat{W}(\Omega)$. In particular when
$\mu_{11}^2/(4\Omega) = \pi$, i.e. at $\Omega = 0.27$
($\mu_{11}=1.84$), we expect a value of $\hat{W}$ close to zero
(not shown in Fig. \ref{WC0}). The first maximum of $\Omega
\sin^2\left[\mu_{11}^2/(4\Omega)\right]$, for values $\Omega >
0.27$ corresponds to  $\Omega = 0.73$. The shift at higher values
of $\Omega$ in Fig. \ref{WC0} is ascribed to contributions from
other modes.

\begin{figure}
\begin{center}
\includegraphics*[width=130mm]{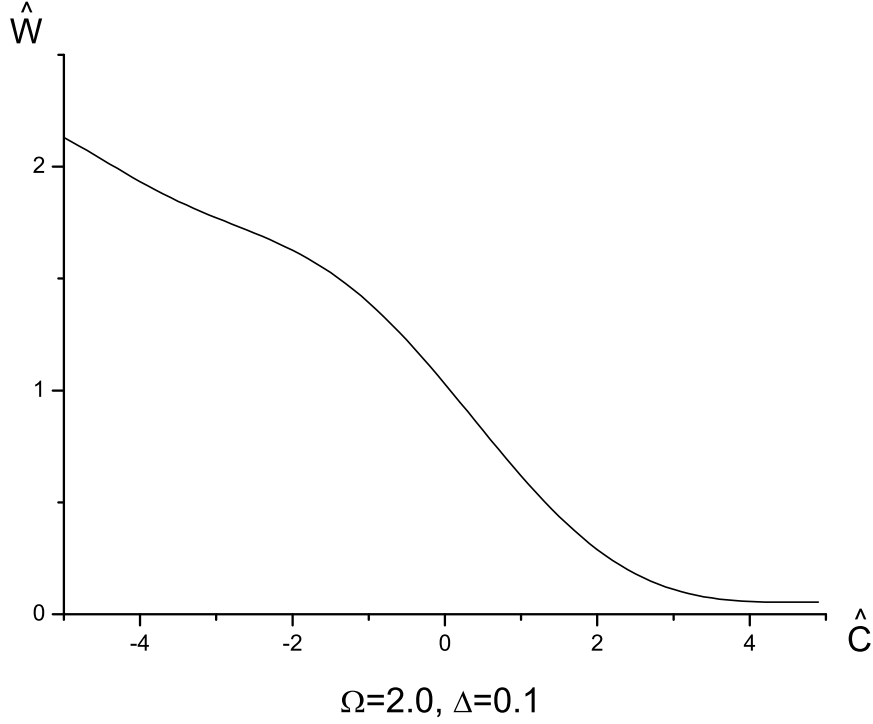}% Here is how to import EPS art
\caption{Plot of $\hat{W}$ as in Eq. (\ref{Wnfinal}) as a function
of $\hat{C} = 2 \pi N_w \Delta \omega/\omega_r$ for $\Omega=2.0$
at $\Delta = 0.1$. \label{WO2D0p1}}
\end{center}
\end{figure}

\begin{figure}
\begin{center}
\includegraphics*[width=130mm]{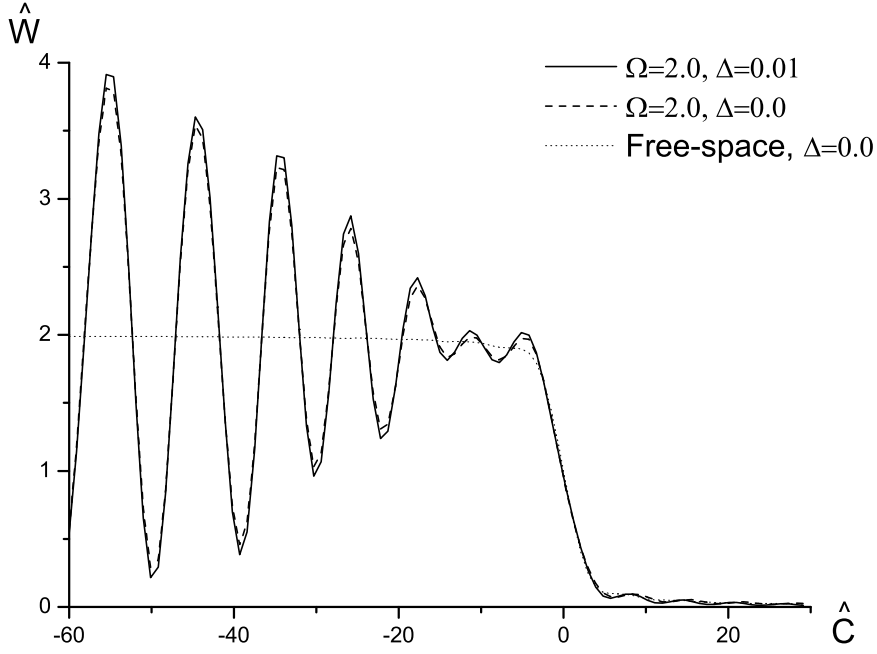}% Here is how to import EPS art
\caption{Plot of $\hat{W}$ as in Eq. (\ref{Wnfinal}) as a function
of $\hat{C}= 2 \pi N_w \Delta \omega/\omega_r$ for $\Omega=2.0$ at
different values of $\Delta=0.01$ and $\Delta=0$. The far-field
limit ($\Omega\longrightarrow \infty$) is also shown for
comparison (dotted line). \label{WO2D0}}
\end{center}
\end{figure}

Another possibility is to set a certain value for $\Omega$ and to
plot the function $\hat{W}(\hat{C})$ for different values of
$\Delta$. Since $\Delta \sim 1/N_w$ it does not make sense to
consider values $\hat{C} \gtrsim 1/\Delta$, because they are
outside of the region of applicability of the resonance
approximation. This explains the range of $\hat{C}$ in Fig.
\ref{WO2D0p1} and Fig. \ref{WO2D0}. In particular we fixed
$\Omega=2.0$ and we plot the case for $\Delta = 0.1$ in Fig.
\ref{WO2D0p1} and values for $\Delta = 0.01$ and $\Delta = 0$ in
Fig. \ref{WO2D0}. It is interesting to remark the characteristic
behavior in Fig. \ref{WO2D0}, where one can distinguish peaks at
certain values of $\hat{C}$. These peaks correspond to those
values of $\hat{C}$ such that $\hat{C}^{\mu,\nu}_k +\hat{C}=0$. At
these values, the $k$-th  TE or TM mode is at resonance. Contrast
of peaks becomes better as $|\hat{C}|$ increases, because
different resonances are further away and peaks overlap less. For
larger values of $\Omega$, as the number of excited modes
increases, overlapping between different resonances hides the
peaks, and one has the free-space limit (also shown for comparison
in Fig. \ref{WO2D0}). Note that by definition of $\hat{W}$, the
integrated power density is normalized to the angle-integrated
power density in free-space \textit{at resonance}, i.e. at
$\hat{C}=0$. This explains why $\hat{W}= 1$ at $\hat{C} = 0$ in
Fig. \ref{WO2D0}.


\begin{figure}
\begin{center}
\includegraphics*[width=150mm]{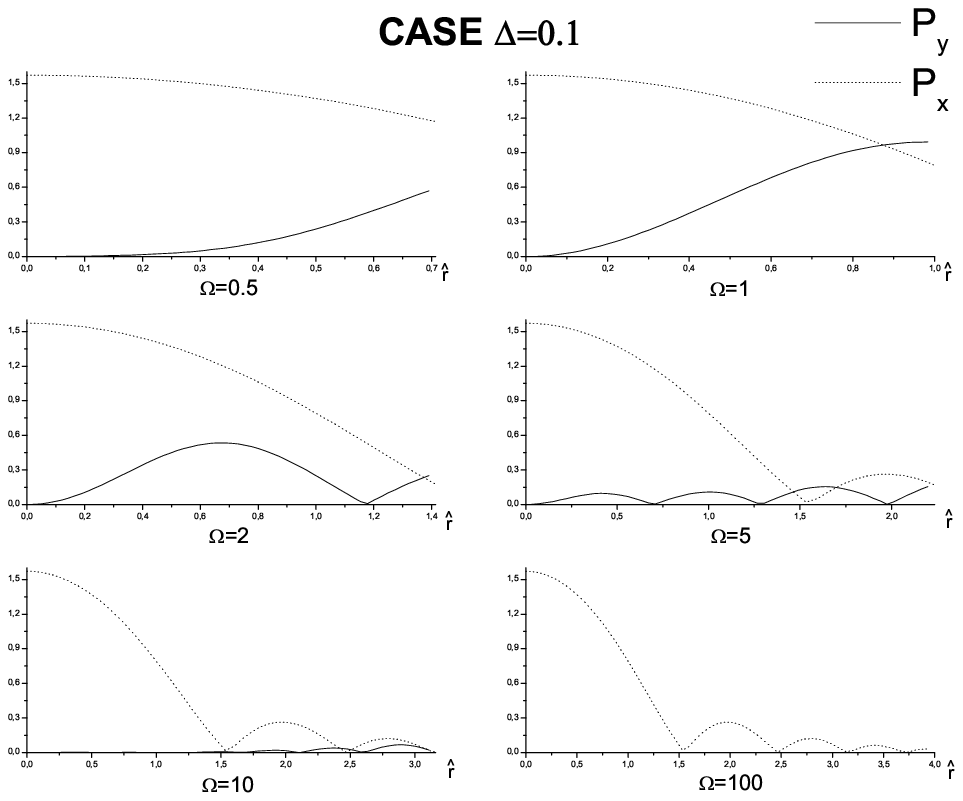}% Here is how to import EPS art
\caption{Comparison of $P_x$ as defined in Eq. (\ref{Pr1}) (dotted
line) and $P_y$ as defined in Eq. (\ref{Pr2}) (solid line) for
$\Delta = 0.1$ at different values of $\Omega=R^2/(\lambdabar
L_w)$ and $\hat{C}=0$ (perfect resonance). Plots refer to the
virtual source position ($\hat{z}=0$). \label{ExEy0p1}}
\end{center}
\end{figure}
\begin{figure}
\begin{center}
\includegraphics*[width=150mm]{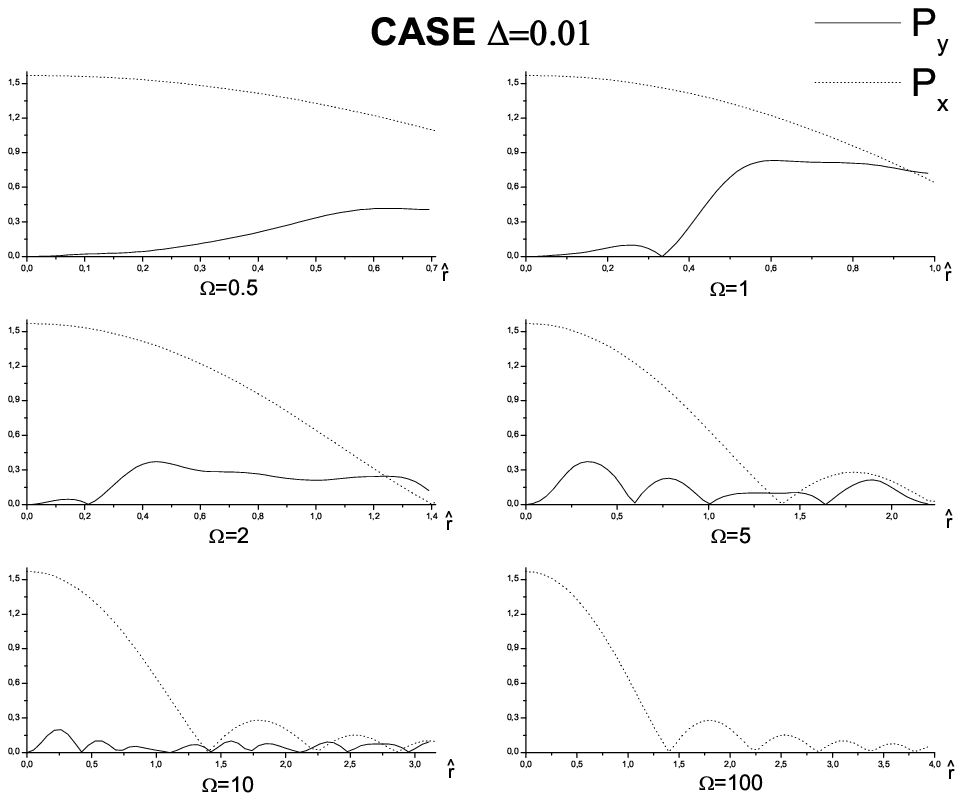}% Here is how to import EPS art
\caption{Comparison of $P_x$ as defined in Eq. (\ref{Pr1}) (dotted
line) and $P_y$ as defined in Eq. (\ref{Pr2}) (solid line) for
$\Delta = 0.01$ at different values of $\Omega = R^2/(\lambdabar
L_w)$ and $\hat{C}=0$ (perfect resonance). Plots refer to the
virtual source position ($\hat{z}=0$).\label{ExEy0p01}}
\end{center}
\end{figure}
\begin{figure}
\begin{center}
\includegraphics*[width=150mm]{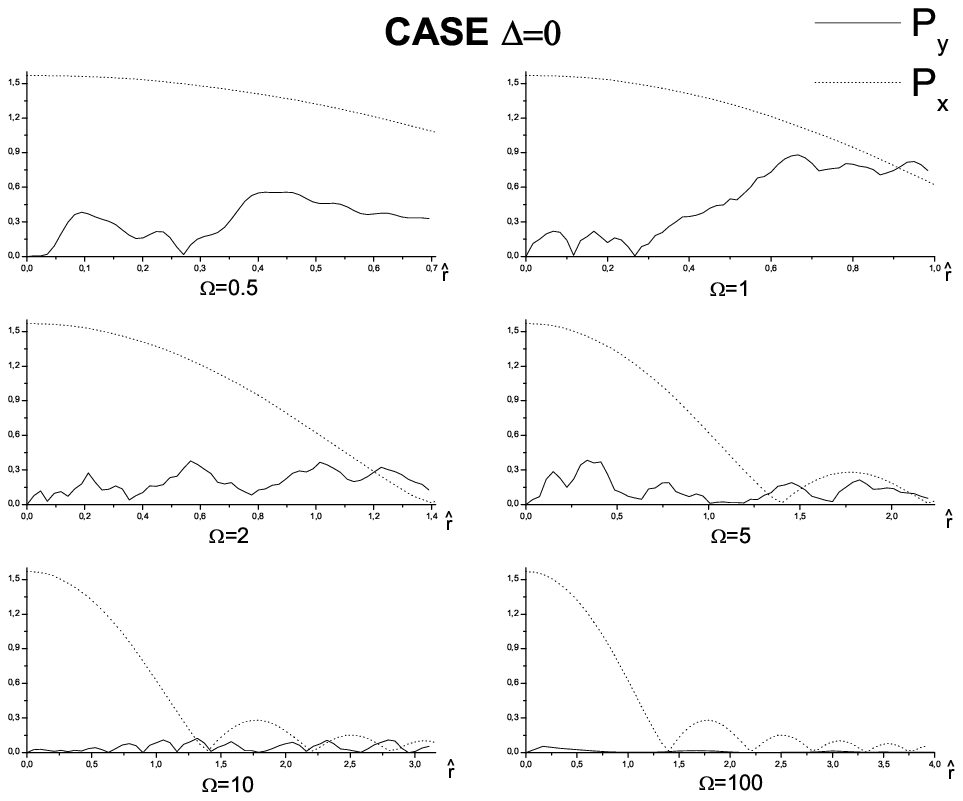}% Here is how to import EPS art
\caption{Comparison of $P_x$ as defined in Eq. (\ref{Pr1}) (dotted
line) and $P_y$ as defined in Eq. (\ref{Pr2}) (solid line) for
$\Delta = 0$ (hard-edge) at different values of $\Omega =
R^2/(\lambdabar L_w)$ and $\hat{C}=0$ (perfect resonance). Plots
refer to the virtual source position ($\hat{z}=0$). \label{ExEy0}}
\end{center}
\end{figure}
Finally, we may asses the influence of the waveguide on the output
radiation by comparing the modulus of the vertical electric field
for different values of $\Omega$ with the modulus of the
horizontally polarized field in free-space. In fact, as has been
said before, the presence of boundary conditions different from
those of free-space destroy the horizontal polarization as well as
the azimuthal symmetry of the undulator radiation. Vertical and
horizontal field are to be considered as functions of the distance
from the $z$ axis. Since the modulus of the vertical electric
field is not azimuthal symmetric we take a cut in the direction
where it is maximal, i.e. at $\phi=\pi/4$, as it can be seen
inspecting Eq. (\ref{fieldeypn}). As done before we limit our
analysis for exemplification purposes to the case $z=0$. We are
thus interested in the functions $P_x$ and $P_y$ respectively
defined as

\begin{eqnarray}
P_x(\hat{r}) =
{\mathrm{Abs}\left[{\hat{E}_x\left(\hat{r},0,0\right)}\right]_{\Big|_{\Omega
\longrightarrow \infty }}} ~.\label{Pr1}
\end{eqnarray}
and

\begin{eqnarray}
P_y(\hat{r},\Omega) =
{\mathrm{Abs}\left[{\hat{E}_y\left(\hat{r},\frac{\pi}{4},0\right)}\right]_{\Big|_{\Omega}}}
~.\label{Pr2}
\end{eqnarray}
We studied $P_y$ as a function of $\hat{r}$ for different values
of $\Omega$ and different values of $\Delta$. We set $\hat{C}=0$.
Comparisons with $P_x$ (studied at $\Omega=1000$) as a function of
$\hat{r}$ at different values of $\Delta$ are presented in Fig.
\ref{ExEy0p1} for $\Delta =0.1$, Fig. \ref{ExEy0p01} for $\Delta
=0.01$ and Fig. \ref{ExEy0} for $\Delta =0$ at different values of
$\Omega$. Note that the maximal value of $\hat{r}$ is
$\sqrt{\Omega}$, and is due to the fact that the pipe poses a
geometrical limit to the transverse region of interest. Figures
underline the role of the vertical polarization component of the
field for values of $\Omega$ around unity. Thus, the level of the
vertical polarization component of the field compared with that of
the horizontal polarization gives a measure of the waveguide
influence.

\begin{figure}
\begin{center}
\includegraphics*[width=150mm]{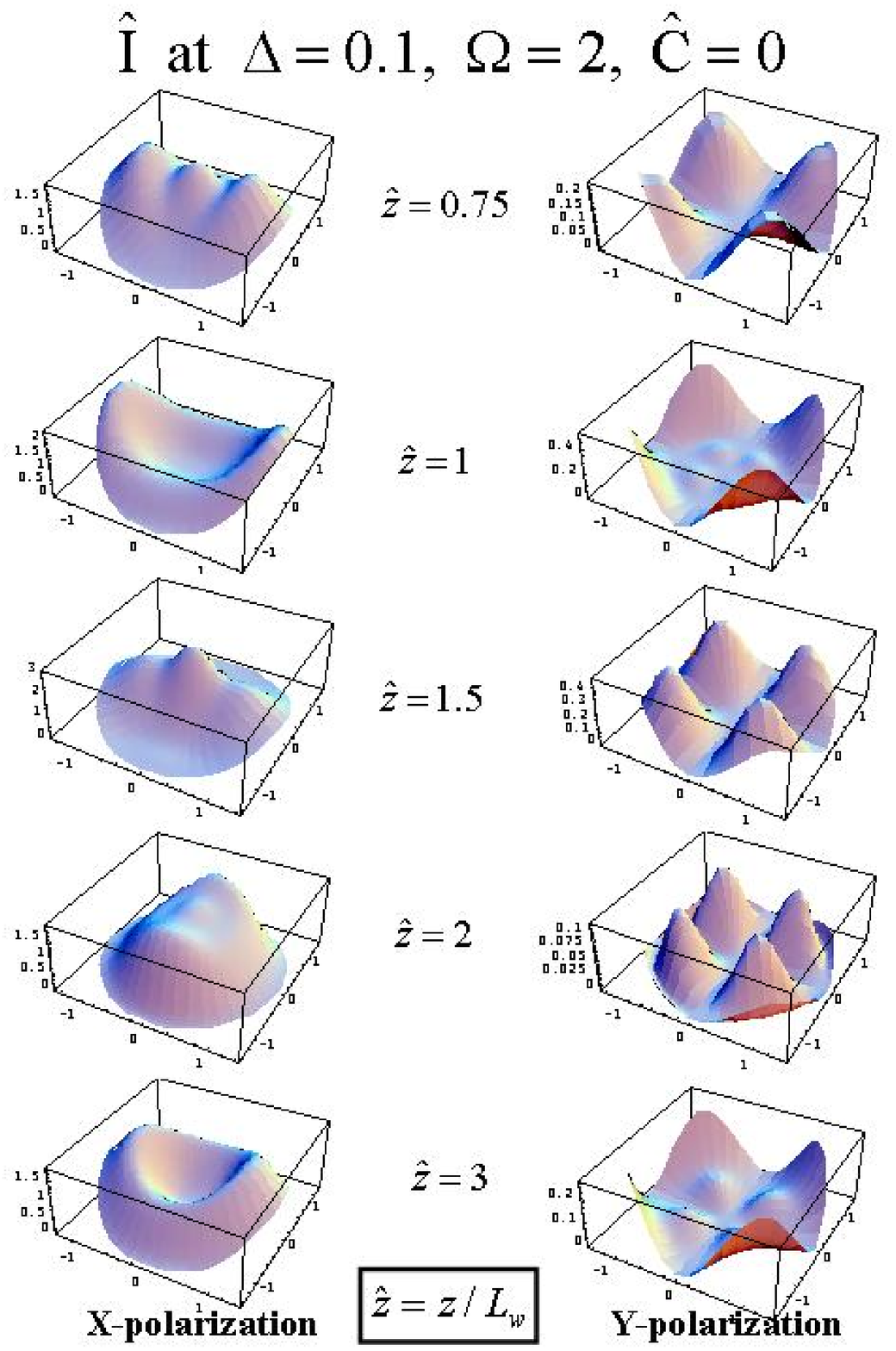}% Here is how to import EPS art
\caption{Intensity profiles at different values of $\hat{z}=z/L_w$
($\hat{z}=0$ at the undulator center) for $\Omega = 2$,
$\hat{C}=0$ (perfect resonance) and $\Delta = 0.1$. \label{Evol}}
\end{center}
\end{figure}

As it is clear by inspection of Eq. (\ref{fieldexp}) and Eq.
(\ref{fieldeyp}), our method allows straightforwardly to study the
evolution of the radiation pulse after the undulator, provided
that the pipe section remains unvaried. In Fig. \ref{Evol} we show
the evolution of the intensity profile as a function of the
normalized distance from the center of the undulator $\hat{z}$. We
take $\Omega = 2.0$, $\hat{C}=0$ and $\Delta = 0.1$.

To conclude this Section we estimate the range of parameters of
interest in the case of the infrared undulator beamline at FLASH.
The wavelength range is between $50~\mu$m and $200~\mu$m. The
radius of the vacuum pipe is $R = 1.8$ cm, the number of undulator
periods $N_w = 9$ with a period length $\lambda_w = 40$ cm,
yielding $L_w = 3.6$ m. Note that the $K$ parameter is always much
larger than unity, increasing when the wavelength is increased by
ramping up the magnetic field in the undulator ($K=20\div 40$ in
the THz-gap operation range). This set of parameters means $\Delta
\sim 1/N_w \sim 0.1$ and $3\lesssim \Omega \lesssim 10$. As it can
be seen in Fig. \ref{delta0p1} and \ref{ExEy0p1}, for $\Delta
=0.1$ in the short wavelength limit ($\Omega \simeq 10$) the
vacuum chamber has a small influence on the field distribution and
polarization. A strong influence is visible, instead, in the long
wavelength range ($\Omega \simeq 3$). Finally, from Fig. \ref{WC0}
we see that the total power at perfect resonance ($\hat{C}=0$) is
practically independent of the wavelength in our range of
interest.

\section{\label{sec:resi} Wall-resistance effects}

Up to now we studied the problem of undulator radiation in a
waveguide assuming perfectly conducting walls, i.e. infinite
conductivity $\sigma \longrightarrow \infty$. In this Section we
follow the approach proposed in \cite{FELB} to describe the case
when the refractive index of the walls is still mainly defined by
the conductivity $\sigma$ (i.e. we are still dealing with a
metal), but $\sigma$ has a finite value. We will conclude that in
practical situations of interest, the presence of wall-resistance
effects introduce important changes to the theory for perfectly
conductive walls.

Let us start our investigations with Maxwell's equations in the
time domain, written in all generality for a medium having
conductivity $\sigma$ and permittivity $\epsilon$:

\begin{eqnarray}
&& \vec{\nabla} \times \vec{H} = \frac{4 \pi \sigma}{c} \vec{E} +
\frac{\epsilon}{c} \frac{\partial \vec{E}}{\partial
t}~,~~\vec{\nabla} \times \vec{E} = -\frac{1}{c} \frac{\partial
\vec{H}}{\partial t} \cr &&  \vec{\nabla} \cdot \vec{H} =
0~,~~~~~~~~~~~~~~~~~~~~~~~~\vec{\nabla}\cdot \vec{E} = 0~.
\label{max1}
\end{eqnarray}
For a monochromatic wave of angular frequency $\omega$ we have

\begin{eqnarray}
&& \vec{\nabla} \times \vec{\bar{H}} = - i \frac{\omega}{c} (n')^2
\vec{\bar{E}}~,~~\vec{\nabla} \times \vec{\bar{E}} =
i\frac{\omega}{c} \vec{\bar{H}} \cr && \vec{\nabla} \cdot
\vec{\bar{E}} = 0~,~~~~~~~~~~~~~~~~~~~~\vec{\nabla}\cdot
\vec{\bar{H}} = 0~, \label{max2}
\end{eqnarray}
where

\begin{eqnarray}
n'=\sqrt{\epsilon+i\frac{4\pi \sigma}{\omega}} \label{np}
\end{eqnarray}
is the complex refractive index of the medium. A single equation
for $\vec{\bar{E}}$ can be written as

\begin{eqnarray}
\nabla^2 \vec{\bar{E}} + (n')^2 \frac{\omega^2}{c^2} \vec{\bar{E}}
=0~. \label{heln}
\end{eqnarray}
A solution of Eq. (\ref{heln}) is a plane wave specified by:

\begin{eqnarray}
&&\vec{\bar{E}} \exp(-i\omega t) + C.C. = \vec{E}_o
\exp[i(\vec{k}'\cdot \vec{r} - \omega t)]+C.C.\cr && \vec{\bar{H}}
= \frac{c}{\omega}(\vec{k}' \times \vec{\bar{E}})~, \label{palne}
\end{eqnarray}
where $\vec{k}' = k'_{x} \vec{e}_x+k'_{y} \vec{e}_y+k'_{z}
\vec{e}_z$ is the (complex) wave vector inside the metal, with
$(k')^2= (k'_{x})^2+(k'_{y})^2+(k'_{z})^2=(n')^2 \omega^2/c^2$ a
complex number.

Let us consider a plane wave at the interface between vacuum (with
unitary refraction index) and a medium described by Eq.
(\ref{np}), and let $\vec{k}$ be the wave vector of the incident
wave in vacuum with $k=|\vec{k}|=\omega/c$. We indicate with
subscripts $t$ and $n$ the tangential and normal components wave
vectors to the metallic surface. Continuity of the tangential
component of the wave vector on the boundary (i.e. $k'_{t}=k_t$)
implies:

\begin{eqnarray}
n' \frac{k'_{t}}{k'} = \frac{k_t}{k}~. \label{cont}
\end{eqnarray}
As a result one obtains

\begin{eqnarray}
k'_n=\sqrt{k'^2-(k'_t)^2} =
k'\sqrt{1-\left(\frac{k'_t}{k'}\right)^2}=
k'\sqrt{1-\frac{1}{(n')^2}\left(\frac{k_t}{k}\right)^2}~.
\label{cont2}
\end{eqnarray}
In a metal, the refractive index is mainly defined by the
conductivity $\sigma$ according to

\begin{eqnarray}
n'=\sqrt{\epsilon+i\frac{4\pi \sigma}{\omega}} \simeq
\sqrt{i\frac{4\pi \sigma}{\omega}}~.\label{np2}
\end{eqnarray}
Since $\sigma \gg 1$, also $|n'|\gg 1$, and Eq. (\ref{cont2})
yields

\begin{eqnarray}
k'_n \simeq k'~. \label{met}
\end{eqnarray}
Thus, the propagation direction of the wave in the metal is almost
perpendicular to the surface, independently of the direction of
the incident wave. This means that electric and magnetic field
into the metal have only tangential components, $\bar{E}_t$ and
$\bar{H}_t$. Moreover, these components must be continuous on the
surface, as follows from Maxwell's equations. Using Eq.
(\ref{palne}) one obtains

\begin{eqnarray}
\frac{\bar{E}_t}{\bar{H}_t} = \frac{1}{n'}~. \label{EHT}
\end{eqnarray}
Eq. (\ref{EHT}) can be considered as an approximate boundary
condition. Note that this is valid not only for plane waves. In
fact, any wave can be decomposed in terms of a linear
superposition of plane waves. Since Eq. (\ref{EHT}) is valid for
each component, it must be valid for their linear superposition.
Eq. (\ref{EHT}) can also be extended for any shape of the boundary
surface, provided that typical value of the curvature radius is
much larger than the wavelength. Eq. (\ref{EHT}) is named after
Leontovich, who first derived it, and can be written in vector
form as

\begin{eqnarray}
(\vec{n} \times \vec{\bar{E}})_{\Big|_S} = \frac{1}{n'} ~\vec{n}
\times (\vec{n} \times \vec{\bar{H}})_{\Big|_S}~, \label{leo}
\end{eqnarray}
where the vector $\vec{n}$, as defined as in Section
\ref{sec:boun} and shown in Fig. \ref{geobou}, is pointing
inwards. Using Leontovich boundary condition gives a good
approximation in the case of a metallic waveguide, and drastically
simplifies the solution of the electrodynamical problem. Namely,
the problem of mode excitation in a waveguide with resistive walls
can be solved following the same approach in Section
\ref{sec:boun} and Section \ref{sec:circ}, where boundary
conditions are now substituted by Eq. (\ref{leo}).

Since we are applying the paraxial approximation, we assume that
the field amplitude does not change much along the $z$ direction
within a wavelength and we are interested about the (main)
transverse components of the field $\widetilde{E}_x$ and
$\widetilde{E}_y$. Using Eq. (\ref{leo}) and Maxwell's equations,
Eq. (\ref{max2}), it is possible to find boundary conditions
involving the transverse field components only.

First, using Eq. (\ref{max2}) in terms of slowly varying envelopes
$\vec{\widetilde{E}}$,  we can write Eq. (\ref{leo}) as

\begin{eqnarray}
\left(\vec{n} \times \vec{\widetilde{E}}\right)_{\Big|_S} =
-\frac{i c}{\omega n'} ~\vec{n} \times \left[\vec{n} \times
\left(\vec{\nabla}\times \vec{\widetilde{E}}
\right)\right]_{\Big|_S}~. \label{leo2}
\end{eqnarray}
The transverse components of Eq. (\ref{leo2}) only involve
$\widetilde{E}_z$, while the component along $\vec{e}_z$ only
involves $\vec{\widetilde{E}}_\bot$. We may thus write

\begin{eqnarray}
\left(\vec{n} \times \vec{\widetilde{E}}_\bot\right)_{\Big|_S} =
-\frac{i c}{\omega n'} ~\vec{n} \times \left[\vec{n} \times
\left(\vec{\nabla}_\bot\times \vec{\widetilde{E}}_\bot
\right)\right]_{\Big|_S}~. \label{leo3}
\end{eqnarray}
Eq. (\ref{leo3}) will substitute the first boundary condition in
(\ref{pb1}).

Second, from Eq. (\ref{leo}) we see that, on $S$, the projections
of $\vec{\bar{H}}$ and $\vec{\bar{E}}$ on the plane orthogonal to
$\vec{n}$, i.e. $\vec{\bar{H}}_t$ and $\vec{\bar{E}}_t$, are
orthogonal and their modulus are related by Eq. (\ref{EHT}). As  a
result, we have $\bar{E}_z = \vec{\bar{E}}_t \cdot \vec{e}_z =
(1/n') \vec{\bar{H}}_t \cdot (\vec{e}_z \times \vec{n})$. Now, Eq.
(\ref{max2}) implies $\vec{\nabla}_\bot\cdot \vec{\bar{E}}_\bot =
- (\omega/c)\bar{E}_z$. Therefore $\vec{\nabla}_\bot\cdot
\vec{\bar{E}}_\bot = - \omega/(n' c) \vec{\bar{H}}_t \cdot
(\vec{e}_z \times \vec{n})=- \omega/(n' c) \vec{\bar{H}} \cdot
(\vec{e}_z \times \vec{n})$. But Eq. (\ref{max2}) also yields
$\vec{\bar{H}} = - i c/\omega \vec{\nabla}\times \vec{\bar{E}}$.
We then obtain

\begin{eqnarray}
\left(\vec{\nabla}_\bot\cdot \vec{\bar{E}}_\bot\right)_{\Big|_S} =
\frac{i}{n'} \left(\vec{\nabla}\times
\vec{\bar{E}}\right)_{\Big|_S} \cdot (\vec{e}_z \times \vec{n}) =
\frac{i}{n'} \vec{n}\cdot \left[\left(\vec{\nabla}\times
\vec{\bar{E}}\right)_{\Big|_S} \times \vec{e}_z
\right]~.\label{uno}
\end{eqnarray}
If we now calculate the expression in $[\cdot]$ brackets in Eq.
(\ref{uno}) and  we account for the fact that, in paraxial
approximation, $\partial_z \bar{E}_{x,y} \simeq (\omega/c)
\bar{E}_{x,y} \gg \partial_{x,y} \bar{E}_z$ we can re-write Eq.
(\ref{uno}) in terms of slowly varying field amplitudes as

\begin{eqnarray}
\left(\vec{\nabla}_\bot\cdot
\vec{\widetilde{E}}_\bot\right)_{\Big|_S} = \frac{i\omega}{c n'}
\left(\vec{n}\cdot \vec{\widetilde{E}}_\bot\right)_{\Big|_S}
~,\label{unob}
\end{eqnarray}
that will substitute the second boundary condition in Eq.
(\ref{pb1}).

Thus, the problem in (\ref{pb1}) should be substituted with

\begin{eqnarray}
\left\{
\begin{array}{l}
\mathcal{D} \left[\vec{\widetilde{E}}_\bot(z,\vec{r}_\bot)\right]
= \vec{f}(z, \vec{r}_\bot)
\\
\left(\vec{n} \times \vec{\widetilde{E}}_\bot\right)_{\Big|_S} =
-{{i c}/{(\omega n')}} ~ \left\{\vec{n} \times \left[\vec{n}
\times \left(\vec{\nabla}_\bot\times \vec{\widetilde{E}}_\bot
\right)\right]_{\Big|_S}\right\} \\ \left(\vec{\nabla}_\bot\cdot
\vec{\widetilde{E}}_\bot\right)_{\Big|_S} = {i\omega}/({c n'})
\left(\vec{n}\cdot \vec{\widetilde{E}}_\bot\right)_{\Big|_S}  ~.
\end{array}\right.\label{pbleo}
\end{eqnarray}
We may now follow the same method used in Section \ref{sec:boun}
to solve (\ref{pbleo}). Namely, after consistent application of
Laplace transformations, we find an eigenvalue problem analogous
to (\ref{pbeigH}):

\begin{equation}
\left\{
\begin{array}{l}
\nabla_\bot^2 \vec{F}_j(\vec{r}_\bot) + \lambda_j
\vec{F}_j(\vec{r}_\bot) = 0
\\
\left(\vec{n} \times \vec{F}_j\right)_{\Big|_S} = -{{i c}/{(\omega
n')}} ~ \left\{\vec{n} \times \left[\vec{n} \times
\left(\vec{\nabla}_\bot\times \vec{F}_j
\right)\right]_{\Big|_S}\right\}
\\
\left(\vec{\nabla}_\bot \cdot \vec{F}_j\right)_{\Big|_S} =
{i\omega}/({c n'}) \left(\vec{n}\cdot \vec{F}_j\right)_{\Big|_S}~.
\end{array}\right.\label{pbeigHleo}
\end{equation}
As for the case of perfectly metallic walls, boundary conditions
are homogeneous, so that the domain of the Laplacian operator is
the vector space of twice differentiable (square integrable)
functions obeying boundary conditions in (\ref{pbeigHleo}).
However, the Laplacian operator defined in this way is not
self-adjoint with respect to the inner product defined in Eq.
(\ref{inn}). This is a result of the fact that vectors $\vec{F}_j$
in Eq. (\ref{pbeigHleo}), at the boundary, are not orthogonal to
the surface $S$. Then, eigenvalues are not real, nor
eigenfunctions are orthogonal with respect to the inner product in
Eq. (\ref{inn}). In general we do not know wether the spectrum is
discrete, completeness is not granted and we cannot prove the
existence of a set of eigenfunctions either. Yet, direct
calculations in \cite{FELB} show that

\begin{equation}
\left<\vec{F}_{j}^*,{\vec{F}}_{i}\right> = \int_S \vec{F}_j\cdot
\vec{F}_i ~d\vec{r}_\bot=\delta_{ji}~. \label{orto}
\end{equation}
It follows that functions $\vec{F}_j$ form a bi-orthogonal set of
eigenfunctions. Bi-orthogonality is often exploited  in different
problems (see e.g. \cite{KRIB,SIEG,PLAS}). In particular we will
take advantage, without proving it, of completeness and
discreteness of the bi-orthogonal set. This allows us to decompose
$\widehat{G}^\alpha_{~\beta}$ as

\begin{equation}
\widehat{G}^\alpha_{~\beta} = \sum_j \frac{F_j^{~\alpha}
F^{}_{j~\beta}}{2 i \omega p/c-\lambda_j}~, ~\label{LGdec2}
\end{equation}
exactly as for the case of perfectly conducting walls. Note that
now eigenvalues and eigenfunctions are complex. In particular, we
write the eigenvalues as

\begin{equation}
\lambda_j = \lambda_j' + i \lambda_j''~. \label{lacom}
\end{equation}
We want to study the problem of wall-resistance in the framework
of a perturbation theory. If $\lambda_{j}^0$ is an unperturbed
eigenvalue (perfectly conductive walls) we require that $|\delta
\lambda_j| = |\lambda_j - \lambda_{j}^0| \ll  |\lambda_k^{0} -
\lambda_{j}^0|$ for any value of $k \ne j$.

In this case we may still formulate an eigenvalue problem for the
two scalar (and complex) functions $\psi_j^{\mathrm{TE}}$ and
$\psi_j^{\mathrm{TM}}$ with complex eigenvalues $\lambda_j$, using
a definition analogous to Eq. (\ref{defpsi}). We get:

\begin{equation}
\left\{
\begin{array}{l}
\nabla_\bot^2 \psi_j^\mathrm{TE,TM}(\vec{r}_\bot) +
\lambda_j^\mathrm{TE,TM} ~\psi_j^\mathrm{TE,TM}(\vec{r}_\bot) = 0
\\
\left[\vec{n} \cdot \vec{\nabla}_\bot
\psi_j^\mathrm{TE}+\left(\vec{e}_z \times \vec{n}\right)\cdot
\vec{\nabla}_\bot\psi_j^\mathrm{TM}\right]_{\Big|_S} = -ic
\left[\left(k_\bot^{\mathrm{TE}}\right)^2
\psi_j^\mathrm{TE}/(\omega n')\right]_{\Big|_S}
\\
\left[\left(k_\bot^{\mathrm{TM}}\right)^2\psi_j^\mathrm{TM}\right]_{\Big|_S}
= i \omega/( c n') \left[\vec{n} \cdot \vec{\nabla}_\bot
\psi_j^\mathrm{TM}-\left(\vec{e}_z \times \vec{n}\right)\cdot
\vec{\nabla}_\bot\psi_j^\mathrm{TE}\right]_{\Big|_S}~,
\end{array}\right.\label{pbeigpsiHo}
\end{equation}
where

\begin{eqnarray}
\int_S d\vec{r}_\bot \left|\vec{\nabla}
\psi_j^{\mathrm{TE,TM}}\right|^2 = 1 ~.\label{normpsi2}
\end{eqnarray}
It should be noted that (\ref{pbeigpsiHo}) is valid only within
the framework of a perturbation theory. In fact, according to
(\ref{pbeigpsiHo}), functions $\psi_j^\mathrm{TE,TM}$ separately
obey Helmholtz's equation, but boundary conditions in
(\ref{pbeigpsiHo}) are not consistent with independency of
$\psi_j^\mathrm{TE,TM}$ required by Eq. (\ref{defpsi}).

In the framework of a perturbation theory though, we may still use
(\ref{pbeigpsiHo}) and formally obtain the same tensor Green's
function in Eq. (\ref{Gfex}). As said above, we want to work in
the framework of perturbation theory. We base such theory on the
small parameter $\omega/((|n' k_\bot | c) \ll 1$. Within this
theory, the problem in (\ref{pbeigpsiHo}) can be written as

\begin{eqnarray}
&&\left\{
\begin{array}{l}
\nabla_\bot^2 \psi_j^\mathrm{TE,TM}(\vec{r}_\bot) +
\lambda_j^\mathrm{TE,TM} ~\psi_j^\mathrm{TE,TM}(\vec{r}_\bot) = 0
\\
\left[\vec{n} \cdot \vec{\nabla}_\bot
\psi_j^\mathrm{TE}\right]_{\Big|_S} = \left\{-ic
\left(k_\bot^{\mathrm{TE}}\right)^2 \psi_j^\mathrm{TE}/(\omega n')
+i \omega/\left[n' c
\left(k_\bot^{\mathrm{TE}}\right)^2\right]\left[\left(\vec{e}_z
\times \vec{n}\right)\cdot
\vec{\nabla}_\bot\right]^2\psi_j^\mathrm{TE}\right\}_{\Big|_S}
\\
\left[\left(k_\bot^{\mathrm{TM}}\right)^2\psi_j^\mathrm{TM}\right]_{\Big|_S}
= i \omega/( c n') \left[\vec{n} \cdot \vec{\nabla}_\bot
\psi_j^\mathrm{TM}\right]_{\Big|_S}~,
\end{array}\right.\cr &&\label{pbeigpsiH2}
\end{eqnarray}
where TE and TM modes are now decoupled. If we restrict our
analysis to a circular waveguide radius $R$, we obtain the
following expression for $\psi_j^\mathrm{TE,TM}$ at the first
order in $\omega/((|n' k_\bot | c)$:

\begin{eqnarray}
\left(
\begin{array}{l}
\psi_{mk1}^\mathrm{TE}
\\
\psi_{mk2}^\mathrm{TE}
\end{array}
\right) = A_{mk}^\mathrm{TE} J_m\left[\left(\mu_{mk} + \delta
\mu_{mk}\right) \frac{r}{R}\right] \left(
\begin{array}{l}
\sin(m\phi)
\\
\cos(m\phi)
\end{array}
\right) ~\label{solHTE}
\end{eqnarray}
and

\begin{eqnarray}
\left(
\begin{array}{l}
\psi_{mk1}^\mathrm{TM}
\\
\psi_{mk2}^\mathrm{TM}
\end{array}
\right) = A_{mk}^\mathrm{TM} J_m\left[\left(\nu_{mk} + \delta
\nu_{mk}\right) \frac{r}{R}\right] \left(
\begin{array}{l}
\sin(m\phi)
\\
\cos(m\phi)
\end{array}
\right) ~,\label{solHTM}
\end{eqnarray}
where $A_{mk}^{\mathrm{TE},\mathrm{TM}}$ should now be calculated
with the help of Eq. (\ref{normpsi2}), while boundary conditions
in (\ref{pbeigpsiH2}) allow to obtain

\begin{eqnarray}
\frac{\delta \mu_{mk}}{\mu_{mk}} = -\frac{i c}{n' \omega R}
\frac{\mu_{mk}^2 + \omega^2 m^2 R^2/(c^2 \mu_{mk}^2)}{\mu_{mk}^2
-m^2} \label{dmu}
\end{eqnarray}
and

\begin{eqnarray}
\frac{\delta \nu_{mk}}{\nu_{mk}} = -\frac{i \omega R}{c
n'\nu_{mk}^2} ~. \label{dnu}
\end{eqnarray}
Note that Eq. (\ref{dmu}) and Eq. (\ref{dnu}) can be written in
terms of our small parameter, since $R\omega/(c |n'|
\{\nu,\mu\}_{1k}) = \omega/(c |n' k_\bot^{1k}|)$, where the
concept of transverse wave number is now applied to each mode $k$.
Actually condition $\omega/((|n' k_\bot | c)\ll 1$ can be
presented as $|\delta \nu_{mk}| \ll 1$ and $|\delta \mu_{mk}| \ll
1$. Thus, we may give an estimation of wall resistance effects
retaining our main results, Eq. (\ref{fieldexpn}) and Eq.
(\ref{fieldeypn}), and substituting $\mu_{mk}$ with
$\mu_{mk}+\delta \mu_{mk}$ and $\nu_{mk}+\delta \nu_{mk}$. In our
case of interest, $m=1$ so that

\begin{eqnarray}
\frac{\delta \mu_{1k}}{\mu_{1k}} = -\frac{i c}{n' \omega R}
\frac{\mu_{1k}^2 + \omega^2  R^2/(c^2 \mu_{1k}^2)}{\mu_{1k}^2 -1}
\label{dmu1}
\end{eqnarray}
and

\begin{eqnarray}
\frac{\delta \nu_{1k}}{\nu_{1k}} = -\frac{i \omega R}{c
n'\nu_{1k}^2} ~. \label{dnu1}
\end{eqnarray}
Addition of $\delta \mu_{1k}$ and $\delta \nu_{1k}$ implies an
exponential damping of each mode, due to the propagation factor
$\exp[-i \hat{C}^{\mu,\nu}_k \hat{z}]$ in the expressions for
$\mathcal{A}^{\mu,\nu}_k(z)$, Eq. (\ref{Amum0}) and Eq.
(\ref{Anun0}). We can estimate the damping exponent as

\begin{equation}
D_{\mu}=\frac{\mu_{1k}|\delta \mu_{1k}|}{\Omega} \hat{z}
=\frac{1}{\mu_{1k}^2-1}\frac{L_w}{ \sqrt{2} |n'| R}\hat{z}
\label{expmu}
\end{equation}
for TE modes, where we used $(\lambdabar/R)^2 \ll 1$ and

\begin{equation}
D_{\nu}=\frac{\nu_{1k}|\delta \nu_{1k}|}{\Omega} \hat{z} =
\frac{L_w}{\sqrt{2} |n'| R }\hat{z}\label{expnu}
\end{equation}
for TM modes. Expressions for the damping exponents $D_\mu$ and
$D_\nu$, Eq. (\ref{expmu}) and Eq. (\ref{expnu}), are known in
literature and may be found, for example, in \cite{LEWI}. Note
that the difference between $D_\mu$ and $D_\nu$ is only in the
factor $(\mu_{1k}^2-1)^{-1}$ in Eq. (\ref{expmu}). $D_{\mu}$
depends on the mode number, $k$, while $D_{\nu}$ is independent of
it.

Let us study a practical case for copper. We know that the skin
depth for copper at wavelength $\lambda = 10$ cm is $\delta_s
\simeq 1.2~ \mu$m and $\delta_s = \sqrt{c \lambda}/(2 \pi\sqrt{
\sigma })$, where $\sigma$ is the conductivity. It follows that
for $\lambda = 200~\mu$m, that is in the range of interest for the
FLASH infrared beamline we obtain $\delta_s \simeq 54$ nm yielding

\begin{eqnarray}
|n'| = \frac{\sqrt{2} \lambda}{2\pi \delta_s}  \simeq 840
~.\label{nnn}
\end{eqnarray}
Let us consider $R=1.8$ cm and $L_w = 3.6$ m (corresponding to
$\Omega = 2.8$), and substitute these values in Eq. (\ref{expmu})
and Eq. (\ref{expnu}). The perturbation parameter is smaller than
unity: $\omega R/(\mu_{11} c |n'|) \simeq 0.37$ ($\mu_{11} \simeq
1.84$), and $\omega R/(\nu_{11} c |n'|) \simeq 0.18 $ ($\nu_{11}
\simeq 3.83$). Therefore, we may apply the perturbation approach
for estimations. It should be noted here that the accuracy of the
first-order perturbation approach becomes better and better for
higher modes, because $\mu_{1k}$ and $\nu_{1k}$ increase while $k$
increases. For the transverse magnetic modes we have a damping
factor common to all modes $D_\nu \simeq 0.17 \hat{z}$. For
transverse electric modes we should specify the mode number. For
the first mode ($k=1$), and we obtain $D_\mu \simeq 0.07~
\hat{z}$. For second mode ($k=2$) we have $\mu_{12} \simeq 5.3$
and thus $D_\mu \simeq 0.006~ \hat{z}$, that gives a negligible
effect. All TM modes are visibly damped. At $\hat{z}=1$, i.e. half
undulator length after the exit of the device, one has a damping
factor $\exp(-0.17)\simeq 0.84$, i.e. modes are dumped by about
$16 \%$. TE modes instead, are left almost unaffected.

We conclude that propagation through the pipe is strongly affected
by wall-resistance effects even for relatively large values of
$\Omega$. Note that this effect strongly depends on the  material
considered. For example steel has a skin depth three times thicker
than copper. Strictly speaking we could not apply a perturbation
approach, since now the perturbation parameter is larger than
unity (although comparable with unity). However, if we still use
the perturbation approach to get a rough estimation, we obtain
that TM modes present a dumping exponent $D_\nu$ that is about
three times larger than copper. This means that TM modes are
dumped by about $40\%$. These estimations allow us to formulate
the following recommendation: the internal part of the vacuum pipe
for the infrared undulator line at FLASH should be
copper-coated\footnote{We underline the fact that these remarks
are only valid in the case of undulator radiation. In particular,
as said before, we will treat edge-radiation setups in a separate
work.}.

\section{\label{sec:conc} Conclusions}

In this article we presented the first exhaustive theory of
undulator radiation within a waveguide and we exemplified it in
the case of the infrared undulator beamline at FLASH. In the
relatively simpler free-space case, paraxial Maxwell's equations
can be solved in terms of a scalar paraxial Green's function. In
that case, the solution for the field, written in reduced form
within the resonance approximation is identical (aside for
polarization properties) for planar and helical undulator
configurations. When a waveguide is present, the relation between
current sources and electromagnetic field is more complicated. The
analysis of the problem is performed by introducing a tensor
Green's function technique, thus complicating the structure of
equations that, in contrast to the free-space case, now depends on
the undulator type and on the waveguide geometry. First we
outlined a solution for a homogeneous waveguide with arbitrary
cross-section. Then we specialized our consideration to the case
of a circular waveguide, and we implemented the planar undulator
case\footnote{We treated the helical case in Appendix A.} within
the applicability region of the resonance approximation. The
electric field was found as a superposition of the waveguide
modes, and was studied for different values of parameters. The
main  parameter involved in the problem is the waveguide parameter
$\Omega$, that can be interpreted as the squared  ratio between
the waveguide radius and the radiation diffraction size, and is a
purely geometrical parameter. When $\Omega$ is comparable, or
smaller than unity, waveguide effects become important, under the
assumption of a perfect conductor. Moreover, we found that
wall-resistance effects play a significant role even for
relatively large values of $\Omega$. To minimize this effect, it
is desirable to coat the internal part of the waveguide with
copper.

\section{\label{sec:graz} Acknowledgements}

The authors are grateful to Martin Dohlus, Michael Gensch and
Oliver Grimm (DESY) for many useful discussions and to Massimo
Altarelli and Jochen Schneider (DESY) for their interest in this
work.

\newpage

\section*{Appendix A: Helical undulator}

We define in all generality the horizontal and vertical velocity
of an electron as a function of the longitudinal position $z'$
along the undulator as

\begin{eqnarray}
&&v_x(z') = - c \theta_s \sin(k_w z') = - \frac{c\theta_s}{2 i}
\left\{\exp[ik_w z']-\exp[-ik_w z']\right\} \cr && v_y(z') = \mp c
\theta_s \cos(k_w z') = \mp \frac{c\theta_s}{2} \left\{\exp[ik_w
z']+\exp[-ik_w z']\right\}~.\cr &&
 \label{vhel}
\end{eqnarray}
The $\mp$ sign in the second equation in (\ref{vhel}) indicates an
electron rotating clockwise ($-$ sign) or counterclockwise ($+$
sign) in the judgement of an observer located after the undulator
and looking towards the device.  The longitudinal Lorentz factor
$\gamma_z$ is now constant.  We can write the phase factor in Eq.
(\ref{efielG}) as

\begin{eqnarray}
\int_0^z \frac{\omega}{2c\gamma_z^2 } d\bar{z}= \frac{ z}{2
\lambdabar \gamma_z^2} = k_w z+C z ~.\label{phaseh2}
\end{eqnarray}
The next step is to substitute Eq. (\ref{vhel}) and Eq.
(\ref{phaseh2}) in Eq. (\ref{efielG}) to calculate the field
components. We can do that within the region of applicability of
the resonance approximation, as in the case of a planar undulator.
Within the region of applicability of the resonance approximation,
condition (\ref{zetar}), and neglecting negligible terms in Eq.
(\ref{efielG}) as specified in Section \ref{sec:wig} we obtain the
field components in cartesian coordinates for the case of a
helical undulator:

\begin{eqnarray}
\widetilde{E}^\alpha_\mp = \frac{2 \pi e \omega \theta_s}{c^2}
\int_{-L_w/2}^{L_w/2} dz' \left(G^\alpha_1 \mp i
G^\alpha_2\right){\Big|_{r'(z')=0}} \exp[i C z']
~.\label{fieldcart}
\end{eqnarray}
By inspection of Eq. (\ref{Gfexcir}), under the resonance
approximation only the terms with $m=1$ survive.  With the help of
Eq. (\ref{Gfexcir}) and neglecting negligible terms we find for
the horizontal field ($\alpha=1$) and for the vertical field
($\alpha=2$) respectively:

\begin{eqnarray}
\widetilde{E}_{x\mp}(r,\phi,z) = -\frac{i\omega e \theta_s}{c^2}
\sum_{k=1}^{\infty} && \left\{ \mathcal{A}^{\mu}_k(z)
\left[J_o\left(\mu_{1k} \frac{r}{R} \right) + J_2\left(\mu_{1k}
\frac{r}{R} \right) \exp[\mp 2 i \phi]\right] + \right.\cr
&&\left. \mathcal{A}^{\nu}_k(z) \left[J_o\left(\nu_{1k}
\frac{r}{R} \right)- J_2\left(\nu_{1k} \frac{r}{R} \right)
\exp[\mp 2 i \phi]\right]\right\} \label{fieldex}
\end{eqnarray}
and

\begin{eqnarray}
\widetilde{E}_{y\mp}(r,\phi,z) = \mp  \frac{\omega e
\theta_s}{c^2} \sum_{k=1}^{\infty} && \left\{
\mathcal{A}^{\mu}_k(z) \left[J_o\left(\mu_{1k} \frac{r}{R} \right)
- J_2\left(\mu_{1k} \frac{r}{R} \right) \exp[\mp 2 i \phi]\right]
+ \right.\cr &&\left. \mathcal{A}^{\nu}_k(z)
\left[J_o\left(\nu_{1k} \frac{r}{R} \right)+ J_2\left(\nu_{1k}
\frac{r}{R} \right) \exp[\mp 2 i \phi]\right]\right\}~.
\label{fieldey}
\end{eqnarray}
One can also write the expressions for the components in polar
coordinates $\widetilde{E}_r$ and $\widetilde{E}_\phi$, that are
related to $\widetilde{E}_x$ and $\widetilde{E}_y$ through

\begin{eqnarray}
\left\{
\begin{array}{l}
{\widetilde{E}_r} = \widetilde{E}_x \cos(\phi) +\widetilde{E}_y
\sin(\phi)
\\
{\widetilde{E}_\phi} = - \widetilde{E}_x \sin(\phi)
+\widetilde{E}_y \cos(\phi)~.\end{array}\right. \label{phitox}
\end{eqnarray}
We thus obtain

\begin{eqnarray}
\widetilde{E}_{r\mp}(r,\phi,z) = -i\frac{\omega e \theta_s}{c^2}
\sum_{k=1}^{\infty} && \left\{ \mathcal{A}^{\mu}_k(z)
\left[J_o\left(\mu_{1k} \frac{r}{R} \right) + J_2\left(\mu_{1k}
\frac{r}{R} \right) \right] \exp[\mp  i \phi]+ \right.\cr &&\left.
\mathcal{A}^{\nu}_k(z) \left[J_o\left(\nu_{1k} \frac{r}{R}
\right)- J_2\left(\nu_{1k} \frac{r}{R} \right) \right] \exp[\mp i
\phi]\right\} \label{fielder}
\end{eqnarray}
and

\begin{eqnarray}
\widetilde{E}_{\phi\mp}(r,\phi,z) = \mp  \frac{\omega e
\theta_s}{c^2} \sum_{k=1}^{\infty} && \left\{
\mathcal{A}^{\mu}_k(z) \left[J_o\left(\mu_{1k} \frac{r}{R} \right)
- J_2\left(\mu_{1k} \frac{r}{R} \right) \right]\exp[\mp  i \phi] +
\right.\cr &&\left. \mathcal{A}^{\nu}_k(z) \left[J_o\left(\nu_{1k}
\frac{r}{R} \right)+ J_2\left(\nu_{1k} \frac{r}{R} \right)
\right]\exp[\mp  i \phi]\right\}~. \label{fieldephi}
\end{eqnarray}
Note that both TM and TE modes are present in the expression for
the field. It is interesting to compare this fact with what can be
found in literature. As already remarked in Section
\ref{sec:intro}, TM modes are neglected in \cite{HART}. For the
readers' commodity we report here words from Section 9.3 of
reference \cite{HART}. There one may find: "It is easily seen that
in the case of a spatially extended charge propagating in a
helical wiggler, the TE modes (...) couple to the wiggler-induced
motion, as the current components (...) By contrast, the TM modes
(...) are driven by the uniform motion of the space-charge
distribution in the cylindrical waveguide. Therefore, in the
remainder of this derivation, we will focus on the TE modes".

Neglecting TM modes is held by us as a misconception. If it was
correct, also undulator radiation produced by planar undulators
should exhibit only TE modes. In fact, as already discussed in
Section \ref{sec:wig}, the field from a planar undulator can be
seen as superposition of fields from two helical trajectories.
Without TM modes we could not even recover the free-space limit
for $R\longrightarrow \infty$ as studied in Section \ref{sec:wig}
for a planar undulator. In particular, we would loose both
horizontal polarization and azimuthal symmetry of the field.

Similarly, one can study the limit $R\longrightarrow \infty$ for
the helical case, following a procedure analogous to that in
Section \ref{sec:wig} for a planar undulator. In analogy with the
planar case, one can show that if TM modes are neglected, both
circular polarization and azimuthal symmetry of the field are
lost.

Neglecting TM modes in \cite{HART} is the result of a more general
misconception introduced when Maxwell's equations are solved with
the Green's function approach. Namely, the misconception is
introduced in the eigenmode decomposition of the four-current (see
paragraph 5.6.4 of \cite{HART}), where TM modes are not coupled
with the transverse current density as they must be.

\end{document}